\documentclass{jfm}
\usepackage{amsmath}
\usepackage{amssymb}
\usepackage{yhmath}
\usepackage[english]{babel}
\usepackage{epstopdf, epsfig}
\usepackage{comment}
\usepackage{color}
\usepackage{setspace}
\usepackage{siunitx}
\usepackage{booktabs}
\usepackage[normalem]{ulem}
\usepackage{comment}
\usepackage[makeroom]{cancel}
\usepackage{xspace}
\usepackage{float}
\usepackage{adjustbox}
\usepackage{makecell}
\usepackage{mathtools}
\usepackage{bbold}
\usepackage{siunitx}
\usepackage{xcolor}
\usepackage{graphicx}
\usepackage{amsmath}
\usepackage{upgreek}
\newcommand{\st}[1]{\textcolor{black}{#1}}

\usepackage[normalem]{ulem}

\renewcommand{\theequation}{\thesection.\arabic{equation}}
\usepackage{mathrsfs}

\DeclareMathSymbol{\Gammaup}{\mathalpha}{operators}{0}

\shorttitle{Drag reduction in superhydrophobic channels}
\shortauthor{S. D. Tomlinson and others}

\title{Drag reduction in surfactant-contaminated superhydrophobic channels at high P\'{e}clet numbers}

\author{Samuel D. Tomlinson\aff{1,\,2}
\corresp{\email{sdt50@cam.ac.uk}},
  Fr\'{e}d\'{e}ric Gibou\aff{3},
  Paolo Luzzatto-Fegiz\aff{3},
  Fernando Temprano-Coleto\aff{4,\,5},
  Oliver E. Jensen\aff{6}
  \and Julien R. Landel\aff{6,\,7}}

\affiliation{
\aff{1}Department of Engineering, University of Cambridge, Trumpington Street, Cambridge, CB2 1PZ, UK
\aff{2}Centre for Climate Repair, Centre for Mathematical Sciences, Wilberforce Road, Cambridge, CB3 0WA
\aff{3}Department of Mechanical Engineering, University of California, Santa Barbara, CA 93106, USA
\aff{4}Andlinger Center for Energy and the Environment, Princeton University, Princeton, NJ 08544, USA
\aff{5} Department of Mechanical and Aerospace Engineering, Princeton University, Princeton, NJ 08544, USA%
\aff{6}Department of Mathematics, University of Manchester, Oxford Road, Manchester, M13 9PL, UK
\aff{7}Universite Claude Bernard Lyon 1, Laboratoire de M\'ecanique des Fluides et d'Acoustique (LMFA), UMR5509, CNRS, Ecole Centrale de Lyon, INSA Lyon, 69622
Villeurbanne, France
}

\begin{document}

\maketitle

\begin{abstract}
Motivated by microfluidic applications, we investigate drag reduction in laminar pressure-driven flows in channels with streamwise-periodic superhydrophobic surfaces (SHSs) contaminated with soluble surfactant.  
We develop a model in the long-wave and weak-diffusion limit, where the streamwise SHS period is large compared to the channel height and the P\'{e}clet number is large. 
Using asymptotic and numerical techniques, we determine the influence of surfactant on drag reduction in terms of the relative strength of advection, diffusion, Marangoni effects and bulk--surface exchange.
In scenarios with strong exchange, the drag reduction exhibits a complex dependence on the thickness of the bulk-concentration boundary layer and surfactant strength.  
Strong Marangoni effects immobilise the interface through a linear surfactant distribution, whereas weak Marangoni effects yield a quasi-stagnant cap. 
The quasi-stagnant cap has an intricate structure with an upstream slip region, followed by intermediate inner regions, and a quasi-stagnant region that is mediated by weak bulk diffusion. 
The quasi-stagnant region differs from the immobile region of a classical stagnant cap, observed for instance in surfactant-laden air bubbles in water, by displaying weak slip. 
As exchange weakens, the bulk and interface decouple: the surfactant distribution is linear when the surfactant is strong, whilst it forms a classical stagnant cap when the surfactant is weak.
The asymptotic solutions offer closed-form predictions of drag reduction across much of the parameter space, providing practical utility and enhancing understanding of surfactant dynamics in flows over SHSs.  
\end{abstract}


\section{Introduction}

Superhydrophobic surfaces (SHSs) offer a promising avenue for decreasing drag in laminar and turbulent flows \citep{lee2016superhydrophobic, park2021superhydrophobic}.
Hydrophobic chemistry and microscopic topography on the SHSs entrap gas bubbles that lubricate the flow and decrease the wall-average shear stress compared to solid walls.
This design presents numerous opportunities for industrial and environmental applications, including enhanced cooling \citep{lam2015analysis} and reduced emissions  \citep{xu2020superhydrophobic}.
Consequently, SHSs have gathered significant attention in the academic literature \citep{schonecker2014influence, park2014superhydrophobic, turk2014turbulent, cheng2015numerical, seo2018effect, rastegari2019drag, kirk2020thermocapillary, landel2020theory, tomlinson2023model}.
However, recent investigations have highlighted potential challenges. 
Interfacial displacement and deflection \citep{ng2009stokes, teo2010flow}, gas-phase flow dynamics \citep{game2017physical} and heat- and surfactant-induced Marangoni stresses \citep{peaudecerf2017traces, kirk2020thermocapillary} have been shown to impede the anticipated drag reduction.
Consequently, in certain applications, SHSs may not outperform solid walls.  
\st{Motivated by these questions, this study investigates drag reduction in a surfactant-contaminated flow confined between a SHS and a solid wall. 
Specifically, we focus on the computationally challenging regime of weak bulk molecular diffusion, a regime which has remained largely unexplored in prior literature.}

Surfactants have been measured in both artificial \citep{hourlier2018extraction, temprano2023single} and natural \citep{pereira2018reduced, frossard2019properties} environments. 
When a liquid flow becomes contaminated with soluble surfactant, these chemical compounds are transported throughout the flow, adsorbing onto interfaces and desorbing into the bulk. 
The surfactant distribution within the bulk is regulated by the bulk P\'{e}clet number ($\Pen$), forming concentration boundary layers when $\Pen$ is large.
As the flow carries surfactant towards the downstream stagnation point of an interface, it accumulates, forming a concentration gradient. 
Depending on factors such as the flow properties, surfactant characteristics and geometry, the resulting adverse Marangoni force can cause the theoretically shear-free interface to behave like a no-slip (or partially no-slip) wall \citep{peaudecerf2017traces}. 
Experimental support for this mechanism can be found in studies by \citet{kim2012pressure}, \citet{bolognesi2014evidence}, \citet{schaffel2016local}, \citet{peaudecerf2017traces}, \citet{song2018effect} and \cite{temprano2023single}.

\st{\citet{peaudecerf2017traces} and \citet{landel2020theory} conducted numerical simulations of a two-dimensional (2D) channel flow featuring streamwise-periodic SHSs using COMSOL, employing surfactant properties that are characteristic of sodium dodecyl sulfate (SDS). 
\citet{temprano2023single} extended this methodology to encompass a three-dimensional (3D) channel flow featuring streamwise and spanwise-periodic SHSs, while \citet{sundin2022slip} investigated 2D channel flow laden with surfactants over streamwise-periodic liquid-infused surfaces (LISs).
In all of these configurations, numerical simulations of the Navier--Stokes equations and advection--diffusion equations for bulk and interfacial surfactant incurred substantial computational costs, particularly in the high-Péclet-number regime, where bulk-concentration boundary layers can be extremely thin. 
To address these challenges of high P\'{e}clet number values, we develop a numerical method based on Chebyshev collocation. 
This numerical method clusters nodes around the bulk-concentration boundary layers, reducing computational demand compared to numerical simulations.
We complement numerical simulations with asymptotic analysis that reveals the underlying flow structures and offers direct predictions of flow properties.}

\st{The distinction between strictly insoluble and very weakly soluble surfactant can be subtle.  
In the former case, there is no exchange between bulk and interface and the amount of surfactant present on an SHS plastron must be prescribed as an input parameter. 
In the latter case, adsorption and desorption processes control bulk and interface exchange and the amount of surfactant present on an SHS plastron is determined by coupled transport processes on the interface and in the bulk.  
In many practical applications, diffusive effects are weak, and might not be expected to influence the manner in which surfactant compromises drag reduction.  
However, the surfactant flux between bulk and interface is mediated by bulk diffusion, enabling it to play a central role in determining drag reduction, even at very high P\'eclet numbers.  
In such circumstances, weak bulk diffusion can be the cause of weak apparent solubility, even when adsorption and desorption processes are very fast.  
Furthermore, while bulk diffusion cannot be neglected, interfacial diffusion can be, but then adsorption and desorption must be accommodated by weak stretching or compression of the interface.  
Taking these factors together, we will see how structures such as the classical stagnant cap, of some prescribed size when surfactant is strictly insoluble, must be replaced by a more mobile structure that we will term a quasi-stagnant cap.}

Scaling theories have been developed to analyse the slip length over SHSs and LISs with streamwise-periodic gratings \citep{landel2020theory, sundin2022slip} and SHSs with streamwise and spanwise-periodic gratings \citep{temprano2023single}. 
These theories typically assume small surfactant concentrations and uniform shear stresses at the interface.
For common surfactants such as SDS, \citet{temprano2023single} demonstrated a significant slip length when the grating length exceeds a modified depletion length and a mobilisation length. 
The modified depletion length depends on the depletion length \citep{manikantan2020surfactant}, the height of the channel and the ratio of interfacial and bulk diffusivities. 
It characterizes the interfacial length above which interfacial diffusion is weak compared to bulk--interface exchanges. 
The mobilisation length depends on the surfactant concentration and surfactant properties through the Marangoni, Damk\"{o}hler and Biot numbers, and the SHS geometry \citep{temprano2023single}. 
The mobilisation length being generally larger than the modified depletion length, it is the key length scale above which an interface can display significant slip.
\citet{landel2020theory}, \citet{temprano2023single} and \citet{sundin2022slip} used numerical simulations to calibrate the empirical coefficients linked to the bulk-concentration boundary-layer thickness. 
\st{Unlike these prior studies, our asymptotic approach
to the problem bypasses the need for empirical coefficient fitting, extending the slip-length scalings
identified in previous studies \citep{landel2020theory,temprano2023single} and
uncovering new drag reduction regimes for surfactant-contaminated SHSs.}

\citet{tomlinson2023laminar} developed an asymptotic theory for laminar channel flow featuring streamwise and spanwise-periodic grooves and then investigated the impact of spatio--temporal fluctuations in surfactant concentration over the SHS \citep{tomlinson2024unsteady}.
Their theory assumed that the channel is long and bulk diffusion is strong enough to eliminate cross-channel concentration gradients. 
They derived a spatially-1D long-wave model that accommodated non-uniform shear stresses at the interface and enabled an analysis of the soluble stagnant-cap regime at low bulk P\'{e}clet numbers.
The numerical simulations performed at moderate bulk P\'{e}clet number by \citet{sundin2022slip} show adsorption at the upstream end of the stagnant cap, but they did not offer a scaling theory for this regime. 
Similar stagnant-cap scenarios, where there is no adsorption at the upstream end of the interface, were studied by \citet{baier2021influence} and \citet{mayer2022superhydrophobic} in the context of insoluble surfactant, with a linear and nonlinear equation of state, respectively. 
These studies were extended to include protrusion at the liquid--gas interface: \citet{baier2022shear} found recirculating interfacial flows for weak protrusions, which \citet{rodriguez2023superhydrophobic} extended to larger protrusions, deriving an effective slip length. 
\citet{crowdy2023fast} investigated a linear extensional flow between two fluids. 
Assuming that $\Pen$ is zero, they showed that solubility and strong exchange with the bulk can reduce the length of the stagnant cap and make the interface less immobile.  
\st{We will show how ``remobilisation'' due to strong bulk--surface exchange can also arise when ${Pe}$ is large, revealing the coupling between the bulk-concentration boundary layer and the underlying interface, which is particularly intricate when Marangoni effects are weak.}

\st{\citet{tomlinson2023laminar} delineated three key asymptotic regions of parameter space: one dominated by Marangoni effects, with minimal drag reduction; and others dominated by advection and diffusion, with significant drag reduction. 
They also identified when their 1D long-wave model first breaks down due to 2D effects.
In this paper, we employ a combination of asymptotic and numerical methods to investigate the impact of soluble surfactant on a laminar pressure-driven channel flow confined between a streamwise-periodic SHS and a solid wall.
We assume that the period of the SHS is significantly longer than the height of the channel and that the bulk and surface P\'{e}clet numbers are large.
\st{To address the numerical challenges of these high-P\'{e}clet-number flows, we develop asymptotic solutions for weak diffusion, which may offer insights into a broader range of surfactant-contaminated flows than flows over SHS. 
For example, numerical simulations of flows involving soluble surfactant-contaminated drops and bubbles exhibit velocity and surfactant distributions reminiscent of stagnant caps at large but finite $\Pen$ \citep{oguz1988effects, tasoglu2008effect}.
However, the structure of these stagnant caps has only been described in the insoluble limit \citep{harper2004, palaparthi2006theory}.
}
Additionally, we construct asymptotic expressions for practical quantities such as the slip length and drag reduction.
These asymptotic expressions not only aid in understanding the physics of surfactant-contaminated flows over SHSs but also facilitate the prediction of the slip length and drag reduction in experiments.}
    
In \S\ref{sec:formulation}, we formulate the 2D problem. 
We non-dimensionalise the fluid and surfactant equations in the long-wave limit and derive 1D and 2D long-wave models for bulk and interfacial surfactant behaviour.  
{\color{black}The calculations underpinning drag reduction predictions are laid out in appendices.}
In \S\ref{sec:results}, we present our main asymptotic and numerical findings. 
We analyse the drag reduction and compare the predicted flow and surfactant fields with results obtained from COMSOL simulations and experiments.
In \S\ref{discussion}, we illustrate the uses of our study and discuss the key asymptotic results that may be generalisable to broader surfactant-contaminated flows. 

\section{Formulation} \label{sec:formulation}

\subsection{Governing equations of the 2D model}

\begin{figure}
    \centering    \includegraphics[width=.9\textwidth]{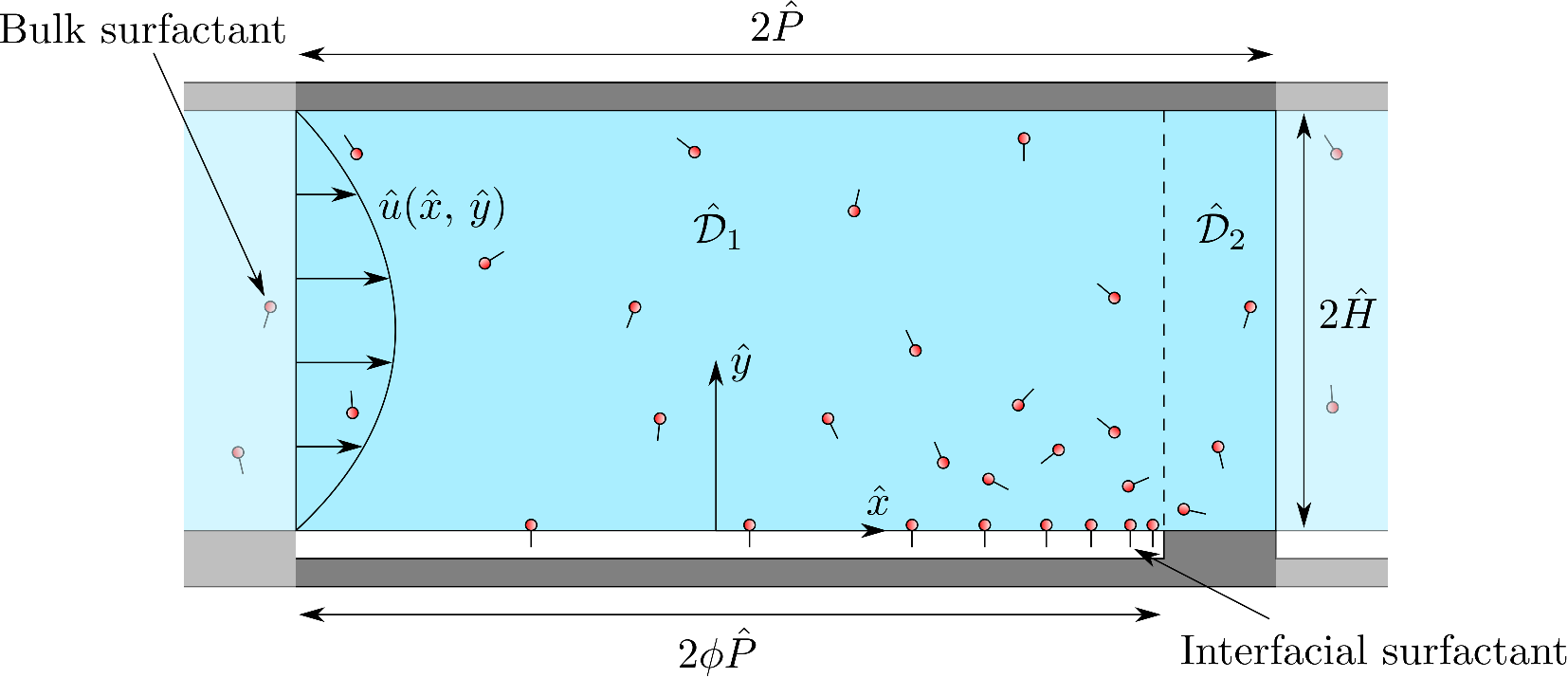}
    \caption{A 2D laminar pressure-driven channel flow transporting soluble surfactant confined between a streamwise-periodic SHS and solid wall. 
    We position the origin of the Cartesian coordinate system, $(\hat{x}, \, \hat{y})$, at the centre of the liquid--gas interface. 
    Each periodic cell has channel height $2\hat{H}$ and period length $2\hat{P}$.
    The SHS, characterised by gas fraction $\phi$, has an interface region of length $2\phi \hat{P}$ and a solid region of length $2(1 - \phi)\hat{P}$; the area above these regions defines subdomains $\hat{\mathcal{D}}_1$ and $\hat{\mathcal{D}}_2$ respectively, as specified in \eqref{eq:dimensional_domain}. 
    }
    \label{fig:my_label}
\end{figure}

We explore a steady 2D laminar pressure-driven channel flow contaminated with soluble surfactant, confined between a streamwise-periodic SHS and a solid wall (see figure \ref{fig:my_label}).
The streamwise and wall-normal directions are represented by $\hat{x}$- and $\hat{y}$-coordinates respectively, where hats indicate dimensional quantities.
Assuming the fluid to be incompressible and Newtonian, we define the velocity field $\hat{\boldsymbol{u}}=(\hat{u}(\hat{x},\,\hat{y}),\,\hat{v}(\hat{x},\,\hat{y}))$, pressure field $\hat{p}=\hat{p}(\hat{x},\,\hat{y})$, bulk surfactant concentration $\hat{c}=\hat{c}(\hat{x},\,\hat{y})$ and interfacial surfactant concentration $\hat{\Gamma}=\hat{\Gamma}(\hat{x})$.
Due to the periodicity of the SHS in the streamwise direction, our analysis focuses on a single periodic cell with dimensions $2\hat{P}$ in length and $2\hat{H}$ in height.
At $\hat{y}=0$, there is a solid ridge of length $2(1-\phi)\hat{P}$ and a liquid-gas interface, or plastron, of length $2\phi\hat{P}$, where $\phi$ is the gas fraction.
The interface is assumed to be flat.
There is a solid boundary at $\hat{y}= 2\hat{H}$. 
The periodic domain is partitioned into two subdomains,
\begin{subequations} 
\label{eq:dimensional_domain}
\begin{align}
    \hat{\mathcal{D}}_1 &= \{\hat{x}\in [-\phi\hat{P},\, \phi\hat{P}]\} \times \{\hat{y}\in [0, \,2 \hat{H}]\}, \\
    \hat{\mathcal{D}}_2 &= \{\hat{x}\in [\phi \hat{P}, \, (2 - \phi)\hat{P}]\} \times \{\hat{y}\in [0,\, 2 \hat{H}]\},
\end{align}
\end{subequations}
as illustrated in figure \ref{fig:my_label}.

A comprehensive discussion of the equations governing fluid and surfactant behaviour in a 3D geometry featuring streamwise- and spanwise-periodic SHSs was presented in \citet{tomlinson2023laminar}.
Here, we provide a succinct overview of the model for a streamwise-periodic SHS in 2D.  Within domains $\hat{\mathcal{D}}_1$ and $\hat{\mathcal{D}}_2$, we have Stokes equations with an advection--diffusion equation for bulk surfactant
\refstepcounter{equation} \label{eq:dimensional_equations}
\begin{equation} 
	\hat{\boldsymbol{\nabla}} \cdot \hat{\boldsymbol{u}} = 0, \quad \hat{\mu} \hat{\nabla}^2 \hat{\boldsymbol{u}}- \hat{\boldsymbol{\nabla}} \hat{p} = \boldsymbol{0}, \quad
	\hat{D} \hat{\nabla}^2 \hat{c} - \hat{\boldsymbol{u}}\cdot \hat{\boldsymbol{\nabla}} \hat{c} = 0, \tag{\theequation\textit{a--c}}
\end{equation}
where $\hat{\mu}$ is the dynamic viscosity and $\hat{D}$ is the bulk diffusivity.  Linearising the equation of state, we assume $\hat{\sigma}_{\hat{x}} = - \hat{A}\hat{\Gamma}_{\hat{x}}$, where $\hat{\sigma}$ is the surface tension and $\hat{A}$ is the surface activity.
For $\hat{x} \in [-\phi \hat{P}, \, \phi \hat{P}]$ and $\hat{y}=0$, we impose the tangential stress balance, no penetration, linear adsorption-desorption kinetics and an advection--diffusion equation for interfacial surfactant  
\refstepcounter{equation} \label{eq:dimensional_interface_bcs}
\begin{multline} 
	\hat{\mu} \hat{u}_{\hat{y}} - \hat{A} \hat{\Gamma}_{\hat{x}} = 0, \quad \hat{v}=0, \quad \hat{D} \hat{c}_{\hat{y}} - \hat{K}_a \hat{c} + \hat{K}_d \hat{\Gamma} =0, \\ \quad \hat{D}_I \hat{\Gamma}_{\hat{x}\hat{x}} + \hat{K}_a \hat{c} - \hat{K}_d \hat{\Gamma} -(\hat{u} \hat{\Gamma})_{\hat{x}} = 0,
    \tag{\theequation\textit{a--d}}
\end{multline}
where $\hat{D}_I$ is the interfacial diffusivity, $\hat{K}_a$ is the adsorption rate and $\hat{K}_d$ is the desorption rate. 
At $\hat{x} = \pm \phi \hat{P}$ and $\hat{y}=0$, there is no interfacial flux of surfactant,
\begin{equation} 
\label{eq:dimensional_interface_bnd_bcs}
    \hat{u}\hat{\Gamma} - \hat{D}_I \hat{\Gamma}_{\hat{x}} = 0.
\end{equation}
For $\hat{x} \in [\phi \hat{P}, \, (2-\phi)\hat{P}]$ and $\hat{y}=0$, and likewise for $\hat{x} \in [-\phi \hat{P}, \, (2 - \phi)\hat{P}]$ and $\hat{y}=2\hat{H}$, we impose no slip, no penetration and no bulk flux of surfactant
\refstepcounter{equation} \label{eq:dimensional_solid_bcs} 
\begin{equation} 
	\hat{u} = 0, \quad \hat{v} = 0, \quad \hat{c}_{\hat{y}} =0.
	\tag{\theequation\textit{a--c}}
\end{equation}
For $\hat{\boldsymbol{q}} \equiv (\hat{\boldsymbol{u}}, \,\hat{p}_{\hat{x}},\,\hat{c})$, continuity and periodicity conditions on bulk quantities are given by
\begin{equation} 
\label{eq:dimensional_periodicity} 
    \hat{\boldsymbol{q}}((\phi\hat{P})^-,\,\hat{y}) = \hat{\boldsymbol{q}}((\phi\hat{P})^+,\,\hat{y}), \quad \hat{\boldsymbol{q}}(-\phi\hat{P},\,\hat{y}) = \hat{\boldsymbol{q}}((2- \phi)\hat{P},\,\hat{y}),
\end{equation}
where the superscript $-$ ($+$) indicates evaluation in $\hat{\mathcal{D}}_1$ ($\hat{\mathcal{D}}_2$).  
The pressure field is continuous across $x=\pm \phi \hat{P}$ but is not periodic.  
Within $\hat{\mathcal{D}}_1$, the bulk surfactant equations and boundary conditions (\ref{eq:dimensional_equations}\textit{c}, \ref{eq:dimensional_interface_bcs}\textit{c}, \ref{eq:dimensional_solid_bcs}\textit{c}) can be combined to give
\begin{equation} 
\label{eq:dim_cons_1a}
    \frac{\text{d}}{\text{d}\hat{x}} \int_{0}^{2 \hat{H}} (\hat{u}\hat{c} - \hat{D} \hat{c}_{\hat{x}}) \, \text{d}\hat{y}  - (\hat{K}_d \hat{\Gamma} - \hat{K}_a \hat{c}\st{(\hat{x},\, 0)}) = 0,
\end{equation}
which, combined with the interfacial surfactant equation (\ref{eq:dimensional_interface_bcs}\textit{d}) leads to an expression for the total flux of surfactant, $\hat{K}$, which must be uniform
\begin{equation} 
\label{eq:dim_cons_1b}
    \int_{0}^{2 \hat{H}} (\hat{u}\hat{c} - \hat{D} \hat{c}_{\hat{x}}) \, \text{d}\hat{y} + (\hat{u}\st{(\hat{x},\, 0)}\hat{\Gamma} - \hat{D}_I \hat{\Gamma}_{\hat{x}}) = \hat{K}.
\end{equation} 
Integrating the interfacial surfactant equation (\ref{eq:dimensional_interface_bcs}\textit{d}) over the plastron and utilizing the no-flux condition \eqref{eq:dimensional_interface_bnd_bcs} yields a constraint on net adsorption and desorption
\begin{equation} 
\label{eq:dim_cons_1c}
    \int_{ - \phi \hat{P}}^{\phi \hat{P}} (\hat{K}_d \hat{\Gamma} - \hat{K}_a \hat{c}\st{(\hat{x},\, 0)}) \, \text{d}\hat{x} = 0.
\end{equation}
Similarly, within $\hat{\mathcal{D}}_2$, (\ref{eq:dimensional_interface_bcs}\textit{c}, \ref{eq:dimensional_solid_bcs}\textit{c}) can be combined to give
\begin{equation} 
\label{eq:dim_cons_2}
    \int_{0}^{2 \hat{H}} (\hat{u}\hat{c} - \hat{D} \hat{c}_{\hat{x}}) \, \text{d}\hat{y} = \hat{K}. 
\end{equation}
From (\ref{eq:dimensional_equations}$a$, \ref{eq:dimensional_interface_bcs}$b$, \ref{eq:dimensional_solid_bcs}$b$), it follows that the volume flux of fluid in $\hat{\mathcal{D}}_1$ and $\hat{\mathcal{D}}_2$, $\hat{Q}$, is also uniform, where
\begin{equation} 
\label{eq:dimensional_velocity_flux}
    \hat{Q} = \int_{0}^{2 \hat{H}} \hat{u} \, \text{d}\hat{y}.
\end{equation}
The fluxes $\hat{Q}$ and $\hat{K}$ are prescribed in this model.

The flow is driven in the streamwise direction by a cross-channel-averaged pressure drop per period, $\Delta \hat{p} \equiv \langle \hat{p}\rangle(-\phi \hat{P}) - \langle \hat{p}\rangle((2-\phi)\hat{P}) > 0$, where $\langle \cdot \rangle \equiv \int_{0}^{2 \hat{H}} \cdot \, \text{d}\hat{y} / (2\hat{H})$.
We define the normalised drag reduction
\begin{equation} 
\label{eq:dimensional_drag}
{DR} = \frac{\Delta \hat{p}_I - \Delta \hat{p}}{\Delta \hat{p}_I - \Delta \hat{p}_U},
\end{equation}
where $\Delta \hat{p} = \Delta \hat{p}_I$ when the interface is immobilised due to surfactant (${DR}=0$) and $\Delta \hat{p} = \Delta \hat{p}_U$ when the interface is unaffected by surfactant (${DR}=1$).
We aim to determine the dependence of ${DR}$ on the dimensional parameters of the problem ($\phi$, $\hat{P}$, $\hat{H}$, $\hat{\mu}$, $\hat{D}$, $\hat{A}$, $\hat{D}_I$, $\hat{K}_a$, $\hat{K}_d$, $\hat{K}$ and $\hat{Q}$) in the singular high-P\'{e}clet-number regime.

\subsection{Non-dimensionalisation} \label{subsec:nd}

We non-dimensionalise the governing equations \eqref{eq:dimensional_domain}--\eqref{eq:dimensional_drag} using $\hat{Q}/\hat{H}$ for the velocity scale, $\hat{K}/\hat{Q}$ for the bulk concentration scale and $\hat{K}_a \hat{K}/(\hat{K}_d\hat{Q})$ for the interfacial concentration scale. 
We write
\refstepcounter{equation} \label{eq:nondimensionalisation}
\begin{multline}  
    x = \frac{\hat{x}}{\hat{P}}, \quad  y = \frac{\hat{y}}{\epsilon \hat{P}}, \quad u = \frac{\hat{u}}{\hat{Q}/\hat{H}}, \quad v = \frac{\hat{v}}{\epsilon \hat{Q}/\hat{H}}, \\ p = \frac{\hat{p}}{\hat{\mu} \hat{Q}/(\epsilon \hat{H}^2)}, \quad c = \frac{\hat{c}}{\hat{K}/\hat{Q}}, \quad \Gamma = \frac{\hat{\Gamma}}{\hat{K}_a\hat{K}/(\hat{K}_d\hat{Q})},
    \tag{\theequation\textit{a--g}}
\end{multline}
where $\epsilon \equiv \hat{H}/ \hat{P}$ is the slenderness parameter. 
The subdomains \eqref{eq:dimensional_domain} become
\begin{subequations} 
    \begin{align}
    \mathcal{D}_1 &= \{x\in [-\phi, \,\phi]\} \times \{y\in [0, \,2]\}, \\ 
    \mathcal{D}_2 &= \{x\in [\phi, \,2 - \phi]\} \times \{y\in [0,\, 2]\}.
    \end{align}
\end{subequations}
In $\mathcal{D}_1$ and $\mathcal{D}_2$, the bulk equations \eqref{eq:dimensional_equations} become
\refstepcounter{equation} \label{eq:nondimensional_equations} 
\begin{multline} 
	u_x + v_{y} = 0, \quad \epsilon^2 u_{xx} + u_{yy} - p_x = 0, \\ \epsilon^4 v_{xx} + \epsilon^2 v_{yy} - p_y = 0, \quad
    (\epsilon^2 c_{xx} + c_{yy})/\Pen - \epsilon u c_x - \epsilon v c_y = 0,
    \tag{\theequation\textit{a--d}}
\end{multline}
with $\Pen = \hat{Q}/\hat{D}$ the bulk P\'{e}clet number. 
For $x\in[-\phi, \, \phi]$ and $y=0$, the interface conditions \eqref{eq:dimensional_interface_bcs} give
\refstepcounter{equation}  \label{eq:nondimensional_interface_bcs}
\begin{multline} 
	u_y - \epsilon \Ma \Gamma_{x} = 0, \quad v =0, \quad c_y - \Da (c -\Gamma)  =0, \\ \quad  \epsilon^2 \Gamma_{xx}/\Pen_I + \Bi ( c - \Gamma)  - \epsilon(u \Gamma)_{x} = 0,
    \tag{\theequation\textit{a--d}}
\end{multline}
with $\Ma = \hat{A}\hat{K}_a \hat{K}\hat{H}/(\hat{\mu}\hat{K}_d\hat{Q}^2)$ the Marangoni number, $\Da= \hat{K}_a \hat{H}/\hat{D}$ the Damk\"{o}hler number, $\Pen_I= \hat{Q}/\hat{D}_I$ the interfacial P\'{e}clet number and $\Bi = \hat{K}_d \hat{H}^2/\hat{Q}$ the Biot number. 
At $x = \pm \phi$ and $y=0$, the no-flux condition \eqref{eq:dimensional_interface_bnd_bcs} becomes
\begin{equation} 
\label{eq:nondimensional_noflux}
    u \Gamma - \epsilon \Gamma_{x}/\Pen_I = 0.
\end{equation}
For $x\in[\phi, \, 2 - \phi]$ and $y=0$ (and for $x\in[-\phi, \, 2-\phi]$ and $y=2$), the solid boundary conditions \eqref{eq:dimensional_solid_bcs} give
\refstepcounter{equation} \label{eq:nondimensional_solid_bcs}
\begin{equation} 
	u = 0, \quad v = 0, \quad \ c_{y} =0.
    \tag{\theequation\textit{a--d}}
\end{equation}
For $\boldsymbol{q} = (\boldsymbol{u}, \, p_x, \,c)$, the matching conditions \eqref{eq:dimensional_periodicity} become
\refstepcounter{equation} \label{eq:nondimensional_periodicity}
\begin{equation} 
    \boldsymbol{q}(\phi^-,\,y) = \boldsymbol{q}(\phi^+,\,y), \quad \boldsymbol{q}(-\phi,\,y) = \boldsymbol{q}(2- \phi,\,y).
    \tag{\theequation\textit{a,\,b}}
\end{equation}
For $x\in[-\phi, \, 2 - \phi]$, the surfactant-flux conditions \eqref{eq:dim_cons_1a}--\eqref{eq:dim_cons_2} yield
\begin{subequations} 
\label{eq:nd_fluxes_d1} 
\begin{align}
    \frac{\text{d}}{\text{d}x} \int_{0}^{2} \left( u c - \frac{\epsilon c_x}{\Pen} \right) \, \text{d} y  - \frac{\Da}{\epsilon \Pen}(\Gamma - c\st{(x,\, 0)}) &= 0 \quad \text{in} \quad \mathcal{D}_1, \\ 
    \int_{0}^{2} \left( u c - \frac{\epsilon c_x}{\Pen}\right) \, \text{d}y + \frac{\Da}{\Bi \Pen} \left( u\st{(x,\, 0)} \Gamma - \frac{\epsilon \Gamma_{x}}{\Pen_I}\right) &= 1 \quad \text{in} \quad \mathcal{D}_1, \\
    \int_{ - \phi}^{\phi} ( \Gamma  -  c\st{(x,\, 0)} ) \, \text{d}x &= 0  \quad \text{in} \quad \mathcal{D}_1, \\
    \int_{0}^{2} \left(u c - \frac{\epsilon c_x}{\Pen} \right) \, \text{d}y &= 1  \quad \text{in} \quad \mathcal{D}_2, 
\end{align}
\end{subequations}
and in $\mathcal{D}_1$ and $\mathcal{D}_2$, the volume-flux condition \eqref{eq:dimensional_velocity_flux} gives
\begin{equation} 
\label{eq:nd_velocity_flux}
    \int_{0}^{2} u \, \text{d}y = 1.
\end{equation}
Finally, the drag reduction \eqref{eq:dimensional_drag} becomes
\begin{equation} 
\label{eq:nondimensional_drag}
{DR} = \frac{\Delta p_I - \Delta p}{\Delta p_I - \Delta p_U},
\end{equation}
where $\Delta {p} \equiv \langle p\rangle(-\phi) - \langle p\rangle(2-\phi)$ and $\langle \cdot \rangle \equiv \int_{0}^{2} \cdot \, \text{d} y / 2$. 

In the limit $\epsilon \ll 1$, we proceed to solve the leading-order boundary-value problem and flux constraints defined in \eqref{eq:nondimensional_equations}--\eqref{eq:nd_velocity_flux} for $u$, $v$, $p$, $c$ and $\Gamma$ using asymptotic and numerical methods. 
Subsequently, we evaluate the drag reduction \eqref{eq:nondimensional_drag} and investigate its dependence on the seven dimensionless groups ($\epsilon$, $\Pen$, $\Ma$, $\Pen_I$, $\Bi$, $\Da$ and $\phi$) that characterise the geometry, flow, liquid and surfactant.
Moreover, full solutions of \eqref{eq:nondimensional_equations}--\eqref{eq:nd_velocity_flux} obtained as COMSOL simulations are presented in \S\ref{sec:results} below.

\subsection{The long-wave limit of the 2D model at high P\'{e}clet numbers} \label{sec:model}

We investigate distinguished limits of \eqref{eq:nondimensional_equations}--\eqref{eq:nd_velocity_flux} to uncover different physical balances.
We assume that $1/\Pen = O(\epsilon)$, $1/\Pen_I = O(\epsilon)$, $\Bi = O(\epsilon)$, $\Da = O(1)$ and $\Ma = O(1/\epsilon)$ as $\epsilon \rightarrow 0$, denoting this as the ``2D long-wave model'' in which streamwise and cross-channel concentration gradients are comparable at leading order.
We rescale $1/\Pen = \epsilon / \mathscr{P}$, $1/\Pen_I = \epsilon /\mathscr{P}_I$, $\Bi = \epsilon \mathscr{B}$ and $\Ma = \mathscr{M}/\epsilon$, assuming that $\mathscr{P}$, $\mathscr{P}_I$, $\mathscr{B}$ and $\mathscr{M}$ remain $O(1)$ as $\epsilon \rightarrow 0$.
We substitute the expansions
\begin{equation} 
\label{eq:asymptotic_expansion_wd}
	(u,\,v, \,p,\, c,\, \Gamma) = (u_0,\,v_0, \,p_0,\, c_0, \,\Gamma_0) + \epsilon^2 (u_1,\,v_1, \,p_1,\, c_1, \,\Gamma_1) + ...
\end{equation}
into the governing equations \eqref{eq:nondimensional_equations}--\eqref{eq:nondimensional_drag} and take the leading-order approximation.

In $\mathcal{D}_1$ and $\mathcal{D}_2$, flow is driven by a streamwise pressure gradient, and wall-normal diffusion is comparable to advection.
The bulk equations \eqref{eq:nondimensional_equations} become
\refstepcounter{equation} \label{eq:nondimensional_equations_0_wd} 
\begin{equation} 
	u_{0x} + v_{0y} = 0, \quad u_{0yy} - p_{0x} = 0, \quad p_{0y} = 0, \quad
    c_{0yy}/\mathscr{P} - u_0 c_{0x} - v_0 c_{0y} = 0.
    \tag{\theequation\textit{a--d}}
\end{equation}
For $x\in[-\phi, \, \phi]$ and $y=0$, the interfacial flow is inhibited by the surfactant gradient and exchange is comparable to advection. 
The interface conditions \eqref{eq:nondimensional_interface_bcs} yield
\refstepcounter{equation}  \label{eq:nondimensional_interface_bcs_0_wd}
\begin{multline} 
	u_{0y} - \mathscr{M} \Gamma_{0x} = 0, \quad v_0 =0, \quad c_{0y} - \Da (c_0 -\Gamma_0) =0, \\ \mathscr{B} ( c_0 - \Gamma_0)  - (u_0 \Gamma_0)_{x} = 0.
    \tag{\theequation\textit{a--d}}
\end{multline}
For $x\in[\phi, \, 2 - \phi]$ and $y=0$ (and for $x\in[-\phi, \, 2-\phi]$ and $y=2$), the solid wall boundary conditions \eqref{eq:nondimensional_solid_bcs} become
\refstepcounter{equation} \label{eq:nondimensional_solid_bcs_0_wd}
\begin{equation} 
	u_0 = 0, \quad v_0 = 0, \quad \ c_{0y} =0.
    \tag{\theequation\textit{a--c}}
\end{equation}
Short 2D Stokes-flow regions arise at the junctions between $\mathcal{D}_1$ and $\mathcal{D}_2$. 
In the present long-wave theory, in which surface diffusion appears at the next order, we impose the no-flux condition \eqref{eq:nondimensional_noflux} at $x = \pm \phi$ and $y=0$,
\begin{equation} 
\label{eq:nondimensional_noflux_0_wd}
    u_0 \Gamma_0 = 0.
\end{equation}
Continuity of bulk concentration \eqref{eq:nondimensional_periodicity} between $\mathcal{D}_1$ and $\mathcal{D}_2$ requires
\refstepcounter{equation} \label{eq:nondimensional_periodicity_0_wd}
\begin{equation} 
    c_0(\phi^-,\,y) = c_0(\phi^+,\,y), \quad c_0(-\phi,\,y) = c_0(2- \phi,\,y).
    \tag{\theequation\textit{a,\,b}}
\end{equation}
In \S\ref{sec:results}, comparison with COMSOL simulations will allow us to investigate the effect of these Stokes-flow regions.
Streamwise surfactant transport in the bulk and interface is dominated by advection and exchange, so \eqref{eq:nd_fluxes_d1} becomes
\begin{subequations} 
\label{eq:nd_fluxes_d1_0_wd} 
\begin{align}
    \frac{\text{d}}{\text{d}x} \int_{0}^{2} u_0 c_0 \, \text{d} y  - \frac{Da}{\mathscr{P}}(\Gamma_0 - c_0\st{(x,\, 0)}) &= 0 \quad \text{in} \quad \mathcal{D}_1, \\ 
    \int_{0}^{2} u_0 c_0 \, \text{d}y  + \frac{\Da}{\mathscr{B}\mathscr{P}} \left( u_0\st{(x,\, 0)} \Gamma_0 \right) &= 1 \quad \text{in} \quad \mathcal{D}_1, \\ 
    \int_{ - \phi}^{\phi} ( \Gamma_0  -  c_0\st{(x,\, 0)} ) \, \text{d}x &= 0  \quad \text{in} \quad \mathcal{D}_1, \\
     \int_{0}^{2} u_0 c_0 \, \text{d}y &= 1 \quad \text{in} \quad \mathcal{D}_2. 
\end{align}
\end{subequations}  
In $\mathcal{D}_1$ and $\mathcal{D}_2$, the volume-flux condition \eqref{eq:nd_velocity_flux} gives
\begin{equation} 
\label{eq:nd_velocity_flux_0_wd}
    \int_{0}^{2} u_0 \, \text{d}y = 1.
\end{equation}
The leading-order drag reduction \eqref{eq:nondimensional_drag} is given by
\begin{equation} 
\label{eq:nondimensional_drag_0}
{DR}_0 = \frac{\Delta p_I - \Delta p_0}{\Delta p_I - \Delta p_U}.
\end{equation}

The leading-order pressure field simplifies to $p_0 = p_0(x)$ using the wall-normal equation (\ref{eq:nondimensional_equations_0_wd}\textit{c}).
The streamwise velocity field is driven by the pressure gradient $p_{0x}$ and is inhibited by the surfactant gradient $\Gamma_{0x}$.
Using linear superposition, we can write 
\refstepcounter{equation} \label{eq:u_def_wd}
\begin{equation} 
	u_0 = \tilde{U} p_{0x} + \mathscr{M} \bar{U}\Gamma_{0x} \quad \text{in} \quad \mathcal{D}_1, \quad u_0 = \breve{U} p_{0x} \quad \text{in} \quad \mathcal{D}_2,
	\tag{\theequation\textit{a,\,b}}
\end{equation}
where $\tilde{U} \equiv y^2/2 - 2$, $\bar{U} \equiv y - 2$ and $\breve{U} \equiv y^2/2 - y$ are velocity contributions satisfying the following boundary-value problems:
\refstepcounter{equation}  \label{eq:tildeu_bvp_wd}
\begin{equation} 
    \tilde{U}_{yy} = 1, \quad \text{subject to} \quad \tilde{U}_y(0)=0, \quad \tilde{U}(2)=0;
	\tag{\theequation\textit{a--c}}
\end{equation}
\refstepcounter{equation} \label{eq:baru_bvp_wd}
\begin{equation} 
\vspace{-.32cm}
    \bar{U}_{yy} = 0, \quad \text{subject to} \quad \bar{U}_y(0)= 1, \quad \bar{U}(2) = 0;
	\tag{\theequation\textit{a--c}}
\end{equation}
\refstepcounter{equation} \label{eq:hatu_bvp_wd} 
\begin{equation} 
    \breve{U}_{yy} = 1, \quad \text{subject to} \quad \breve{U}(0)= 0, \ \quad \breve{U}(2)=0.
	\tag{\theequation\textit{a--c}}
\end{equation}
Substituting \eqref{eq:u_def_wd} into \eqref{eq:nd_velocity_flux_0_wd}, the volume flux constraint gives
\refstepcounter{equation}  \label{eq:fluxes_wd}
\begin{equation} 
    \tilde{Q} p_{0x} + \mathscr{M} \bar{Q} \Gamma_{0x} =1 \quad \text{in} \quad \mathcal{D}_1, \quad \breve{Q} p_{0x} = 1 \quad \text{in} \quad \mathcal{D}_2,
	\tag{\theequation\textit{a,\,b}}
\end{equation}
where $\tilde{Q} \equiv -8/3$, $\bar{Q} \equiv -2$ and $\breve{Q} \equiv -2/3$. 
Using the continuity equation (\ref{eq:nondimensional_equations_0_wd}\textit{a}), the wall-normal velocity field is forced by $p_{0xx}$ and $\Gamma_{0xx}$, with velocity contributions that complement those given in \eqref{eq:u_def_wd}--\eqref{eq:baru_bvp_wd},
\refstepcounter{equation} \label{eq:v_def_wd}
\begin{equation} 
	v_0 = \tilde{V} p_{0xx} + \mathscr{M} \bar{V} \Gamma_{0xx} \quad \text{in} \quad \mathcal{D}_1, \quad v_0 = 0 \quad \text{in} \quad \mathcal{D}_2,
	\tag{\theequation\textit{a,\,b}}
\end{equation}
where $\tilde{V} \equiv -y^3/6 + 2 y$ and $\bar{V} \equiv -y^2/2 + 2 y$ (so that $\tilde{V}_{y} = -\tilde{U}$, $\tilde{V}(0)=0$, $\bar{V}_{y} = -\bar{U}$, $\bar{V}(0)= 0$).
The no-penetration condition (\ref{eq:nondimensional_solid_bcs_0_wd}\textit{b}) is satisfied, as $v_0(2) = \tilde{V}(2) p_{0xx} + \mathscr{M}\bar{V}(2)\Gamma_{0xx} = -\tilde{Q} p_{0xx} - \mathscr{M}\bar{Q}\Gamma_{0xx} = 0$, using the volume-flux condition (\ref{eq:fluxes_wd}\textit{a}).

\setlength{\tabcolsep}{0.75em}
\begin{table}
\resizebox{\columnwidth}{!}{%
\centering
    \begin{tabular}{c c c c}
    Coefficient & Name & Definition & Physical interpretation \\[6pt]
    $\alpha$ & Bulk diffusion coefficient & $\displaystyle \hspace{.2cm} \frac{1}{\epsilon \Pen} \equiv \frac{\hat{D}\hat{P}}{\hat{Q}\hat{H}}$ & $ \frac{\textmd{Bulk diffusion}}{\textmd{Bulk advection}}$ \\[10pt]
    $\beta$ & Partition coefficient & $\displaystyle \hspace{-0cm} \frac{\Da}{\Bi\Pen} \equiv \frac{\hat{K}_a}{\hat{H}\hat{K}_d}$& $\frac{\textmd{Adsorption}}{\textmd{Desorption}}$ \\[10pt]
    $\gamma$ & Surfactant strength & $\displaystyle \hspace{.25cm} \frac{\epsilon \Ma\Da}{\Bi\Pen} \equiv \frac{\hat{A}\hat{H}\hat{K}\hat{K}_a^2}{\hat{\mu} \hat{K}_d^2 \hat{P} \hat{Q}^2}$ & $\frac{\textmd{Marangoni effects}}{\textmd{Interfacial advection}}$ \\[10pt]
    $\delta$ & Interfacial diffusion coefficient & $\displaystyle \hspace{-.3cm} \frac{\Da}{\epsilon \Bi\Pen\Pen_I}\equiv \frac{\hat{D}_I \hat{K}_a \hat{P}}{\hat{H}^2\hat{K}_d\hat{Q}}$ & $ \frac{\textmd{Interfacial diffusion}}{\textmd{Interfacial advection}}$ \\[10pt]
    $\nu$ & Exchange strength & $\displaystyle \hspace{.2cm} \frac{\Da}{\epsilon \Pen}\equiv \frac{\hat{K}_a \hat{P}}{\hat{Q}}$ & $\frac{\textmd{Adsorption}}{\textmd{Bulk advection}}$ \\[10pt]
    \end{tabular}%
    }
    \caption{A summary of the transport coefficients in the 2D long-wave model, \eqref{eq:c_bvp_1_wd}--\eqref{eq:c_bvp_6_wda}, with their definition and physical interpretation. 
    The transport coefficients $\alpha$, $\beta$, $\gamma$, $\delta$ and $\nu$ are written in terms of $\Pen$, $\Da$, $\Bi$, $\Ma$, $\Pen_I$ and $\epsilon$ defined in \S\ref{subsec:nd}.
    Compared to the 3D transport coefficients in \citet{tomlinson2023laminar}, $\alpha_\textsl{3D}\propto \epsilon^2\alpha$, $\beta_\textsl{3D}\propto \beta$, $\gamma_\textsl{3D}\propto \gamma$, $\delta_\textsl{3D}\propto \epsilon^2\delta$ and $\nu_\textsl{3D}\propto \epsilon^2\nu$.}
    \label{tab:1}
\end{table}

We substitute (\ref{eq:u_def_wd},\,\ref{eq:v_def_wd}) into the bulk and interfacial surfactant equations \eqref{eq:nondimensional_equations_0_wd}--\eqref{eq:nd_fluxes_d1_0_wd} to derive the 2D long-wave model. 
To facilitate its numerical computation, we retain $O(\epsilon^2)$ bulk and interfacial streamwise diffusion operators to make the problem more elliptic, simplifying the continuity conditions and smoothing the flow between $\mathcal{D}_1$ and $\mathcal{D}_2$, later seeking the limit in which these effects are weak.
The bulk surfactant equation (\ref{eq:nondimensional_equations_0_wd}\textit{c}) gives 
\begin{subequations} 
\label{eq:c_bvp_1_wd} 
\begin{align}
	\alpha (\epsilon^2 c_{0xx} + c_{0yy}) - \bigg(\frac{3}{4} - \frac{3y^2}{16}\bigg) c_{0x} \nonumber \hspace{4.5cm} \\ - \frac{\gamma}{\beta} \bigg(-\frac{1}{2} + y - \frac{3y^2}{8}\bigg) \Gamma_{0x} c_{0x} - \frac{\gamma}{\beta} \bigg(\frac{y}{2} - \frac{y^2}{2} + \frac{y^3}{8}\bigg) \Gamma_{0xx} c_{0y} &= 0 \quad \text{in} \quad \mathcal{D}_1,  \\
    \alpha (\epsilon^2 c_{0xx} + c_{0yy}) - \bigg(\frac{3y}{2} - \frac{3y^2}{4}\bigg) c_{0x} &= 0 \quad \text{in} \quad \mathcal{D}_2, 
\end{align}
\end{subequations}
where $\alpha$, $\beta$ and $\gamma$ are given in table \ref{tab:1}.
This table defines the five primary parameter combinations that we use below and gives their physical interpretation.
For $x\in[-\phi, \, \phi]$ and $y=0$, the boundary condition specifying continuity of surfactant flux and the interfacial surfactant equation (\ref{eq:nondimensional_interface_bcs_0_wd}\textit{c},\,\textit{d}) become
\refstepcounter{equation} 
\label{eq:c_bvp_2_wd}
\begin{equation} 
    c_{0y} - \frac{\nu}{\alpha}(c_{0} - \Gamma_0) = 0, \quad \nu ( c_{0} - \Gamma_0)  - \tfrac{3}{4}\beta \Gamma_{0x} + \tfrac{1}{2}\gamma (\Gamma_{0x}\Gamma_0)_x + \epsilon^2 \delta \Gamma_{0xx} = 0, \tag{\theequation\textit{a,\,b}}
\end{equation} 
where $\nu$ and $\delta$ are given in table \ref{tab:1}. 
At $x= \pm \phi$ and $y=0$, no flux of interfacial surfactant \eqref{eq:nondimensional_noflux_0_wd} gives
\begin{equation} 
\label{eq:c_bvp_3_wd} 
    \tfrac{3}{4}\beta \Gamma_{0} - \tfrac{1}{2}\gamma \Gamma_{0x}\Gamma_0 - \epsilon^2 \delta \Gamma_{0x}  = 0.
\end{equation}
For $x\in[\phi, \, 2 - \phi]$ and $y=0$ (and for $x\in[-\phi, \, 2-\phi]$ and $y=2$), no flux of bulk surfactant (\ref{eq:nondimensional_solid_bcs_0_wd}\textit{c}) becomes
\begin{equation} 
\label{eq:c_bvp_4_wd} 
	c_{0y} =0.
\end{equation}
The surfactant-flux conditions (\ref{eq:nd_fluxes_d1_0_wd}\textit{a,\,b,\,d}) become
\begin{subequations} 
\label{eq:c_bvp_6_wda} 
\begin{align}
    \frac{\text{d}}{\text{d}x} \int_{0}^{2} \left(\bigg(\frac{3}{4} - \frac{3y^2}{16}\bigg) c_{0} + \frac{\gamma}{\beta} \bigg(-\frac{1}{2} + y - \frac{3y^2}{8}\bigg) \Gamma_{0x} c_{0} - \epsilon^2 \alpha c_{0x} \right)\text{d} y \hspace{-.3cm} \nonumber \\  - \nu(\Gamma_0 - c_0\st{(x,\, 0)}) &= 0   \quad \text{in} \quad \mathcal{D}_1, \\
    \int_{0}^{2} \left(\bigg(\frac{3}{4} - \frac{3y^2}{16}\bigg) c_{0} + \frac{\gamma}{\beta} \bigg(-\frac{1}{2} + y - \frac{3y^2}{8}\bigg) \Gamma_{0x} c_{0} - \epsilon^2 \alpha c_{0x}  \right)\text{d}y \hspace{-.3cm}  \nonumber \\ 
    + \frac{3}{4}\beta \Gamma_{0} - \frac{1}{2}\gamma \Gamma_{0x}\Gamma_0 - \epsilon^2 \delta \Gamma_{0x} &= 1   \quad \text{in} \quad \mathcal{D}_1, \\
    \int_{0}^{2} \left( \bigg(\frac{3y}{2} - \frac{3y^2}{4}\bigg) c_0  - \epsilon^2 \alpha c_{0x} \right) \, \text{d}y &= 1  \quad \text{in} \quad \mathcal{D}_2, 
\end{align}
\end{subequations}
and in $\mathcal{D}_1 \cup \mathcal{D}_2$, the volume-flux condition \eqref{eq:nd_velocity_flux_0_wd} becomes
\refstepcounter{equation} 
\label{eq:c_bvp_8_wd} 
\begin{equation} 
    \tilde{Q} p_{0x} + (\gamma / \beta) \bar{Q} \Gamma_{0x} = 1 \quad \text{in} \quad \mathcal{D}_1, \quad \breve{Q} p_{0x} = 1 \quad \text{in} \quad \mathcal{D}_2. \tag{\theequation\textit{a,\,b}}
\end{equation}
\st{To summarise, the seven-parameter Stokes-flow system \eqref{eq:nondimensional_equations}–\eqref{eq:nd_velocity_flux} involving five variables $(\hat{u}, \hat{v}, \, \hat{p}, \, \hat{c}, \, \hat{\Gamma})$ has been reduced to the seven-parameter long-wave system \eqref{eq:c_bvp_1_wd}–\eqref{eq:c_bvp_6_wda} involving two variables $(\hat{c}, \, \hat{\Gamma})$.
Further, by neglecting streamwise diffusion in the bulk, $-\epsilon^2 \alpha c_{0x}$, and at the interface, $-\epsilon^2 \delta \Gamma_{0x}$, the system reduces to a five-parameter system ($\alpha$, $\beta$, $\gamma$, $\nu$, $\phi$).  
This long-wave model disregards short (but passive) inner regions associated with the small parameters $\epsilon$ and $\delta$.}

The numerical solution of \eqref{eq:c_bvp_1_wd}--\eqref{eq:c_bvp_6_wda} (Appendix \ref{sec:numerical}) is generally less expensive (in terms of memory and runtime) than COMSOL simulations.  
The numerical method iteratively solves linearised bulk- and interfacial-concentration problems until convergence is achieved.  Convergence in the nonlinear problem can be slow, necessitating small tolerances.  
Accurate resolution of singularities at $x=\pm\phi$ is essential to achieving convergence; accordingly, we employ domain decomposition and Chebyshev collocation techniques that cluster nodes around $y=0,\,2$ and $x = -\phi,\, \phi,\,2-\phi$.
Once $c_0$ and $\Gamma_0$ are determined, $u_0$ and $v_0$ can be calculated using \eqref{eq:u_def_wd} and \eqref{eq:v_def_wd} respectively. 

Following \citet{tomlinson2023laminar}, we express the leading-order drag reduction (${DR}_0$) in terms of the transport parameters and geometry ($\beta$, $\gamma$ and $\phi$) as follows.
When the interface is unaffected by surfactant, $\Delta p_U = -2\phi/\tilde{Q} - 2(1-\phi)/\breve{Q}$.
When the interface is immobilised by surfactant, $\Delta p_I = -2/\breve{Q}$.
Between these two limits, $\Delta p_0 = -2\phi/\tilde{Q} - 2(1-\phi)/\breve{Q} + \gamma \bar{Q} \Delta \Gamma_0 / (\beta \tilde{Q})$, where $\Delta \Gamma_0 = \Gamma_0(\phi) - \Gamma_0(-\phi)$. 
Substituting these expressions into \eqref{eq:nondimensional_drag_0} gives the leading-order drag reduction as
\begin{equation} 
\label{eq:dr_def}
        {DR}_0 = 1 - \frac{\gamma \Delta \Gamma_0}{3\phi \beta}. 
\end{equation}
When evaluated using COMSOL, we write ${DR}_0 = {DR}_{NS}$.
An alternative measure used commonly in the SHS literature is the effective slip length, $\lambda_0$, which can be evaluated from ${DR}_0$ using \citep{tomlinson2023laminar}
\begin{equation} 
\label{eq:slip}
    \lambda_0 = \frac{{DR}_0(\Delta p_I - \Delta p_U)}{\Delta p_I[\Delta p_U {DR}_0 + \Delta p_I(1 - {DR}_0)]}.
\end{equation}
Before presenting the results of the 2D long-wave model, it is helpful to recall key features of the problem at low P\'eclet numbers, when a 1D description applies.

\subsection{The 1D long-wave model}
\label{subsec:Strong cross-channel diffusion and moderate exchange}

\begin{figure}
    \centering
    \includegraphics[width = .49\textwidth]{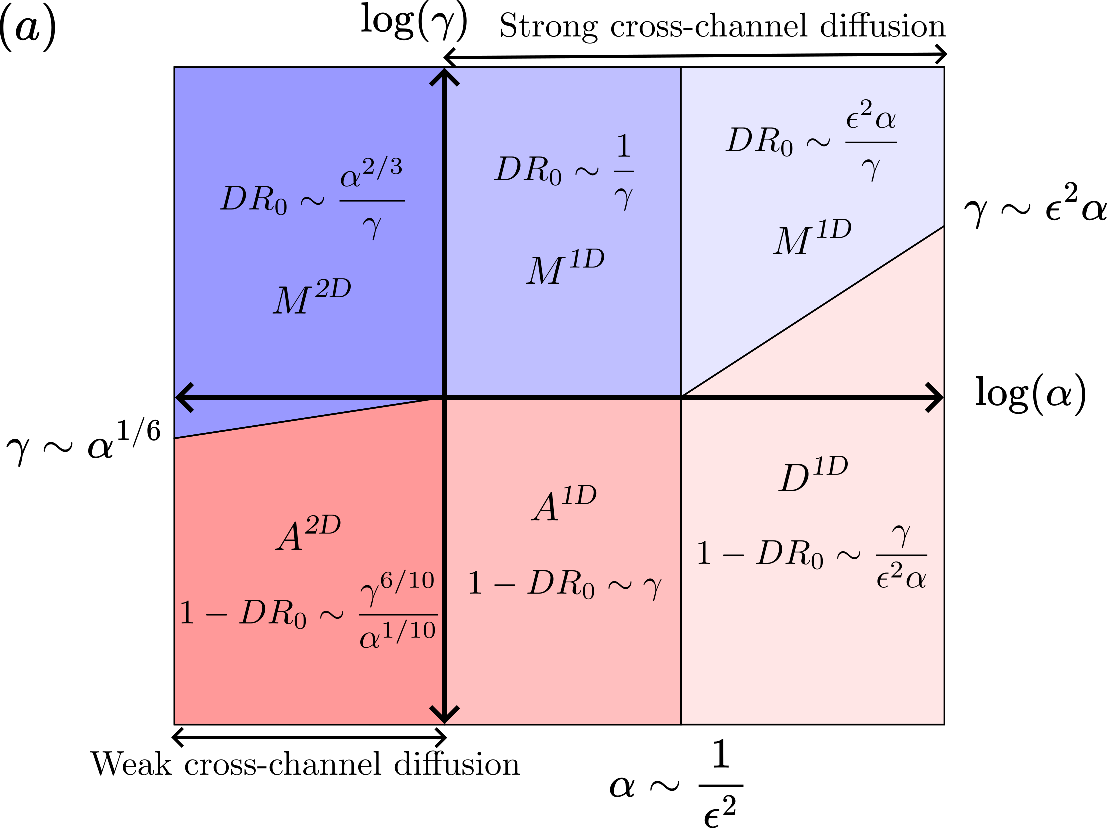} \includegraphics[width = .49\textwidth]{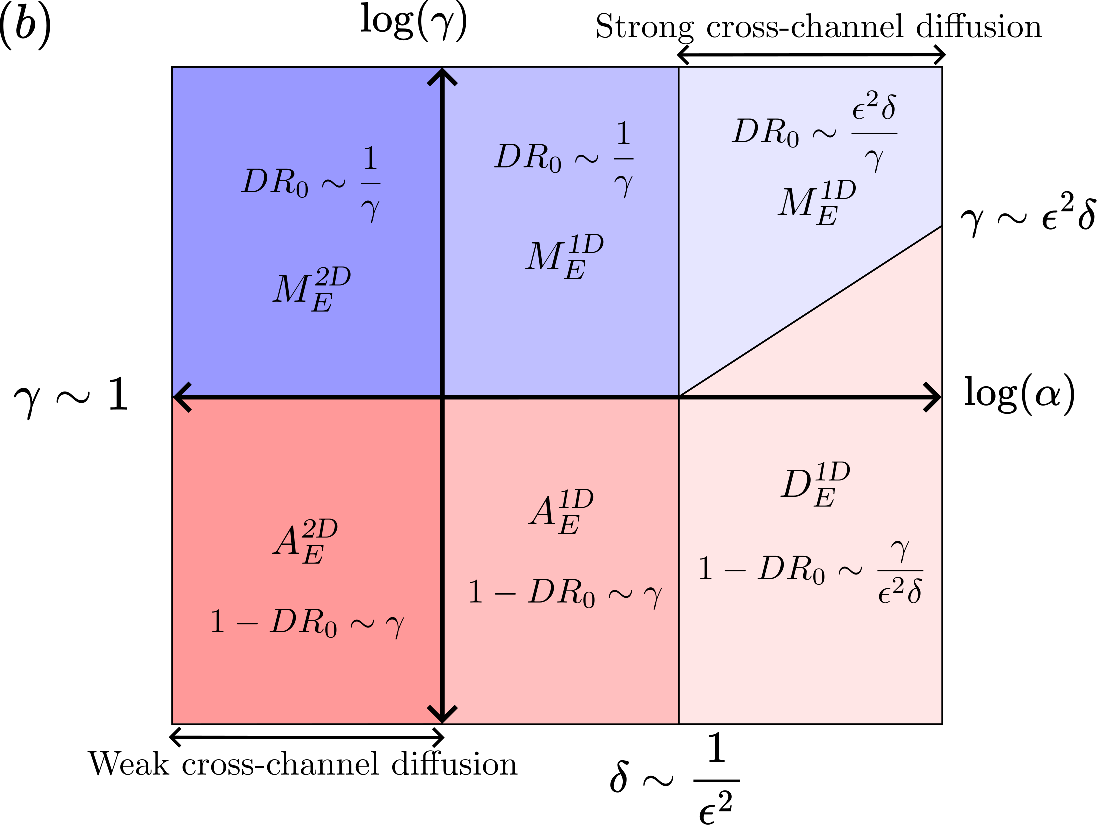} 
    \caption{Maps of ($\alpha$,\,$\gamma$)-parameter space for $\epsilon\ll 1$, using parameters given in table~\ref{tab:1}, showing the magnitude of drag reduction ($DR_0$) across different asymptotic regions.  Existing predictions from the 1D long-wave model of \cite{tomlinson2023laminar} are combined with predictions derived in Appendix~\ref{sec:asymptotic} from the 2D long-wave model for (\textit{a}) strong exchange and (\textit{b}) weak exchange.  The Marangoni-dominated region ($M$) incorporates sub-regions $M^\textsl{1D}$, $M^\textsl{1D}_E$, $M^\textsl{2D}$ and $M^\textsl{2D}_E$, for which ${DR}_0 \ll 1$; the advection-dominated region ($A$) incorporates sub-regions $A^\textsl{1D}$, $A^\textsl{1D}_E$, $A^\textsl{2D}$ and $A^\textsl{2D}_E$, for which $1-{DR}_0 \ll 1$; and the diffusion-dominated region ($D$) incorporates sub-regions $D^\textsl{1D}$ and $D^\textsl{1D}_E$, for which $1-{DR}_0 \ll 1$. 
    Here it is assumed that $\alpha \sim \delta$ and $\beta \sim 1$.
    }
    \label{fig:dr_shems}
\end{figure}

Figure \ref{fig:dr_shems} shows maps of $(\alpha,\,\gamma)$-parameter space, illustrating how the present 2D model relates to existing 1D results.  In the strong-cross-channel-diffusion limit, for which $\alpha \gg 1$ (with $\epsilon^2\alpha=O(1)$ and $\epsilon^2\delta=O(1)$), cross-channel concentration gradients are small.
Expanding the bulk and interfacial concentration fields
\begin{equation} 
    c_0 = \bar{c}_0(x) + \bar{c}_1(x,\,y)/\alpha + ..., \quad \Gamma_0 = \bar{\Gamma}_0(x) + \bar{\Gamma}_1(x)/\alpha + ...,
\end{equation}
\eqref{eq:c_bvp_1_wd}--\eqref{eq:c_bvp_6_wda} reduce to the ``1D long-wave model'', given by
\begin{subequations} 
\label{eq:c_bvp_6_sd} 
\begin{align}
    \bar{c}_{0x} - 2 \epsilon^2\alpha \bar{c}_{0xx} - \nu(\bar{\Gamma}_0 - \bar{c}_0) &= 0 \quad \text{in} \quad \mathcal{D}_1, \\
    \tfrac{3}{4} \beta \bar{\Gamma}_{0x} - \tfrac{1}{2} \gamma (\bar{\Gamma}_{0x} \bar{\Gamma}_0)_x - \epsilon^2\delta \bar{\Gamma}_{0xx}  + \nu (\bar{\Gamma}_0 - \bar{c}_0) &= 0  \quad \text{in} \quad \mathcal{D}_1, \\ 
    \bar{c}_0 - 2 \epsilon^2\alpha \bar{c}_{0x} +\tfrac{3}{4} \beta \bar{\Gamma}_0 - \tfrac{1}{2} \gamma \bar{\Gamma}_{0x} \bar{\Gamma}_0 - \epsilon^2\delta \bar{\Gamma}_{0x}  &= 1 \quad \text{in} \quad \mathcal{D}_1, \\
    \bar{c}_0 - 2 \epsilon^2\alpha \bar{c}_{0x}  &= 1 \quad \text{in} \quad \mathcal{D}_2.
\end{align}
\end{subequations}
The system in \eqref{eq:c_bvp_6_sd} is solved subject to continuity of bulk surfactant, continuity of bulk-surfactant flux and no-flux of interfacial surfactant
\begin{subequations} 
\label{eq:model3}
\begin{align}
        \bar{c}_0(\phi^-) = \bar{c}_0(\phi^+), \\ 
        \bar{c}_0(-\phi) = \bar{c}_0(2-\phi), \\
        \bar{c}_0(\phi^-) - 2\epsilon^2\alpha \bar{c}_{0x}(\phi^-) = \bar{c}_0(\phi^+) - 2\epsilon^2\alpha \bar{c}_{0x}(\phi^+), \\ 
        \bar{c}_0(-\phi) - 2\epsilon^2\alpha \bar{c}_{0x}(-\phi) = \bar{c}_0(2-\phi) - 2\epsilon^2\alpha \bar{c}_{0x}(2-\phi), \\
        \tfrac{3}{4}  \beta \bar{\Gamma}_0(\pm\phi) - \tfrac{1}{2} \gamma \bar{\Gamma}_0(\pm\phi) \bar{\Gamma}_{0x}(\pm\phi) - \epsilon^2\delta \bar{\Gamma}_{0x}(\pm\phi) = 0.
\end{align}
\end{subequations}
The system of ODEs in \eqref{eq:c_bvp_6_sd}--\eqref{eq:model3} can be solved numerically using the method outlined in Appendix A of Tomlinson \textit{et al.} (2023). 
The drag reduction can be evaluated using \eqref{eq:dr_def}.
As illustrated in figure \ref{fig:dr_shems}, the 1D long-wave model \eqref{eq:c_bvp_6_sd}--\eqref{eq:model3} exhibits multiple asymptotic regimes in the $(\alpha,\,\gamma)$-parameter space, which provides useful context for the results to follow.  {\color{black}We highlight regions dominated by Marangoni effects ($M$), interfacial advection ($A$) and interfacial diffusion ($D$); a subscript $E$ denotes cases in which $\Gamma_0(x)$ and $c_0(x,0)$ can be out of equilibrium.} 
We summarise key asymptotic results from \citet{tomlinson2023laminar} that apply in the strong- ($\bar{c}_0 \approx \bar{\Gamma}_0$, large $\nu$) and weak-exchange ($\bar{c}_0 \neq \bar{\Gamma}_0$, small $\nu$) limits in Appendix \ref{sec:1d_asym}, {\color{black}in regions labelled with superscript $\textsl{1D}$.} 
We now proceed to investigate the emergence of 2D flow structures as bulk diffusion weakens, i.e. for $\alpha$ small ($\ll 1$) or at intermediate values $\sim 1$. 

\section{Results} \label{sec:results}

In \S\ref{sec:strong exchange}, we investigate the leading-order drag reduction (${DR}_0$), interfacial and bulk surfactant concentrations ($\Gamma_0$ and $c_0$), as well as the streamwise and wall-normal velocity components ($u_0$ and $v_0$), for strong bulk--surface exchange ($\nu = 100$). 
We vary the bulk diffusion $(\alpha)$ and surfactant strength ($\gamma$), exploring both strong- and weak-cross-channel-diffusion regimes by solving the 2D long-wave model \eqref{eq:c_bvp_1_wd}--\eqref{eq:c_bvp_6_wda}.
We focus primarily on the weak-cross-channel-diffusion regime ($\alpha = \delta \ll 1$), as the strong-cross-channel-diffusion regime (\S\ref{subsec:Strong cross-channel diffusion and moderate exchange}) was addressed in \citet{tomlinson2023laminar}. 
For simplicity, we constrain the partition coefficient ($\beta$) and interfacial diffusion ($\delta$), assuming that $\alpha = \delta$, $\beta = 1$ and $\phi=0.5$ throughout.  
However, asymptotic solutions are derived in Appendix \ref{sec:asymptotic} for generic variables $\alpha$, $\beta$, $\delta$ and $\phi$, which we use to validate numerical simulations and identify dominant physical mechanisms. 
In \S\ref{sec:weak exchange}, we consider the weak-exchange limit ($\nu = 0.01$).  
As illustrated in figure \ref{fig:dr_shems}, in both strong- and weak-exchange limits, we identify regions of parameter space dominated by Marangoni effects ($M$), advection ($A$) and diffusion ($D$).
Additionally, we compare numerical solutions of the 2D long-wave model (\ref{eq:c_bvp_1_wd}--\ref{eq:c_bvp_6_wda}) with numerical solutions of the Stokes and advection--diffusion equations (\ref{eq:nondimensional_equations}--\ref{eq:nd_velocity_flux}) in the weak-cross-channel-diffusion limit.

\subsection{Strong exchange} \label{sec:strong exchange}

\begin{figure}
\centering\includegraphics[width=.49\textwidth]{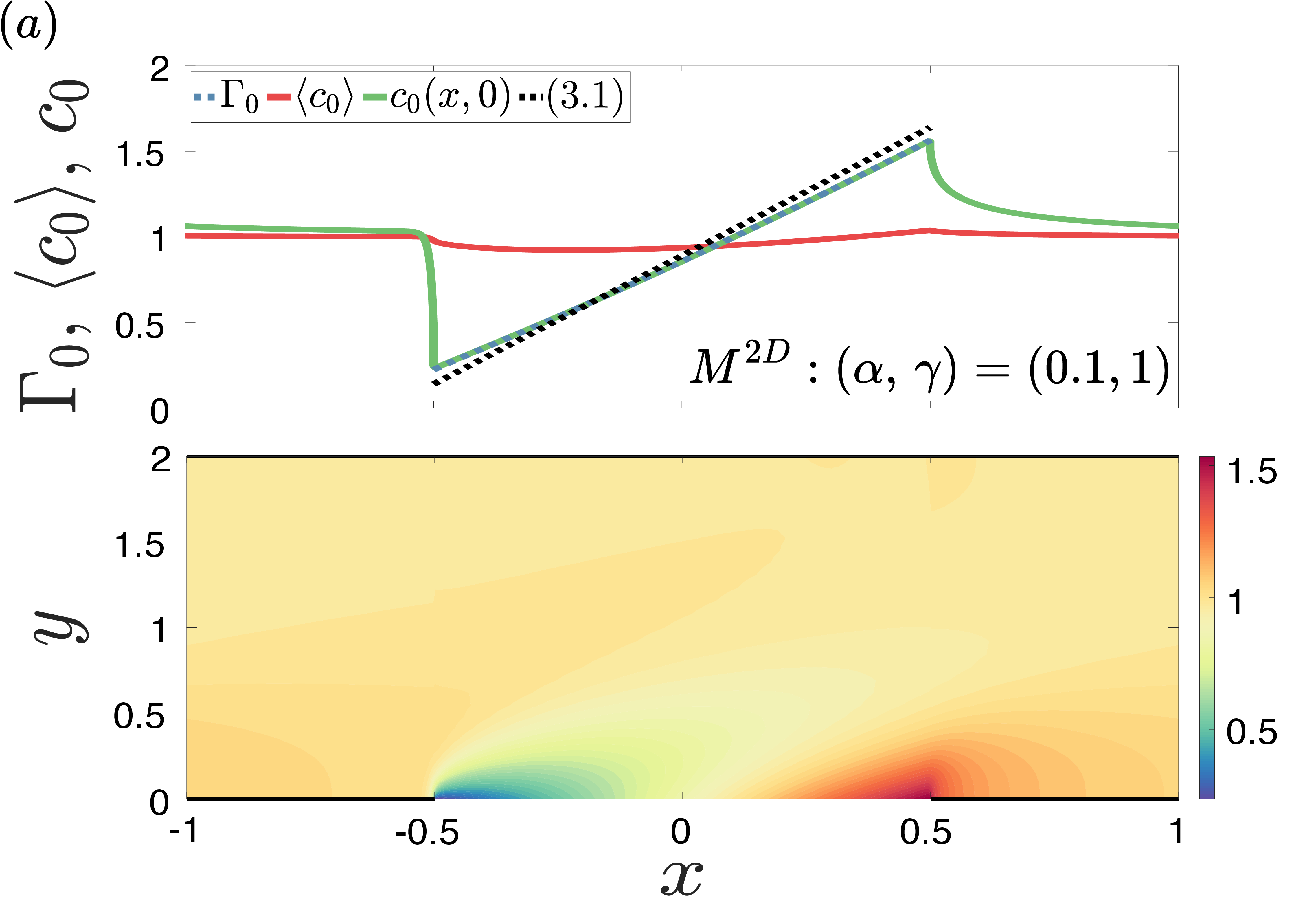} \hfill \includegraphics[width=.48\textwidth]{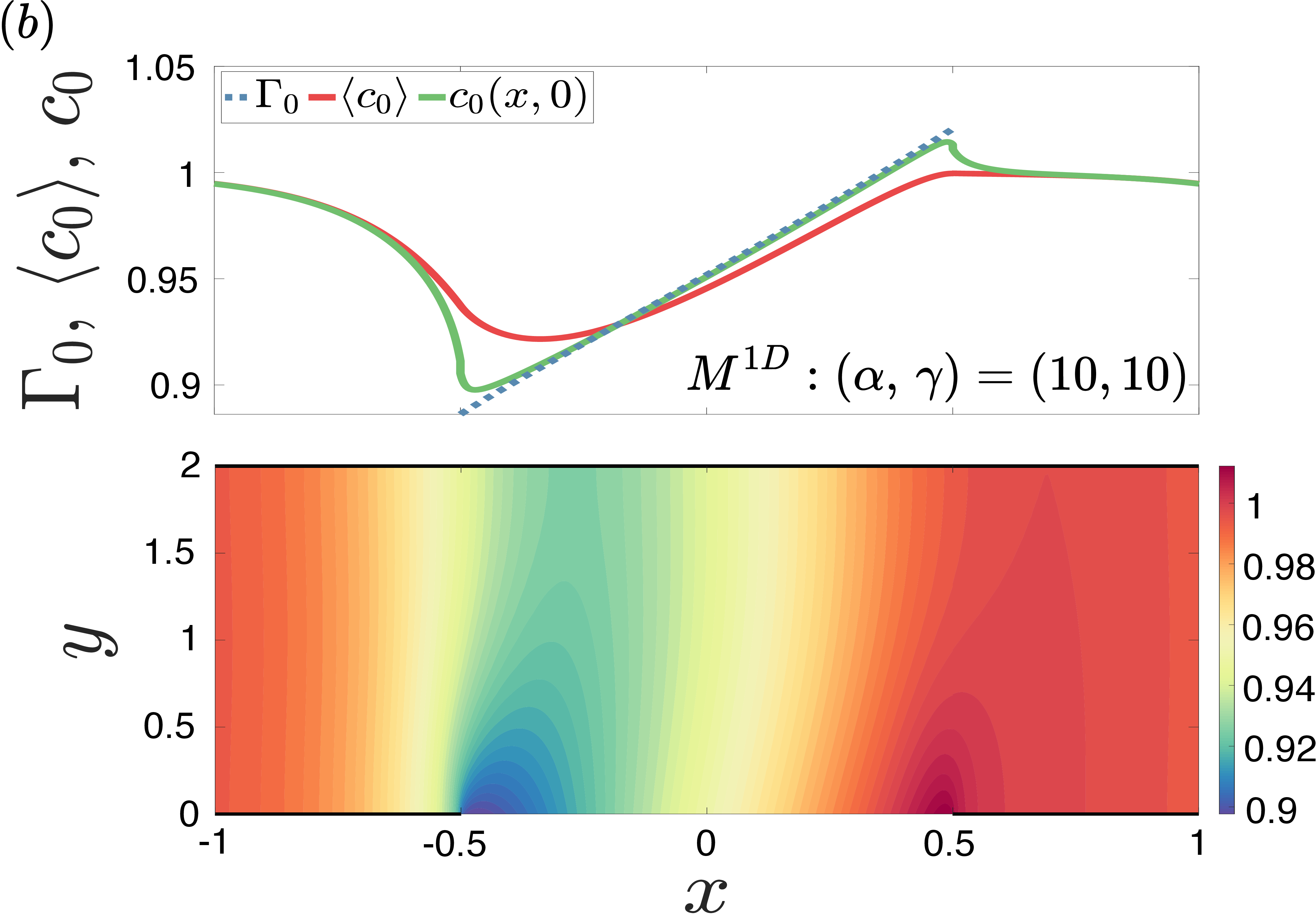} \hfill\includegraphics[,width=.32\textwidth]{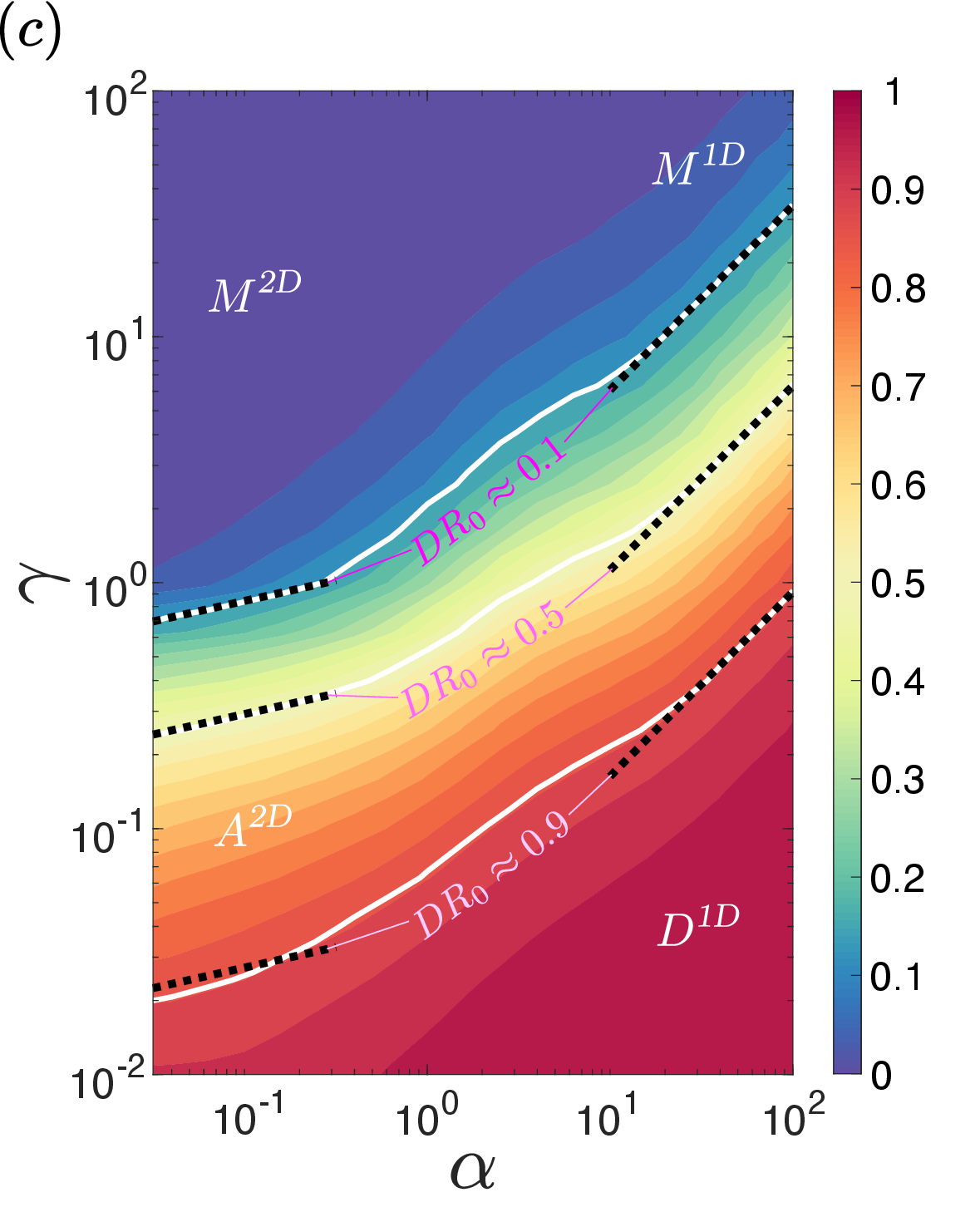} \hfill \includegraphics[width=.32\textwidth]{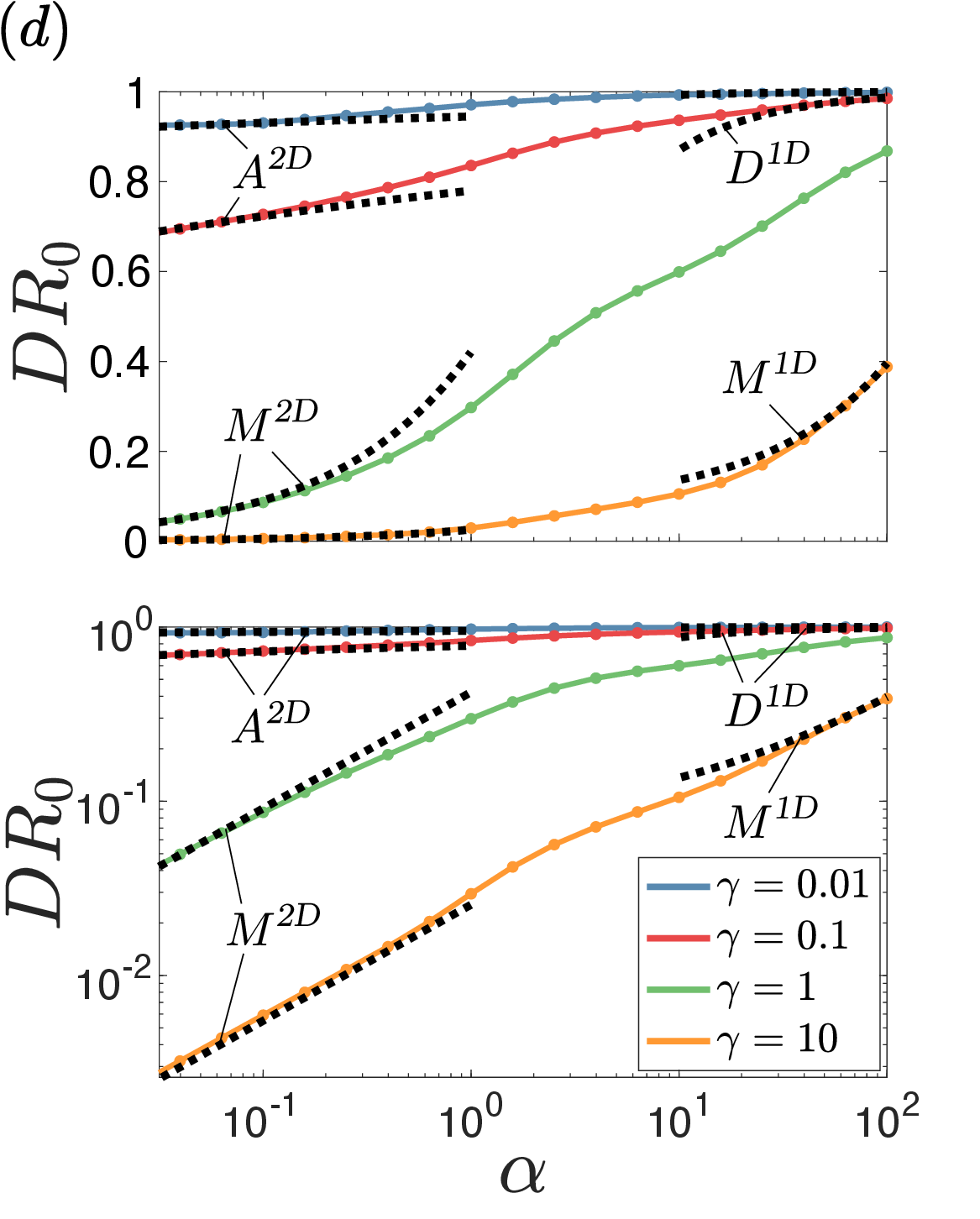} \hfill \includegraphics[width=.32\textwidth]{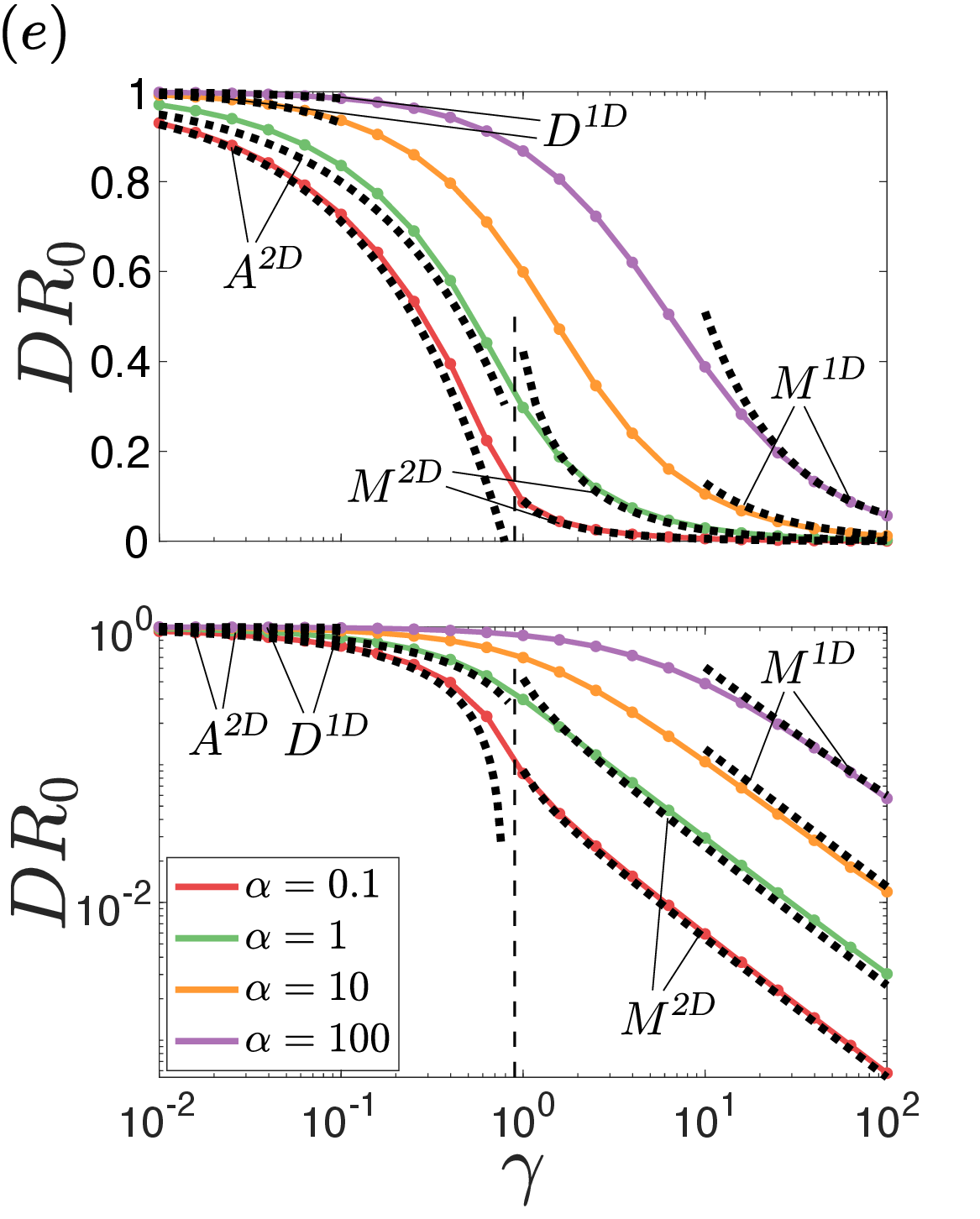} \\ \includegraphics[width=.48\textwidth]{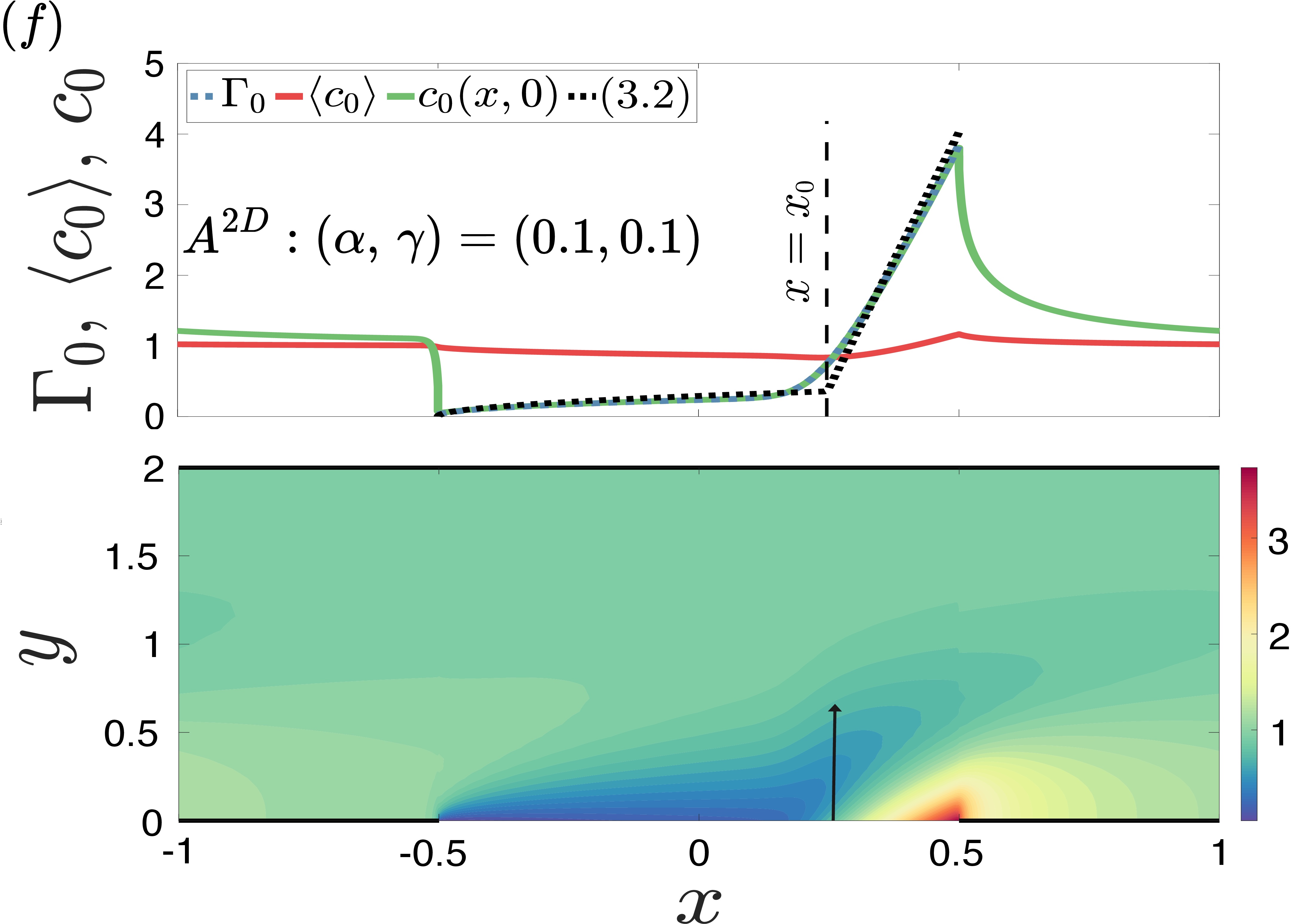} \hfill \includegraphics[width=.49\textwidth]{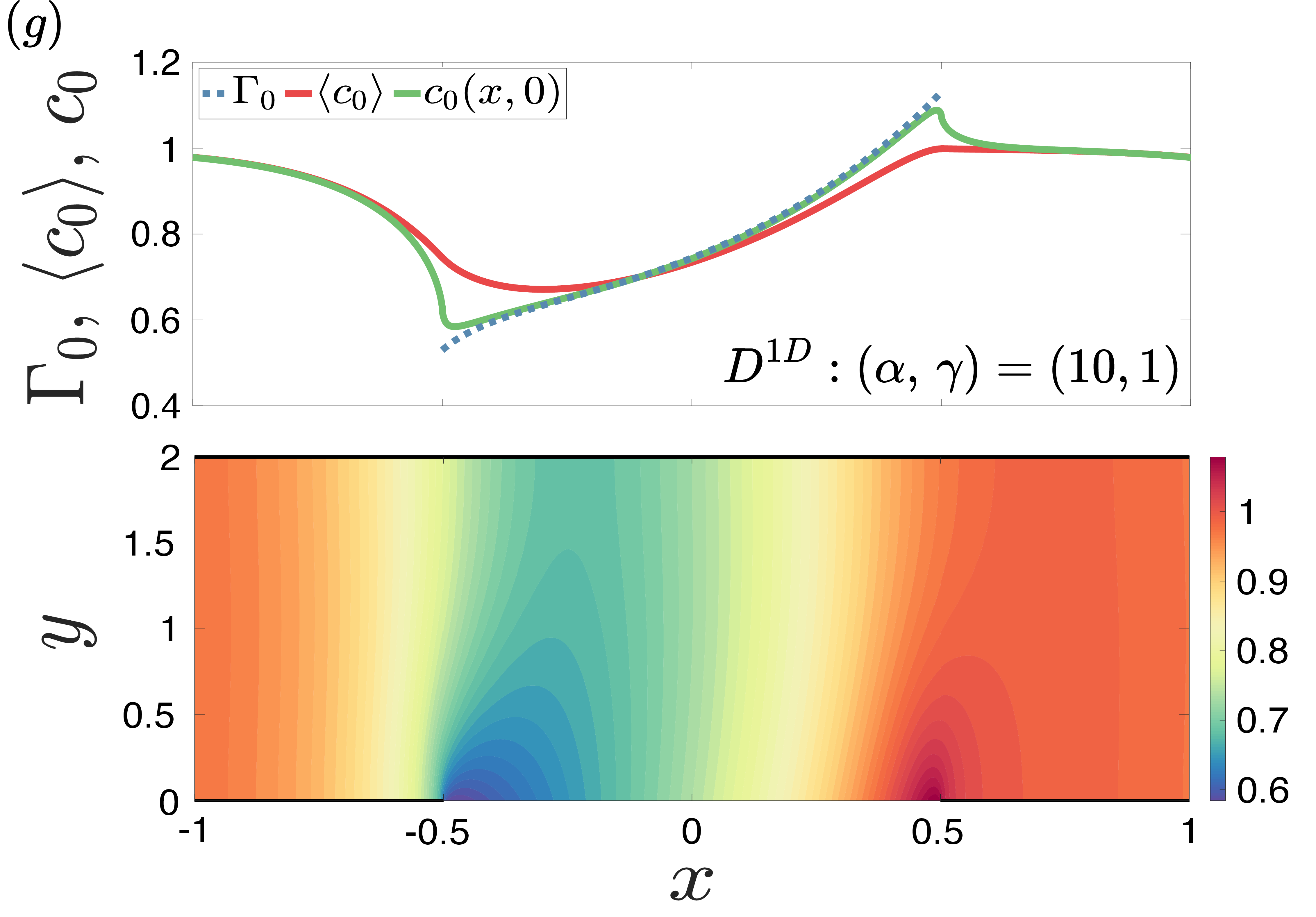} \hfill
\caption{\st{The leading-order drag reduction (${DR}_0$), bulk ($c_0$) and interfacial ($\Gamma_0$) surfactant concentration fields for $\beta = 1$, $\epsilon = 0.1$, $\nu = 100$ and $\phi = 0.5$, computed using \eqref{eq:c_bvp_1_wd}--\eqref{eq:c_bvp_6_wda} when bulk--surface exchange is strong. In the Marangoni-dominated ($M$) region, the SHS is no-slip (${DR}_0\ll 1$), and in the advection- ($A$) and diffusion-dominated ($D$) regions, the interface is shear-free (${DR}_0\approx 1$).
    $c_0$, $\Gamma_0$, $\langle c_0 \rangle$ and $c_0(x, \, 0)$ are plotted in (\textit{a}) $M^\textsl{2D}$ and (\textit{b}) $M^\textsl{1D}$.
    (\textit{c}) Contours of ${DR}_0$, (\textit{d}) plots of ${DR}_0$ for different $\gamma$ and (\textit{e}) plots of ${DR}_0$ for different $\alpha$, where (\textit{c,\,d,\,e}) are compared to asymptotic predictions (\ref{eq:m}\textit{b}) in $M^\textsl{1D}$, (\ref{eq:d}\textit{b}) in $D^\textsl{1D}$, (\ref{eq:bl_m_gamma}\textit{b}) in $M^\textsl{2D}$ and (\ref{eq:a_e}\textit{b}) in $A^\textsl{2D}$.
    The dashed line in (\textit{e}) represents the largest $\gamma$ for which $\Gamma_0(-\phi)=0$ when $\alpha \ll 1$.
    $c_0$, $\Gamma_0$, $\langle c_0 \rangle$ and $c_0(x, \, 0)$ are plotted in (\textit{f}) $A^\textsl{2D}$ (notice the \textit{quasi-stagnant} cap profile) and (\textit{g}) $D^\textsl{1D}$, where $x=x_0$ is plotted using \eqref{eq:gamma_qs_cap}.}
    }
    \label{fig:12}
\end{figure}

Figure \ref{fig:12} summarises the surfactant concentration profiles and drag reduction in the strong-exchange limit, computed using the 2D long-wave model, \eqref{eq:c_bvp_1_wd}--\eqref{eq:c_bvp_6_wda}.  
For large values of $\gamma$, with strong Marangoni effect, the interface is almost immobilised (figure \ref{fig:12}\textit{a--b}), the $\Gamma_0$ profile is almost linear, and the bulk surfactant concentration transitions from a 2D (figure \ref{fig:12}\textit{a}) to a 1D (figure \ref{fig:12}\textit{b}) distribution as $\alpha$ increases.  
Surfactant adsorbs onto the interface across the upper half of the plastron and desorbs from it in the lower half.  
In contrast, for small $\gamma$ with weaker Marangoni effects, interfacial surfactant accumulates near the downstream end of the plastron, lengthening the adsorption region and compressing the desorption region (figure \ref{fig:12}\textit{f}).  

Figure \ref{fig:12}(\textit{c--e}) illustrates the variation of ${DR}_0$ with $\alpha$ and $\gamma$.  
In the present strong-exchange limit, drag reduction decreases as $\alpha$ decreases and the bulk concentration becomes more 2D. 
Reduced bulk diffusion promotes the formation of bulk-concentration boundary layers, while slowing fluxes between the bulk and interface.  
We characterise this transition by comparing the established large-$\alpha$ asymptotes (in regions $M^\textsl{1D}$ and $D^\textsl{1D}$ of the $(\alpha,\gamma)$-plane, see (\ref{eq:m}, \ref{eq:d})) with new small-$\alpha$ limits (derived below) in regions $M^\textsl{2D}$ and $A^\textsl{2D}$.    
We use these limits to characterise the immobilisation of the interface as $\gamma$ increases, highlighting in particular the transition from region $A^\textsl{2D}$ to region $M^\textsl{2D} $ for small $\alpha$. 
The central asymptotic sub-regions $M^\textsl{1D}$ and $A^\textsl{1D}$ for $1\ll \alpha \ll 1/\epsilon^2$ in figure \ref{fig:dr_shems}(\textit{a}), which formally provide a bridge between the 1D and 2D long-wavelength models, are smoothed out by transitional effects associated with the nominally small geometric parameter $\epsilon$ taking the value 0.1 in Figure~\ref{fig:12}.
We now discuss these new regions, starting with $M^\textsl{2D}$.

\begin{figure}
    \centering
    \includegraphics[trim={3cm 0cm 2cm 0cm},clip, width = .49\textwidth]{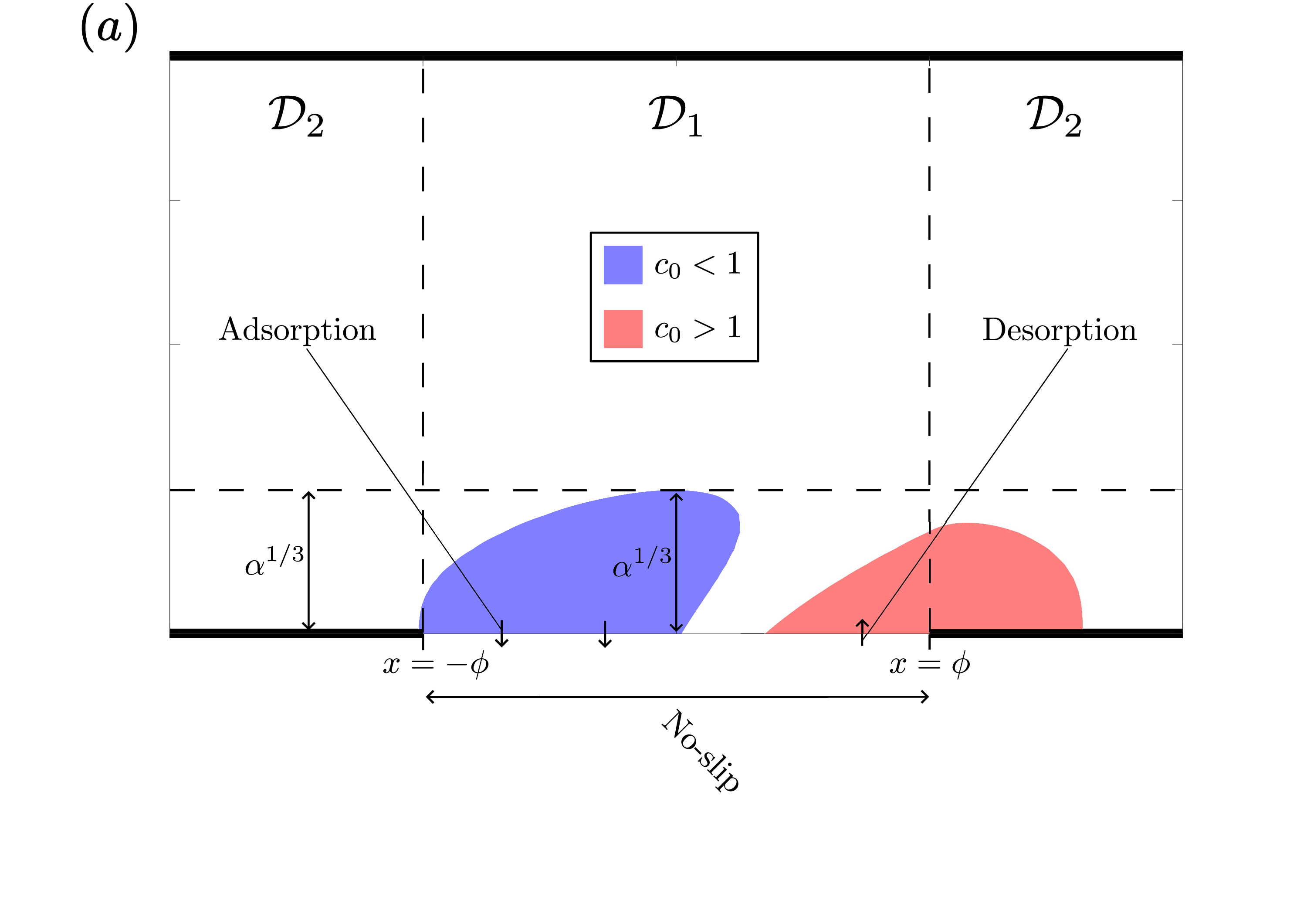} \includegraphics[trim={3cm 0cm 2cm 0cm},clip,width = .49\textwidth]{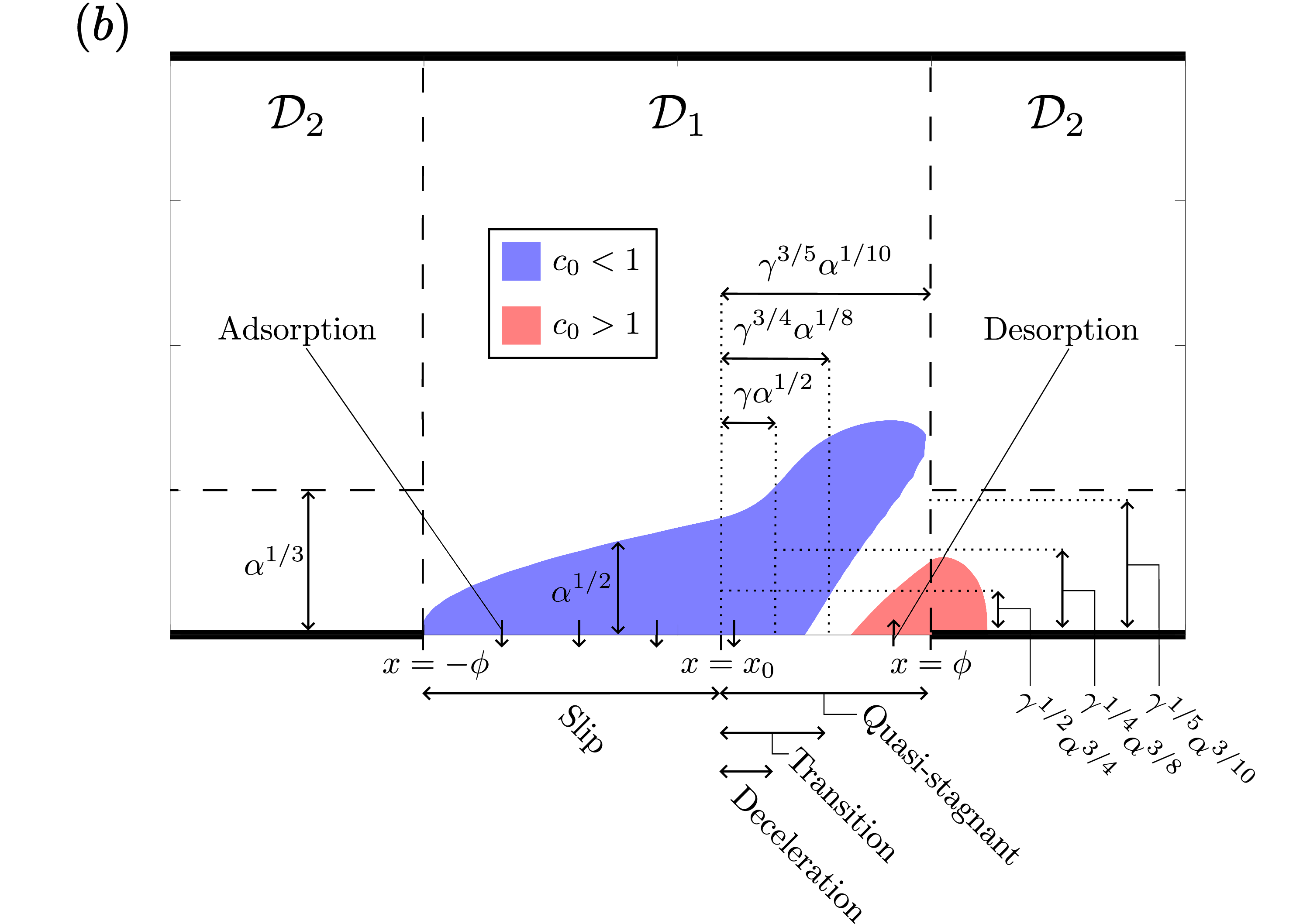} 
    \caption{Schematics of the asymptotic structure of the bulk-concentration boundary layer.
    Weak diffusion ensures that $c_0$ is approximately uniform in the core of the channel, varying primarily in a thin concentration boundary layer near the SHS.  Blue (pink) regions illustrate regions where surfactant is drawn from (released into) the bulk onto (from) the interface.  
    $(\textit{a})$ The bulk-concentration boundary layer when Marangoni effects are strong (region $M^\textsl{2D}$) and the surfactant distribution almost immobilises the interface. 
    $(\textit{b})$ The bulk-concentration boundary layer when Marangoni effects are weak (region $A^\textsl{2D}$), creating a slip region with low surfactant concentration upstream of a quasi-stagnant region in which interfacial surfactant accumulates.  {\color{black}Details of each asymptotic region are provided in Appendix~\ref{sec:asymptotic}.}
    }
    \label{fig:asym_shems}
\end{figure}

Region $M^\textsl{2D}$ (figures \ref{fig:dr_shems}\textit{a} and \ref{fig:12}\textit{a}) emerges when $\alpha \ll 1$ and $\gamma$ is large.  
In this region, Marangoni effects dominate, with weak bulk diffusion resulting in a bulk-concentration boundary layer near $y=0$.  
\st{We provide a scaling argument for ${DR}_0$ as follows, with a detailed derivation given in Appendix \ref{sec:asymptotic_m} and a schematic of the asymptotic structure in figure~\ref{fig:asym_shems}(\textit{a}).}
At the interface, $\Gamma_0$ is in equilibrium with $c_0$, and the interface is nearly immobile, exhibiting a linear surfactant distribution with a slope of size $1/\gamma$. 
The bulk concentration is close to unity, and $\Gamma_0<1$ ($\Gamma_0>1$) in the upstream (downstream) half of the plastron, leading to adsorption (desorption).
Perturbations to the bulk-concentration boundary layer are driven by the linear surface concentration distribution, giving the boundary layer a self-similar structure. 
The thickness of the boundary layer is $\alpha^{1/3} \ll 1$ (characteristic of a shear flow), resulting in fluxes of size $\alpha c_{0y} \sim \alpha^{2/3}$ onto and off the interface. 
Adsorption and desorption are accommodated by weak stretching and compression of the interface, generating smaller contributions to $\Gamma_0$ of size $\alpha^{2/3}/\gamma^2$ and to ${DR}_0$ of size $\alpha^{2/3}/\gamma$, using \eqref{eq:dr_def}.  
The constraint $\alpha \ll 1 $ ensures that the boundary layer is thin and we require $\gamma\gg 1$ and $\alpha^{2/3} \ll \gamma$ for the correction to $\Gamma_0$ to be small (although a more precise condition will emerge below).  The leading-order surfactant distribution and drag reduction are given by 
\refstepcounter{equation} \label{eq:m_e}
\begin{equation} 
    c_0(x, \, 0) \approx \Gamma_0 \approx 1 + \frac{3\beta}{2\gamma}\left(x - \frac{\phi}{5}\right), \quad {DR}_0 \approx \frac{m_1\alpha^{2/3}\phi^{5/3}}{\gamma}, \tag{\theequation\textit{a,\,b}}
\end{equation}
where the coefficient $m_1 \approx 0.79$ is given explicitly in \st{Appendix \ref{sec:asymptotic_m}}.  
The range of validity of the leading-order solution in \eqref{eq:m_e} is extended from $\gamma \gg 1$ to $\gamma \sim 1$ in \eqref{eq:bl_m_gamma} below, where the extension must be evaluated numerically.  
Therefore, we give the simpler formula above, but compare \eqref{eq:bl_m_gamma} with numerical results from the 2D long-wave model at small $\alpha$ in figure \ref{fig:12}(\textit{c--e}), which capture numerical results successfully.
In particular, decreasing $\gamma$ leads to a steepening of the interfacial surfactant gradient until $\Gamma_0$ approaches zero at the upstream contact line. 
This transition provides a lower bound on region $M^\textsl{2D}$, namely $\Gamma_0(-\phi)=0$ when $\gamma = 9\phi\beta/5$ (see vertical dashed lines in figure \ref{fig:12}\textit{e}). 
Hence, we plot asymptotic predictions up to this limit in figure \ref{fig:12}(\textit{e}). 

When $\alpha \ll 1$ and $\gamma$ is sufficiently small, we transition into region $A^\textsl{2D}$ (figures \ref{fig:dr_shems}\textit{a} and \ref{fig:12}\textit{f}), where advection dominates Marangoni effects at the interface, while interacting with a strongly coupled bulk-concentration boundary layer. 
We distinguish two primary regions along the interface: (i) a slip region, $x\in[-\phi,\,x_0)$ (for some $x_0$), at the upstream end of the interface, where Marangoni effects are weak and surfactant adsorbs from the bulk; and (ii)  a \textit{quasi}-stagnant region, $x\in(x_0,\,\phi]$, at the downstream end of the interface, where surfactant gradients decelerate the interface through Marangoni effects without completely immobilising the interface (in contrast with a \textit{classic} stagnant cap), while desorbing the accumulated surfactant.  
\st{We now provide a scaling argument for ${DR}_0$ that is supported by a detailed derivation in Appendix \ref{sec:asymptotic_a} below; a schematic of the asymptotic structure is given in figure~\ref{fig:asym_shems}(\textit{b}).}
In the slip region, of length $x_0+\phi=O(1)$, $c_0$ varies by $O(1)$ across a boundary-layer thickness $y\sim \alpha^{1/2}$, yielding a flux per unit length of size $\alpha c_{0y}\sim \alpha^{1/2}$ onto the interface.  
This allows $\Gamma_0$ to grow along the slip region from zero to $O(\alpha^{1/2})$.
Although $c_0$ is effectively zero at the interface at leading order, it is coupled to $\Gamma_0$ through an $O(\alpha^{1/2})$ correction.  
The slip region therefore delivers an interfacial flux of size $O(\alpha^{1/2})$ at $x\approx x_0$, which is carried through short regions across which the interface decelerates rapidly at the leading edge of the quasi-stagnant region (i.e. at $x\approx x_0$).  \st{The sizes of the short regions are indicated in figure~\ref{fig:asym_shems}(\textit{b}).}
Across the quasi-stagnant region, of length $L=\phi-x_0$ (to be determined), $\Gamma_0$ is approximately linear with a slope of size $1/\gamma$, setting $c_0$ and $\Gamma_0$ to be of size $L/\gamma$. 
In the bulk, diffusion balances shear-flow transport over a vertical length $y \sim (\alpha L)^{1/3}$ and therefore $\alpha c_{0y} \sim (\alpha L)^{2/3}/\gamma$.  
Integrated over $L$, this accommodates the $O(\alpha^{1/2})$ interfacial flux, resulting in $L \sim \gamma^{3/5}/\alpha^{1/10}$.  
Consequently, the drag reduction arising from the surfactant adsorbed and desorbed in the slip and quasi-stagnant regions is proportional to ${DR}_0 \sim \gamma \Delta \Gamma \sim L$.

In \st{Appendix \ref{sec:asymptotic_a}, we present a more detailed analysis showing that}
\refstepcounter{equation} \label{eq:a_e}
\begin{multline} 
    c_0(x, \, 0) \approx \Gamma_0 \approx
    \begin{cases}
        \displaystyle \frac{4(\alpha(x+\phi))^{1/2}}{\sqrt{3\pi}\beta} \quad \text{for} \quad x\in [-\phi, \, x_0), \\
    \displaystyle\frac{3\beta}{2\gamma}(x-x_0) \hspace{.92cm} \text{for} \quad x\in (x_0,\,\phi],
    \end{cases} \quad {DR}_0 = 1 - \frac{a_1 \gamma^{3/5}}{\alpha^{1/10}}, \\
{\color{black}    u_0(x, \, 0) \approx
\begin{cases}3/4 \hspace{6.06cm} \text{for} \quad x\in [-\phi,x_0), \\
        \displaystyle \frac{3}{4} \left[G\left(\frac{x-x_0}{\gamma\alpha^{1/2}}\right)\right]^{-1} \hspace{3.6cm} \text{for} \quad x\approx  x_0, \\
        \displaystyle \frac{3\alpha^{2/3} C'(0)}{5\beta} \Bigg((x-x_0)^{2/3} - \frac{(\phi-x_0)^{5/3}}{x-x_0}\Bigg) \quad \text{for} \quad x\in (x_0,\,\phi].
    \end{cases} }\hfill
\tag{\theequation\textit{a--c}}
\end{multline} 
The coefficient $a_1 = (-5\lambda/(6C'(0)))^{3/5}/(2\phi)$ in (\ref{eq:a_e}$b$) where $\lambda = (16(x_0+\phi)/(3\pi\beta^2))^{1/2}$; $C'(0) \approx -0.92$ in (\ref{eq:a_e}$c$) is given explicitly in \st{Appendix \ref{sec:asymptotic_a}; the function $G$ in (\ref{eq:a_e}$c$) is given implicitly in \eqref{eq:GGG}}. 
The length of the quasi-stagnant cap is 
\begin{equation}
    L=\phi-x_0= \frac{2 \phi a_1 \gamma^{3/5}}{\alpha^{1/10}}.
\end{equation}
{\color{black}In (\ref{eq:a_e}), $x\in [-\phi,x_0)$, $x\approx x_0$ and $x\in(x_0,\phi]$ are shorthand for the slip, deceleration and quasi-stagnant regions shown in figure~\ref{fig:asym_shems}(\textit{b}); the leading-order approximations in (\ref{eq:a_e}) match together as explained in Appendix~\ref{sec:asymptotic}. 
Equation (\ref{eq:a_e}$c$) highlights the role of interfacial compression in the quasi-stagnant region,  accommodating desorption.}

Asymptotes for ${DR}_0$, $c_0(x,\,0)$ and $\Gamma_0$ in figure \ref{fig:12}(\textit{c--f}) are evaluated numerically for a given $\alpha$ and  $\gamma$ and capture the behaviour exhibited by the numerical simulations of the 2D long-wave model. 
Increasing $\gamma$ from low values causes lengthening of the quasi-stagnant region, pushing $x_0$ towards the upstream contact line and reducing $DR_0$. 
Figure \ref{fig:12}(\textit{e}) shows how (\ref{eq:a_e}$b$) applies almost until $x_0$ approaches $-\phi$.  
For small $\alpha$, $c_0(-\phi, \, 0) \approx \Gamma_0(-\phi) \approx 0$ for $\gamma \ll 1$, causing the $c_0$ field to develop a singular first derivative at the upstream contact point. 
This concentration-gradient singularity significantly impacts the numerical solution (both in terms of the required resolution and run time) of the 2D long-wave model and COMSOL simulations (below).

\begin{figure}
    \centering
    \hfill 
    \includegraphics[trim={0cm 0cm 0cm 0cm},clip,width=.49\textwidth]{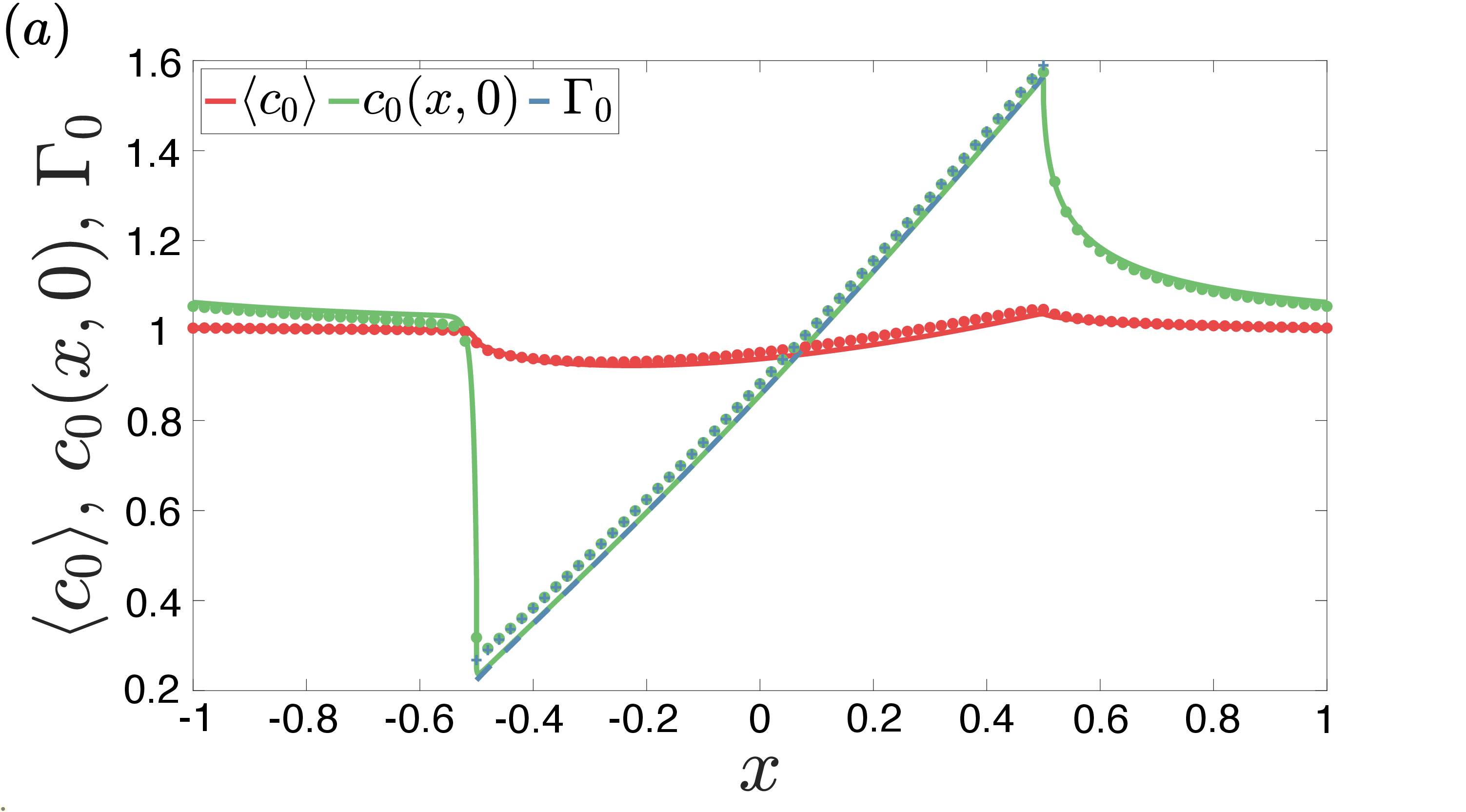}  \hfill \includegraphics[trim={0cm 0cm 0cm 0cm},clip,width=.49\textwidth]{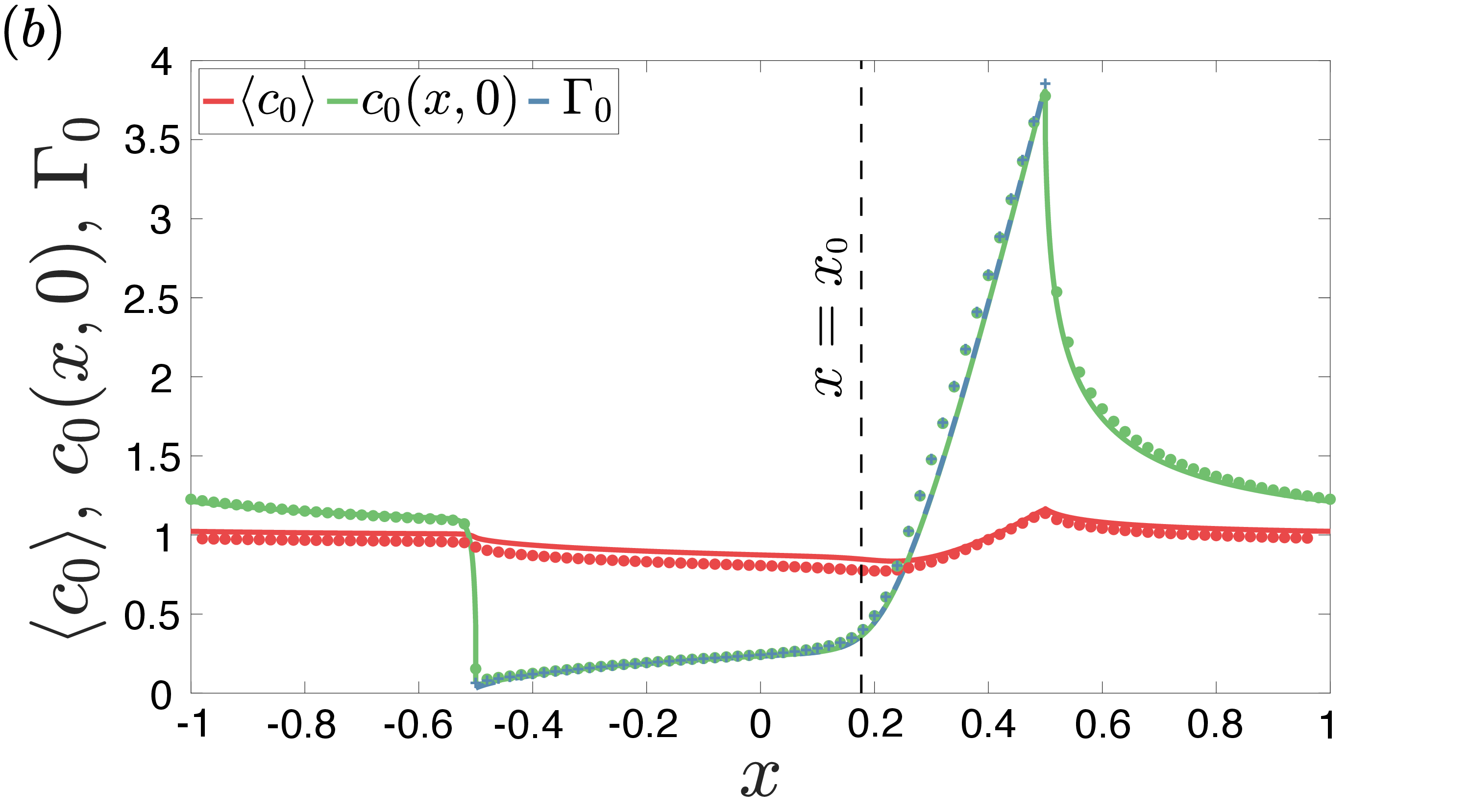} \hfill \hfill \hfill \\
    \hfill \includegraphics[trim={0cm 0cm 0cm 0cm},clip,width=.49\textwidth]{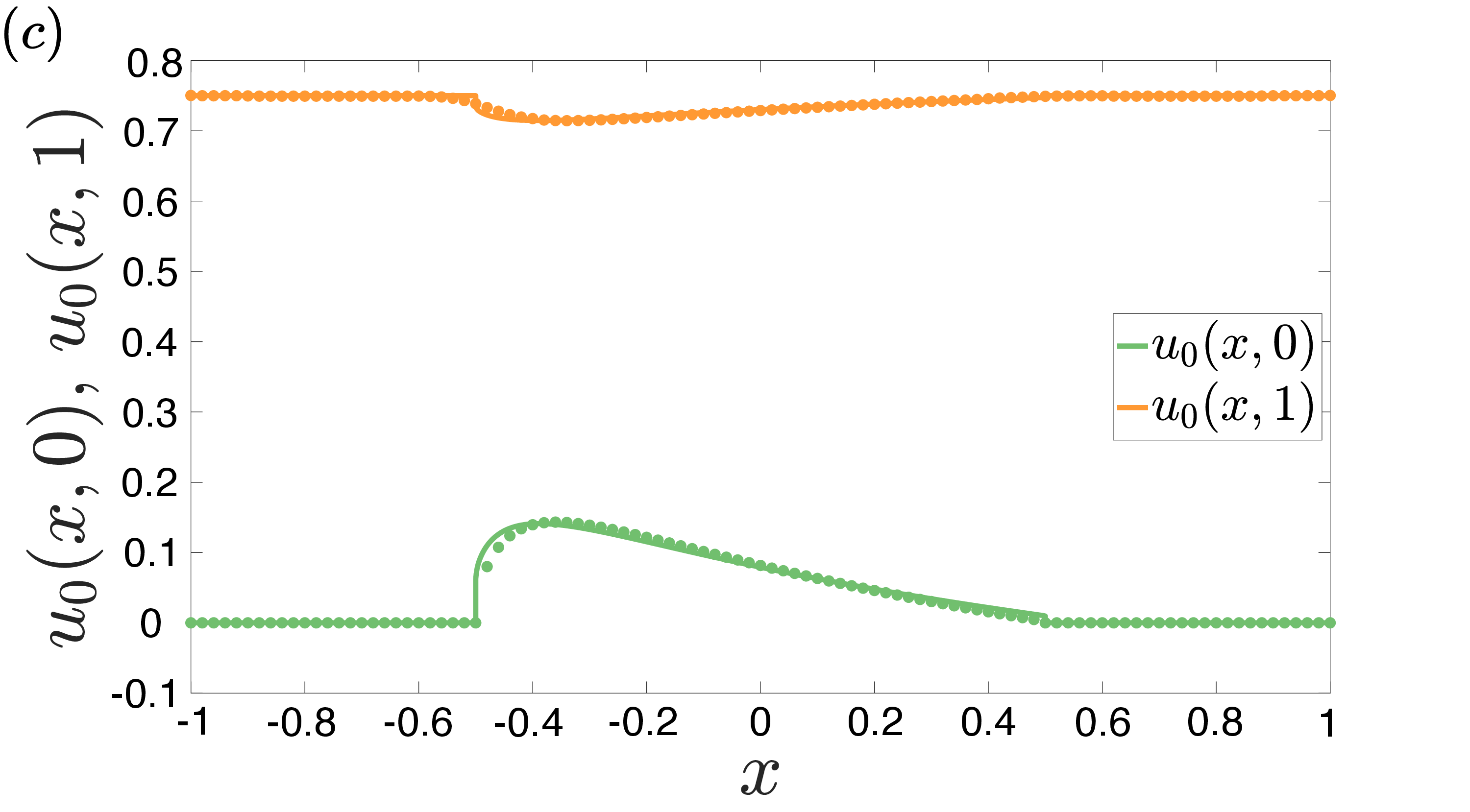} \hfill \includegraphics[trim={0cm 0cm 0cm 0cm},clip,width=.49\textwidth]{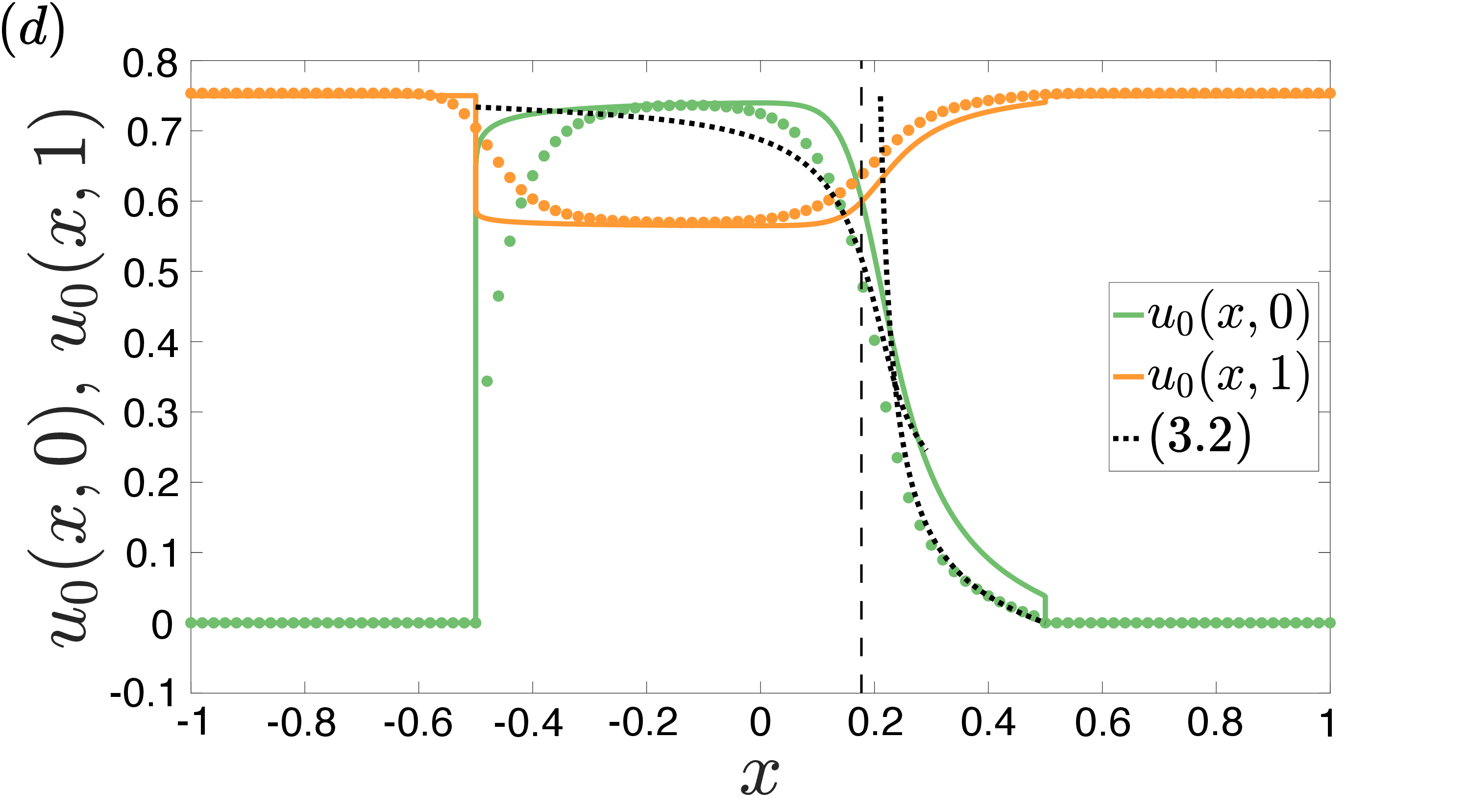} \hfill \hfill \hfill \\  
    \caption{The leading-order streamwise velocity field ($u_0$) and surfactant concentration fields ($c_0$ and $\Gamma_0$) for $\beta = 1$, $\nu = 100$, $\epsilon = 0.1$ and $\phi = 0.5$ evaluated using \eqref{eq:c_bvp_1_wd}--\eqref{eq:c_bvp_6_wda} (lines) and COMSOL simulations \eqref{eq:nondimensional_equations}--\eqref{eq:nd_velocity_flux} (symbols), when bulk--surface exchange is strong. 
    (\textit{a,\,c}) Plots of $c_0$ and $u_0$, respectively, for $\alpha = 0.1$ and $\gamma = 1$, when the flow is in the Marangoni--dominated region with weak cross-channel diffusion ($M^\textsl{2D}$).
    (\textit{b,\,d}) Plots of $c_0$ and $u_0$, respectively, for $\alpha = 0.1$ and $\gamma = 0.1$, when the flow is in the advection--dominated region with weak cross-channel diffusion ($A^\textsl{2D}$), where $x=x_0$ is plotted using \eqref{eq:gamma_qs_cap}.
    }
    \label{fig:12_b}
\end{figure}

Further insight into the $M^\textsl{2D}$ and $A^\textsl{2D}$ solutions is provided in figure \ref{fig:12_b}, which compares $u_0$, $c_0$ and $\Gamma_{0}$ computed using the 2D long-wave model, \eqref{eq:c_bvp_1_wd}--\eqref{eq:c_bvp_6_wda}, to the Stokes and advection-diffusion equations in COMSOL, \eqref{eq:nondimensional_equations}--\eqref{eq:nd_velocity_flux}, with the same parameters as the examples shown in figure \ref{fig:12}(\textit{a,\,f}).
For the solution in the $M^\textsl{2D}$ region, $u_0$ exhibits a slight deviation from no-slip at the liquid--gas interface (figure \ref{fig:12_b}\textit{c}), generating a weak vertical flow towards the interface. 
The concentration profiles in the two simulations match well (figure \ref{fig:12_b}\textit{a}), with the leading-order drag reduction predicted from the 2D long-wave model, ${DR}_0 = 0.106$, in close agreement with the COMSOL simulations, ${DR}_{NS} = 0.109$.  
For the solution in the region $A^\textsl{2D}$, depicted in figure \ref{fig:12_b}(\textit{b,\,d}), the liquid--gas interface is nearly shear-free and $u_0\approx 3/4$ in the slip region, where the interfacial surfactant concentration is low, before falling towards zero across the quasi-stagnant region, where the gradient of $\Gamma_0$ is large. 
The asymptotic predictions (\ref{eq:a_e}\textit{c}) capture respectively the slight decrease of the slip velocity in the deceleration region and then the rapid fall towards zero across the quasi-stagnant region; {\color{black}the shift parameter in (\ref{eq:GGG}) was chosen as $X_0 = 0.5$ to minimise the error against the COMSOL simulations}.

Figure~\ref{fig:12_b} illustrates the distinction between a \textit{classical} stagnant cap of a strictly insoluble surfactant, for which $u_0$ vanishes abruptly in the cap region where $\Gamma_0$ has a large gradient, and the present \textit{quasi}-stagnant cap structure, where desorption of the soluble surfactant in the bulk boundary layer is mediated by molecular diffusion, and which leads to a small but non-zero slip velocity (\ref{eq:a_e}$c$) across the quasi-stagnant cap region.
The predictions for the drag reduction and concentration field from the 2D long-wave model and numerical simulations in COMSOL agree, with ${DR}_0 = 0.745$ and ${DR}_{NS} = 0.747$.  
Both methods capture weak interfacial stretching (allowing adsorption) in the slip region ($u_{0x}>0$ for $x\in[-\phi, \, x_0]$ in figure \ref{fig:12_b}\textit{d}) and stronger compression (allowing desorption) in the quasi-stagnant region ($u_{0x}<0$ for $x\in[x_0,\,\phi]$ in figure \ref{fig:12_b}\textit{d}).
There is some disagreement in $u_0$ at the upstream contact line:  the 2D long-wave theory does not capture the Stokes-flow region where $u_0$ rapidly transitions from zero to the slip velocity, and at the downstream contact line: here, the retention of weak surface diffusion leads to a small spurious non-zero value of $u_0$ which ensures zero surfactant flux in the 2D long-wave theory, whilst in the COMSOL simulation the slip velocity tends to zero at the contact line.
However, these short inner regions, which cannot be captured by 2D long-wave theory by construction, are of secondary importance for the drag-reduction calculation.  
Further features of the surface velocity profile at the tip of the quasi-stagnant region, which explain the dramatic thickening of the boundary layer in figures~\ref{fig:12}(\textit{f}) and \ref{fig:asym_shems}(\textit{b}), are discussed in \st{Appendix \ref{sec:asymptotic_a}}.

\subsection{Weak exchange} \label{sec:weak exchange}

\begin{figure}
\centering
    \includegraphics[width=.49\textwidth]{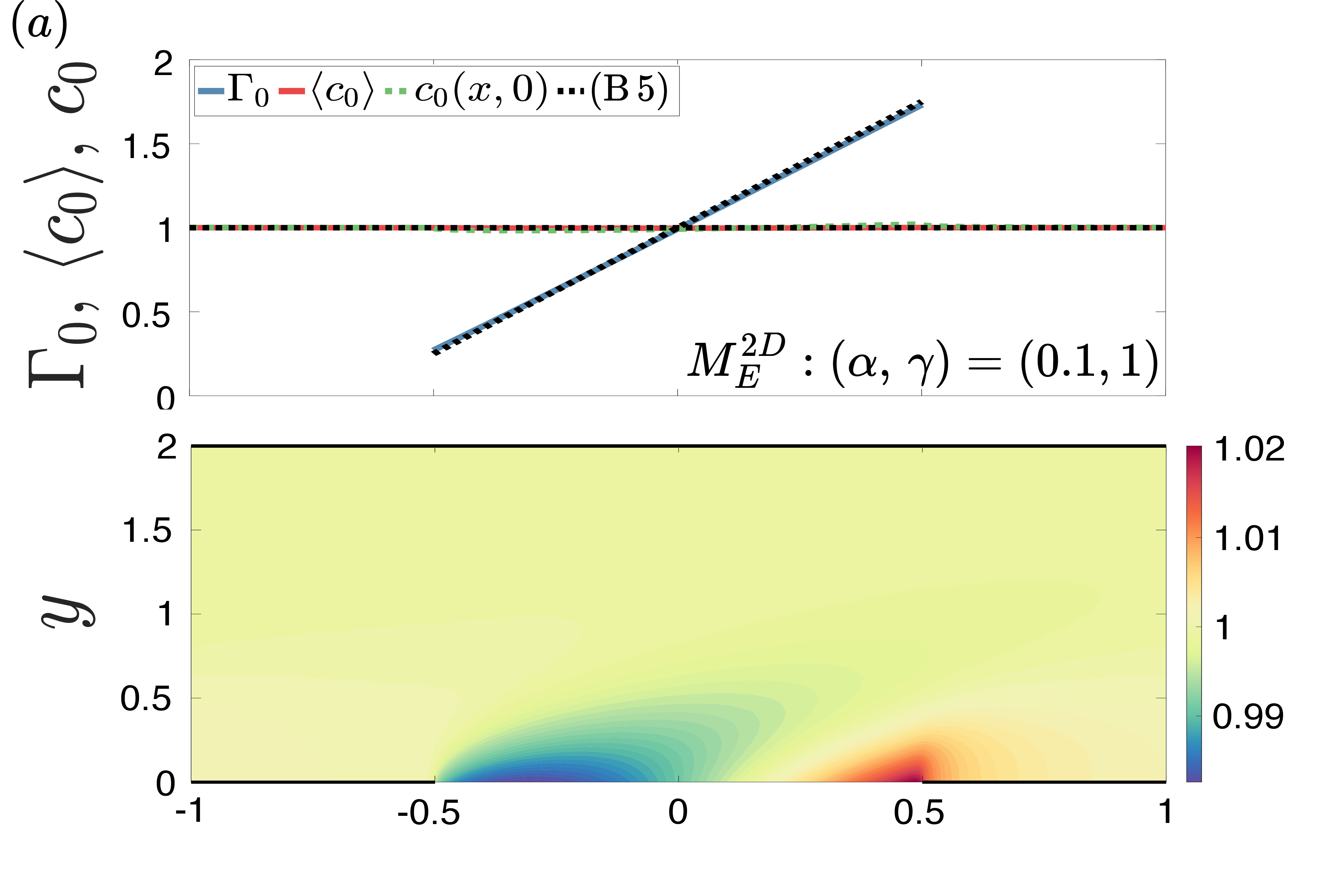} \hfill \includegraphics[width=.49\textwidth]{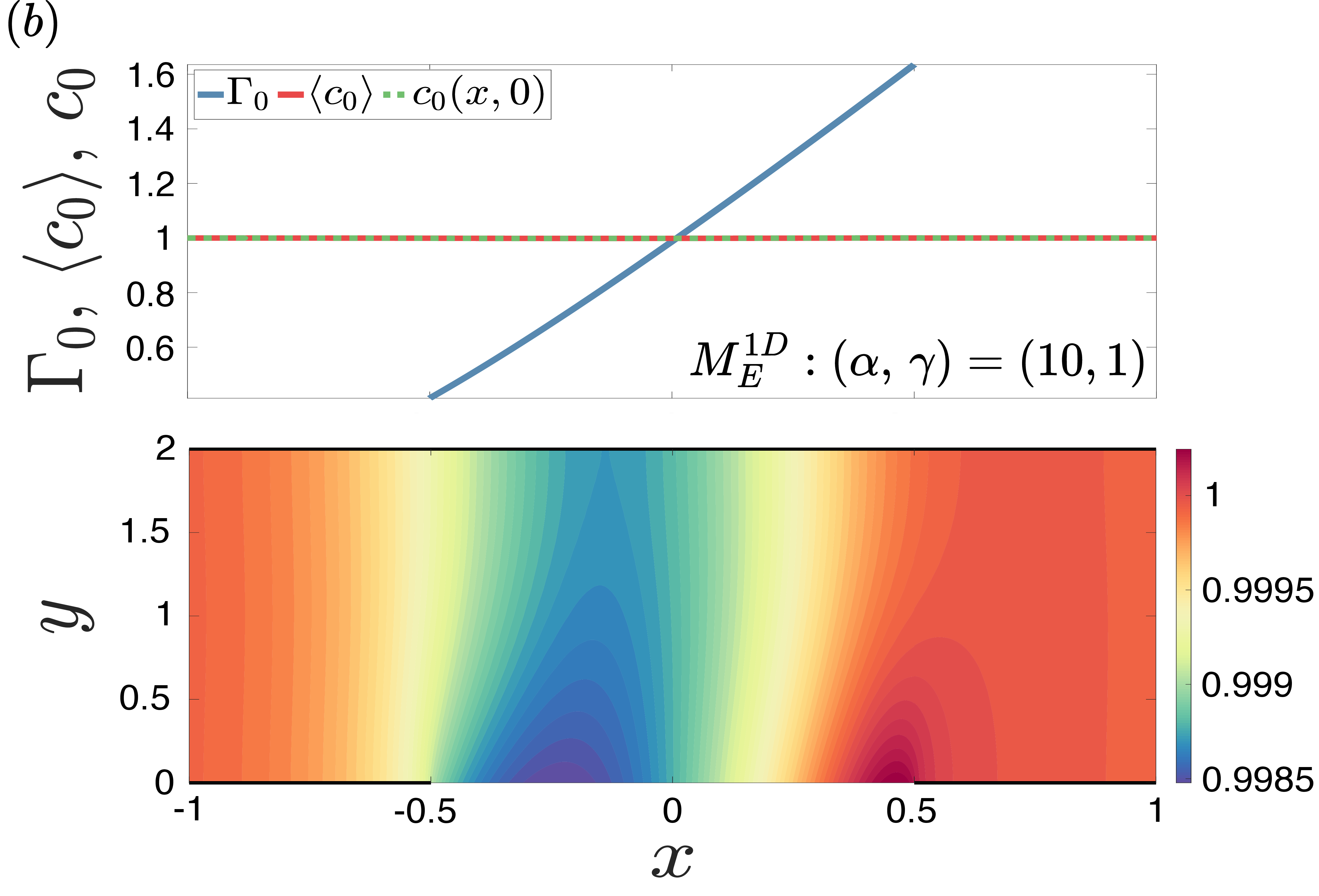} \\ 
    \includegraphics[width=.32\textwidth]{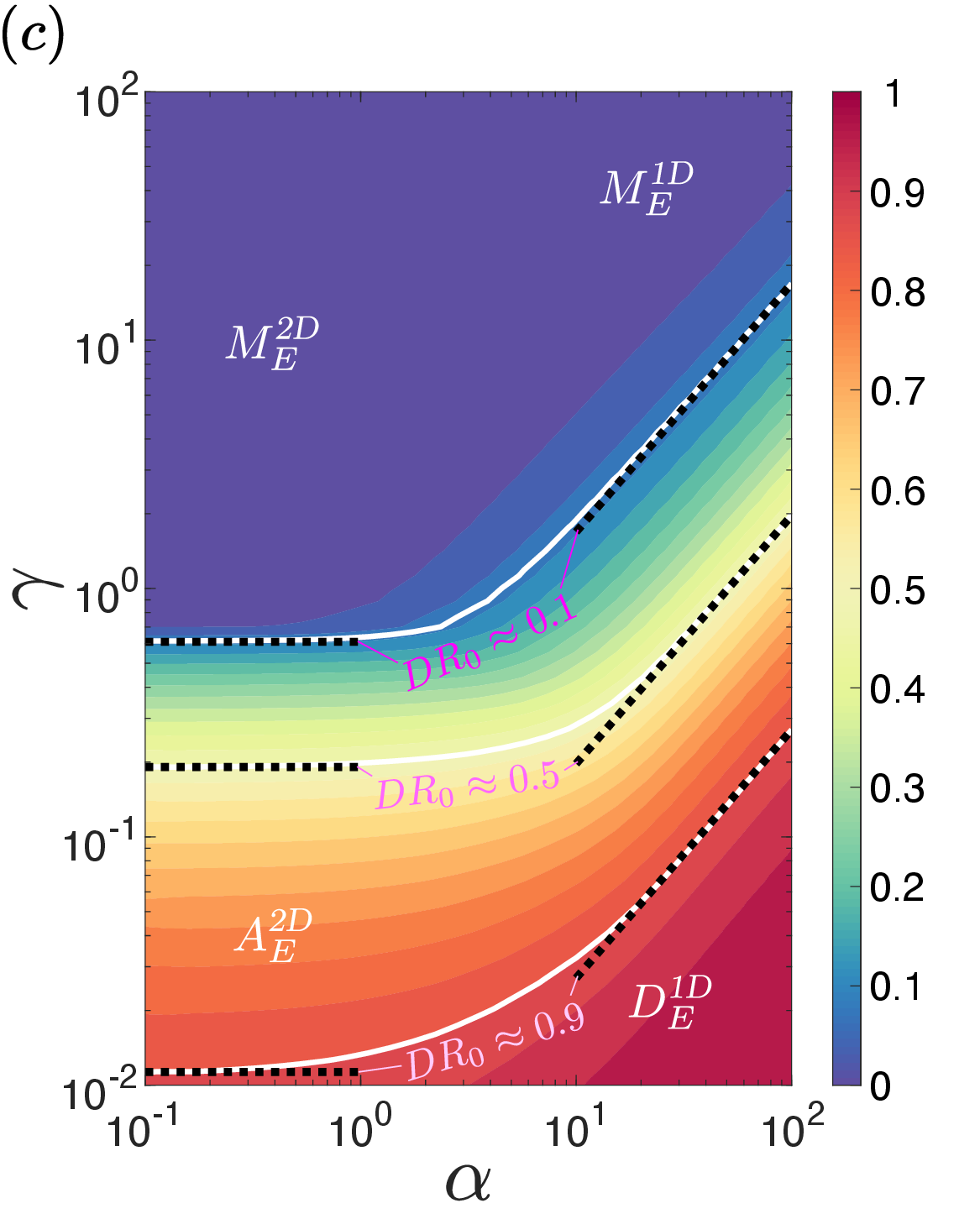} \hfill \includegraphics[width=.32\textwidth]{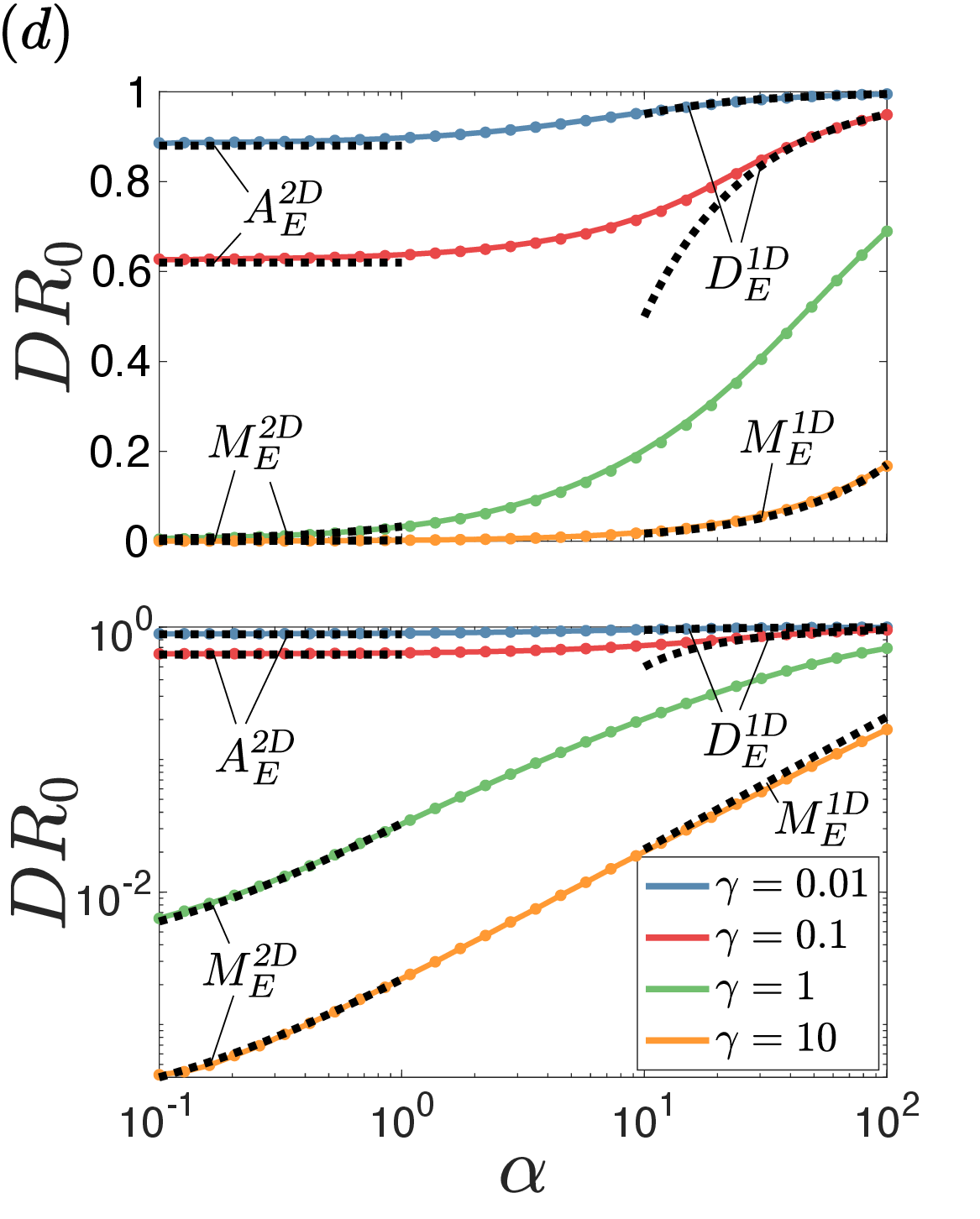} \hfill \includegraphics[width=.32\textwidth]{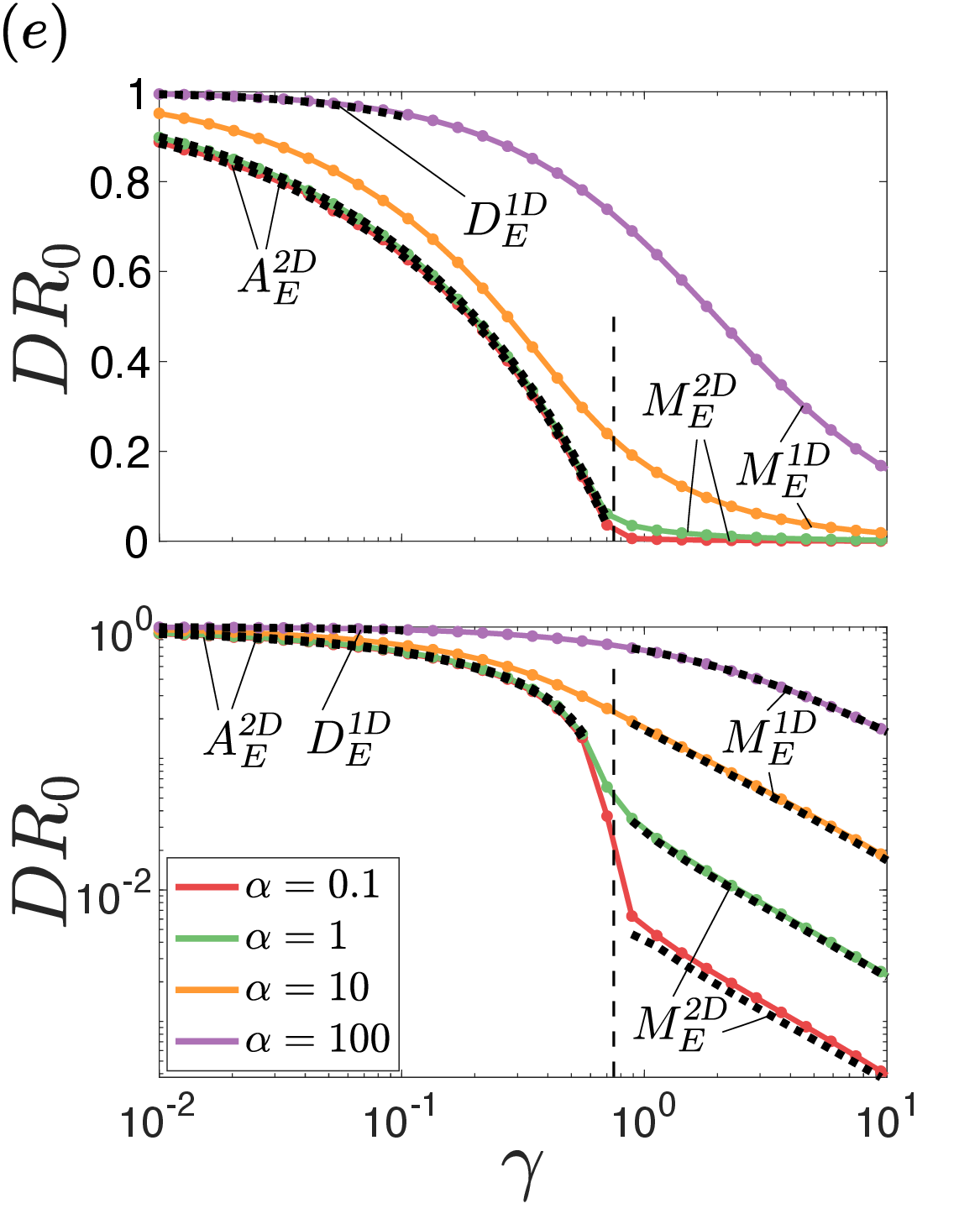} \\ 
    \includegraphics[width=.49\textwidth]{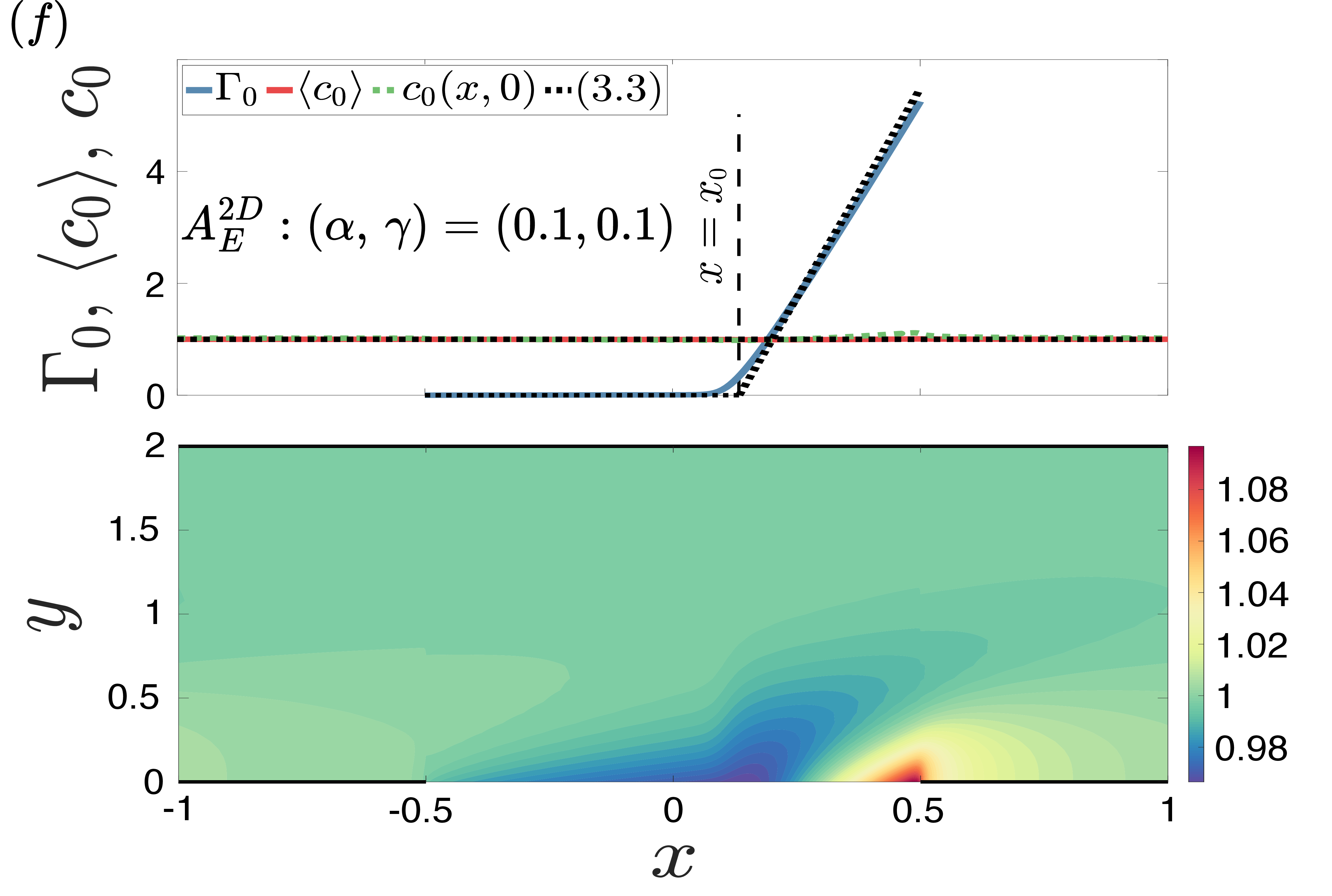} \hfill \includegraphics[width=.48\textwidth]{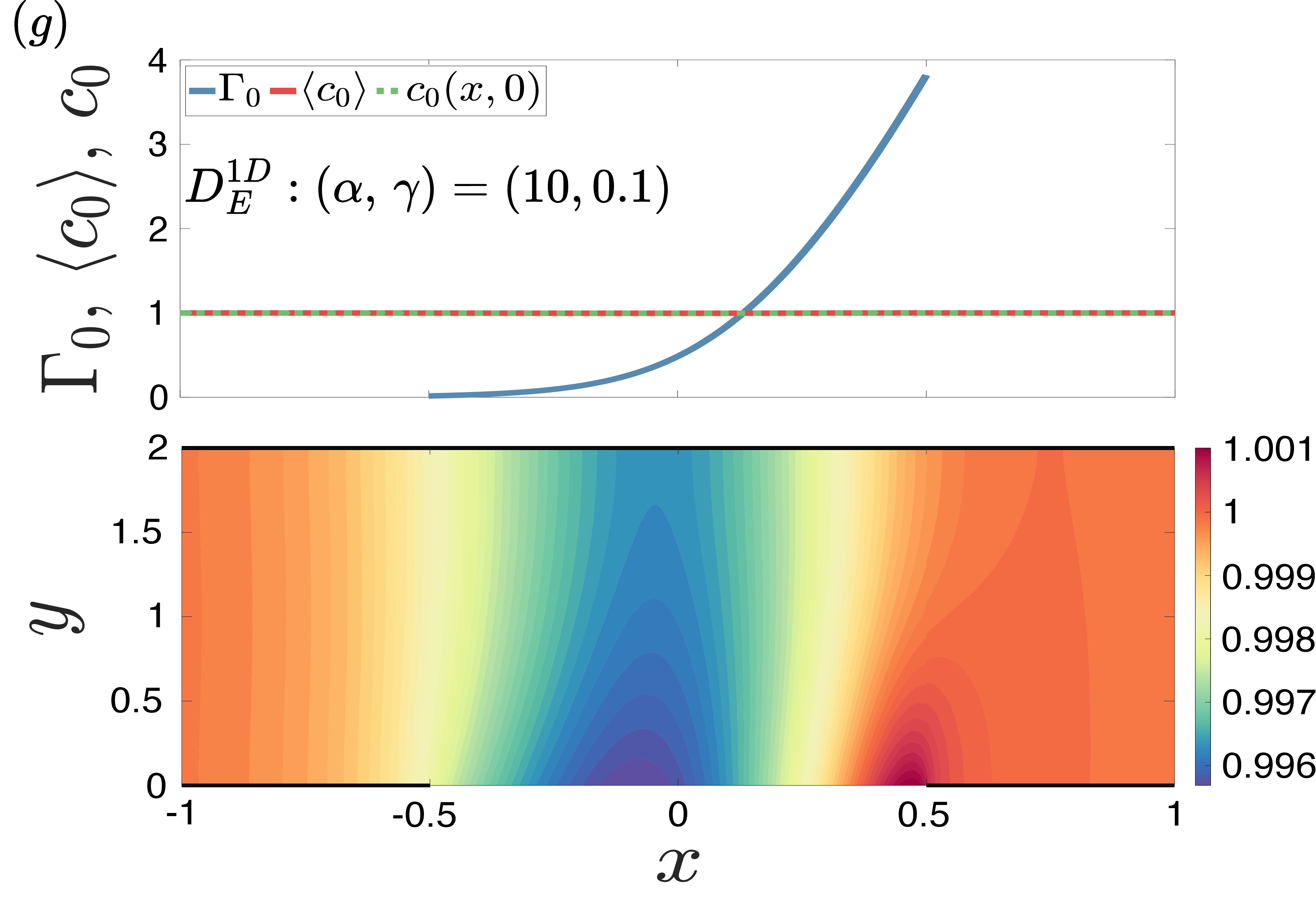}   
    \caption{\st{The leading-order drag reduction (${DR}_0$), bulk ($c_0$) and interfacial ($\Gamma_0$) surfactant concentration fields for $\beta = 1$, $\epsilon = 0.1$, $\nu = 0.01$ and $\phi = 0.5$, computed using \eqref{eq:c_bvp_1_wd}--\eqref{eq:c_bvp_6_wda} when bulk--surface exchange is weak.
    In the Marangoni-dominated ($M_E$) region, the SHS is mostly no-slip (${DR}_0\ll 1$), and in the advection- ($A_E$) and diffusion-dominated ($D_E$) regions, the interface is mostly shear-free (${DR}_0\approx1$).
    $c_0$, $\Gamma_0$, $\langle c_0 \rangle$ and $c_0(x, \, 0)$ are plotted in (\textit{a}) $M^\textsl{2D}_E$ and (\textit{b}) $M^\textsl{1D}_E$.
    (\textit{c}) Contours of ${DR}_0$, (\textit{d}) plots of ${DR}_0$ for different $\gamma$ and (\textit{e}) plots of ${DR}_0$ for different $\alpha$, where (\textit{c,\,d,\,e}) are compared to (\ref{eq:m_we}\textit{b}) in $M^\textsl{1D}_E$, (\ref{eq:d_we}\textit{b}) in $D^\textsl{1D}_E$, (\ref{eq:m_we}\textit{b}) in $M^\textsl{2D}_E$ and (\ref{eq:a_we}\textit{b}) in $A^\textsl{2D}_E$.
    The dashed line in (\textit{e}) represents the largest $\gamma$ for which $\Gamma_0(-\phi)=0$ when $\alpha \ll 1$.
    $c_0$, $\Gamma_0$, $\langle c_0 \rangle$ and $c_0(x, \, 0)$ are plotted in (\textit{f}) $A^\textsl{2D}_E$ (notice the \textit{classical} stagnant cap profile) and (\textit{g}) $D^\textsl{1D}_E$ (where streamwise diffusion has smoothed the stagnant cap), where $x=x_0$ is plotted using \eqref{eq:we_x_0}.}
    }
    \label{fig:11}
\end{figure}

When bulk--surface exchange is weak ($\nu = 0.01$), 2D effects are again prevalent in bulk concentration profiles at small $\alpha$ (figure \ref{fig:11}). 
However, the impact of bulk boundary layers on the drag reduction is more modest than in the strong-exchange limit, in that they have less of an effect on the leading-order drag reduction. 
Solutions at large $\gamma$ again have approximately linear interfacial surfactant profiles that immobilise the interface (figure \ref{fig:11}\textit{a--b}).
These profiles are almost decoupled from the bulk.
They act as a weak sink/source combination driving adsorption/desorption.  
Solutions with weak Marangoni effects (small $\gamma$) show the formation of a more classical stagnant cap at small $\alpha$ (figure \ref{fig:11}\textit{f}), which is smoothed by streamwise diffusion as $\alpha=\delta$ increase (figure \ref{fig:11}\textit{g}).

Figure \ref{fig:11}(\textit{c--e}) illustrates the dependency of ${DR}_0$ on $\alpha$ and $\gamma$, evaluated against existing limits $M_E^\textsl{1D}$ and $D_E^\textsl{1D}$ (see (\ref{eq:m_we}, \ref{eq:d_we})) for large $\alpha$ and introducing new limits $M_E^\textsl{2D}$ and $A_E^\textsl{2D}$ at small $\alpha$. 
Importantly, weak coupling between the bulk and the interface suppresses the dependence of $DR_0$ on $\alpha$ in the limit of very small bulk diffusivity, making the drag-reduction contours horizontal in figure~\ref{fig:11}(\textit{c}) for $\alpha\to 0$.

The following scaling argument applies to region $A_E^{2D}$ (figure \ref{fig:dr_shems}\textit{b} and \ref{fig:11}\textit{f}), where Marangoni effects and bulk diffusion are weak. 
Here, the interfacial concentration is approximated by the classical distribution for nearly insoluble surfactant, which is close to zero along the interface except at the downstream end where it forms a stagnant cap of length $L$ and slope $1/\gamma$.  
Thus, $\Gamma_0$ is of magnitude $L/\gamma$, and its integral over the whole interface is $O(L^2/\gamma)$, while the bulk concentration remains close to unity. 
The surfactant flux-balance constraint (\ref{eq:nd_fluxes_d1_0_wd}\textit{c}), which matches net absorption to net desorption over the interface, therefore requires $L^2/\gamma=O(1)$, implying that $\Gamma_0$ is $O(\gamma^{-1/2})$ and $DR_0$ is $O(\gamma^{1/2})$.  
This scaling argument, which mirrors the 1D case, is supported by asymptotic calculations in \S\ref{subsec:MA_WE} below, which yield 
\refstepcounter{equation} \label{eq:a_we}
\begin{equation} 
    c_0 \approx 1, \quad \Gamma_0 \approx \begin{cases} \ 0 \hspace{1.92cm} \text{for} \quad x\in[-\phi,\, x_0], \\ \displaystyle \ \frac{3\beta}{2\gamma}(x-x_0) \quad \text{for} \quad x\in[x_0,\,\phi], \end{cases} {DR}_0 \approx 1 - \left(\frac{2\gamma}{3\phi \beta}\right)^{1/2}, \tag{\theequation\textit{a--c}}
\end{equation}
where $\phi - x_0 =  (8 \phi\gamma /(3\beta))^{1/2}$ is the length of the stagnant cap.
Bulk diffusion has a higher-order effect, influencing exchange via adsorption and desorption, and hence weak stretching and compression of the interface.
In figure \ref{fig:11}(\textit{c--f}), dashed black asymptotes for $c_0$, $\Gamma_0$ and ${DR}_0$ match those results from the 2D long-wave model, almost up to $\gamma = 3\beta\phi/2$ in figure \ref{fig:11}(\textit{e}) (see vertical dashed lines), which denotes the conditions where the stagnant cap reaches $x=-\phi$ and the slip region of this non-linear distribution vanishes. 
The disappearance of the stagnant cap distribution with increasing $\gamma$ is linked to a rapid decrease in $DR_0$, owing to the loss of the slip region.
The same rapid transition is recovered if we approach region $A^{2D}$ from region $M_E^{2D}$, using (\ref{eq:m_we}\textit{b}), i.e. for decreasing $\gamma$.

\begin{figure}
    \centering
    \hfill \includegraphics[trim={0cm 0cm 0cm 0cm},clip,width=.49\textwidth]{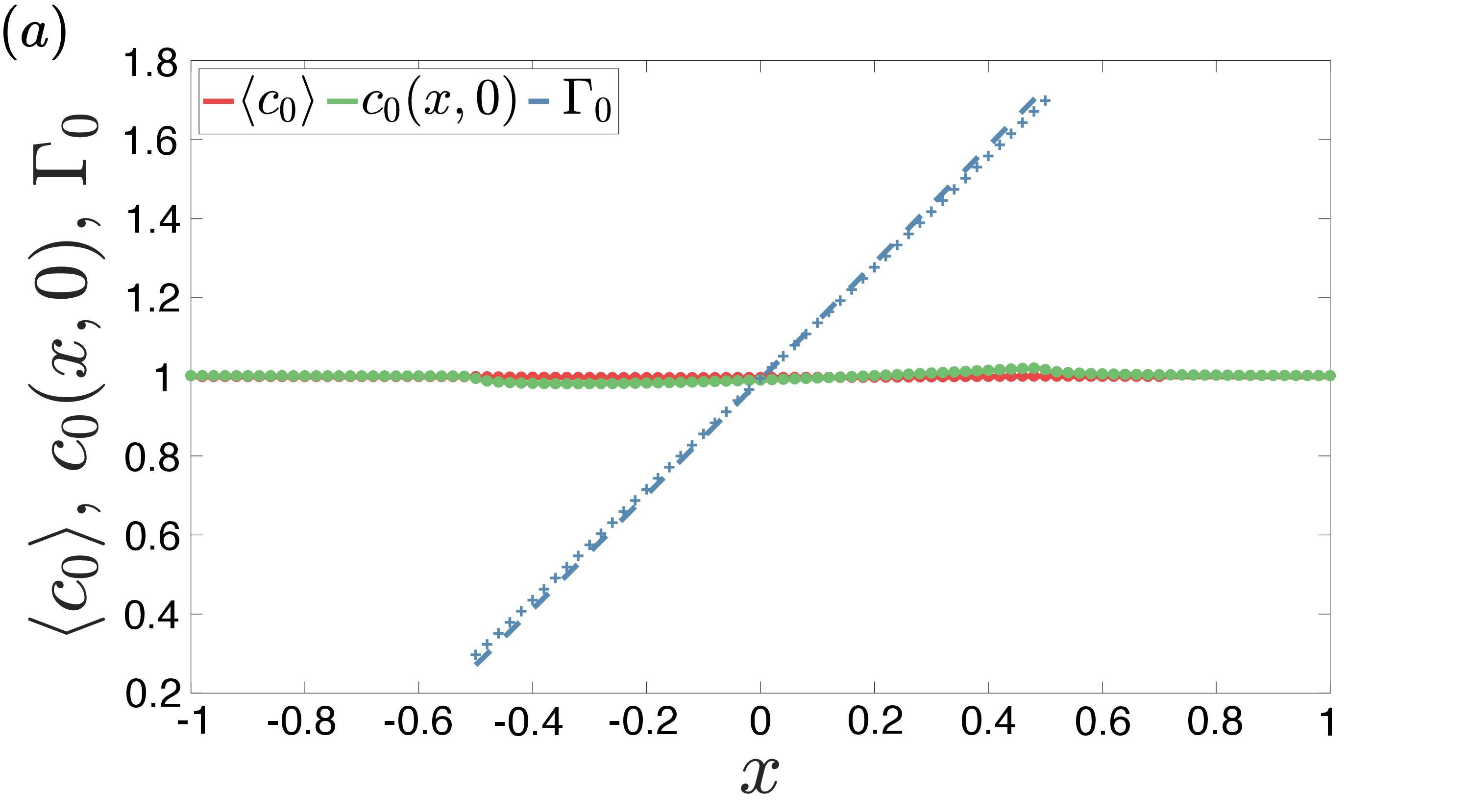}  \hfill \includegraphics[trim={0cm 0cm 0cm 0cm},clip,width=.49\textwidth]{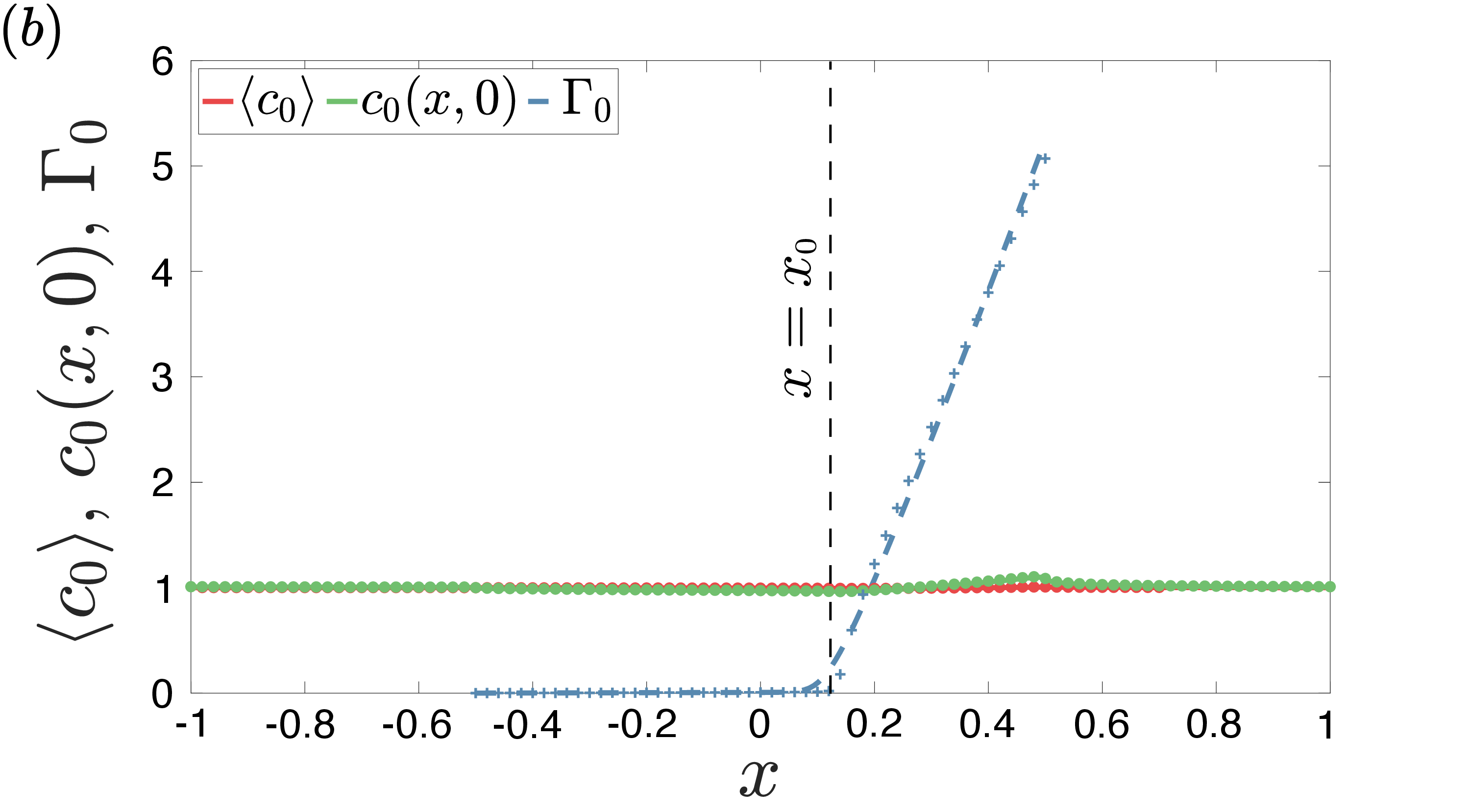} \hfill \hfill \hfill \\  
    \hfill \includegraphics[trim={0cm 0cm 0cm 0cm},clip,width=.49\textwidth]{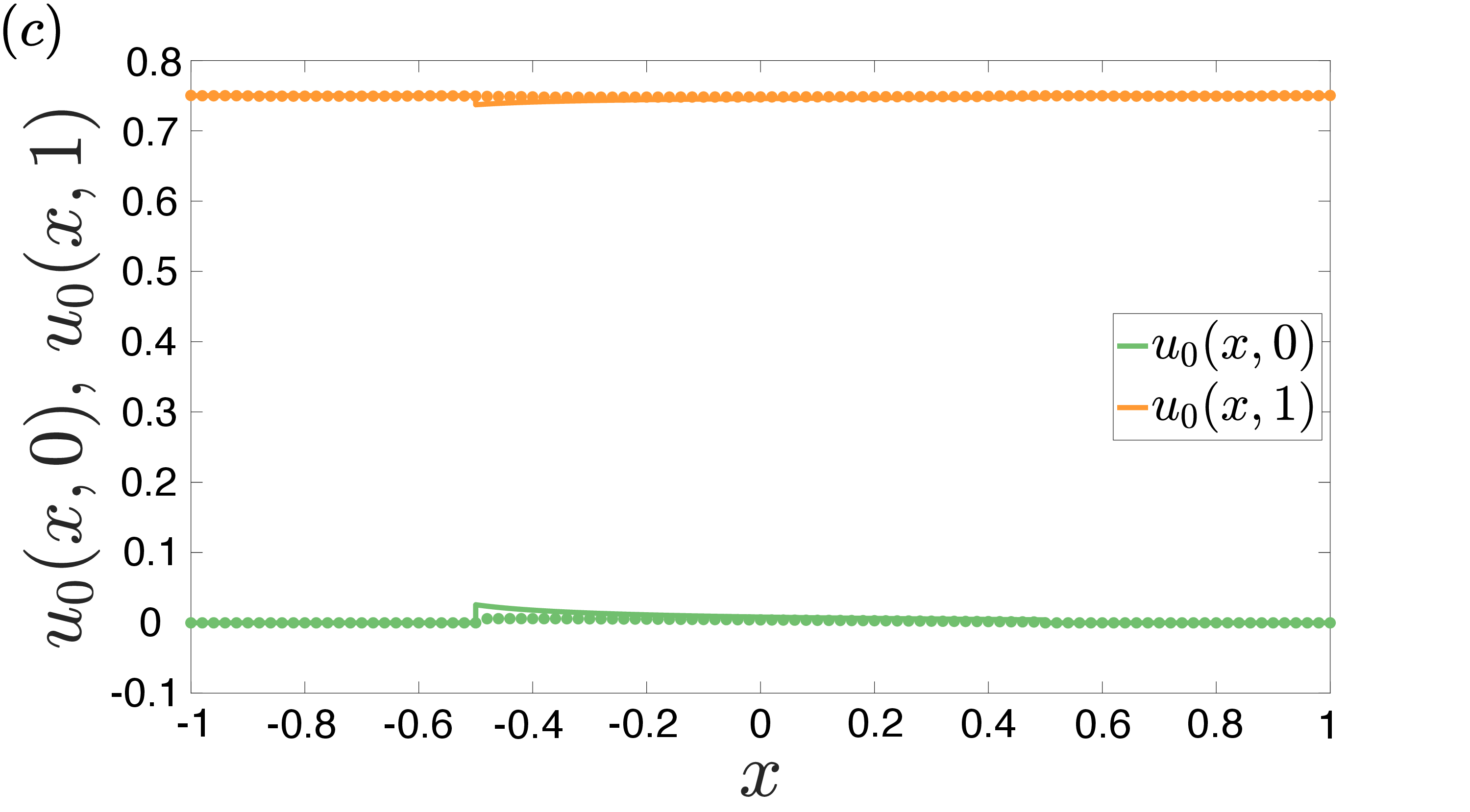} \hfill \includegraphics[trim={0cm 0cm 0cm 0cm},clip,width=.49\textwidth]{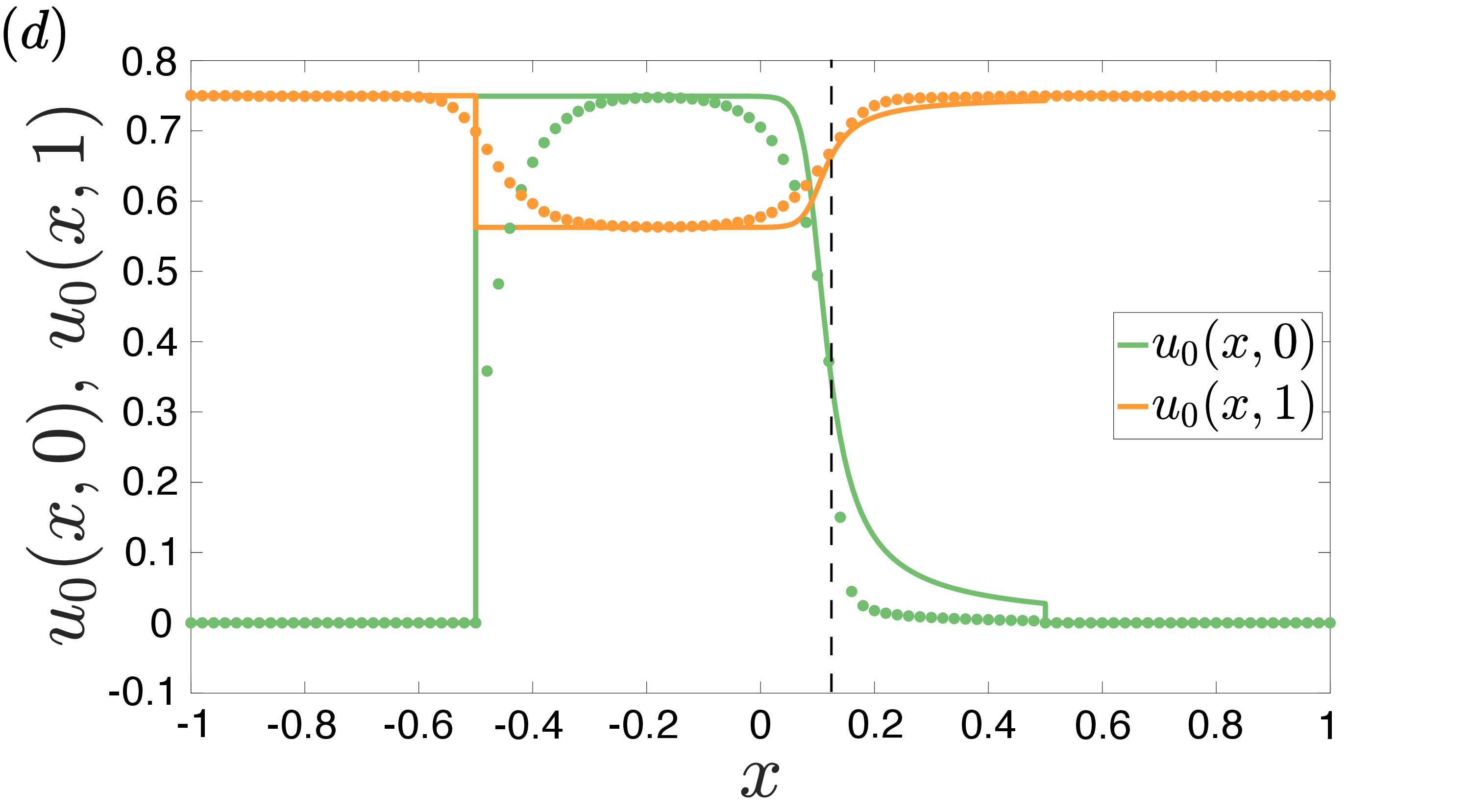} \hfill \hfill \hfill \\ 
    \caption{The leading-order streamwise velocity field ($u_0$) and surfactant concentration fields ($c_0$ and $\Gamma_0$) for $\beta = 1$, $\nu = 0.01$, $\epsilon = 0.1$ and $\phi = 0.5$ evaluated using \eqref{eq:c_bvp_1_wd}--\eqref{eq:c_bvp_6_wda} (lines) and COMSOL simulations \eqref{eq:nondimensional_equations}--\eqref{eq:nd_velocity_flux} (symbols) when bulk--surface exchange is weak. 
    (\textit{a,\,c}) Plots of $c_0$ and $u_0$, respectively, for $\alpha = 0.1$ and $\gamma = 1$, when the flow is in the Marangoni--dominated region with weak cross-channel diffusion ($M^\textsl{2D}_E$).
    (\textit{b,\,d}) Plots of $c_0$ and $u_0$, respectively, for $\alpha = 0.1$ and $\gamma = 0.1$, when the flow is in the advection--dominated region with weak cross-channel diffusion ($A^\textsl{2D}_E$) and $x = x_0$ is plotted using \eqref{eq:we_x_0}.
    }
    \label{fig:11_b}
\end{figure}

A comparison between the solutions evaluated using the 2D long-wave model, \eqref{eq:c_bvp_1_wd}--\eqref{eq:c_bvp_6_wda}, and COMSOL simulations is given in figure \ref{fig:11_b}, focussing on the examples discussed in figure \ref{fig:11}(\textit{a,\,f}).
Figure \ref{fig:11_b}(\textit{a,\,c}) shows $c_0$, $\Gamma_0$ and $u_0$ in region $M_E^\textsl{2D}$.
The slip velocity is approximately zero for both symbols and lines, as the interfacial surfactant gradient effectively immobilises the interface.
The 2D long-wave model predicts ${DR}_0 = 0.027$, while the numerical simulations yield ${DR}_{NS} = 0.056$ (due to the 4\% difference in $\Delta \Gamma_0$ shown in figure \ref{fig:11_b}\textit{a}).
Figure \ref{fig:11_b}(\textit{b,\,d}) depicts stagnant-cap solutions in region $A_E^\textsl{2D}$.
The upstream end of the liquid--gas interface ($x\in[-\phi,\, x_0]$) is shear-free, resulting in a slip velocity $u_0 = 3/4$ (substituting $\Gamma_{0x} = 0$ into (\ref{eq:u_def_wd}\textit{a}, \ref{eq:fluxes_wd}\textit{a}), $u_0 = \tilde{U} p_{0x}$, where $\tilde{U}(0) = - 2$ and $p_{0x} = - 3 / 8$).
Conversely, the downstream end of the liquid--gas interface ($x\in[x_0,\, \phi]$) does not exhibit slip, leading to a near-zero streamwise velocity at the downstream stagnation point.
The transitional and downstream regions induce a strong wall-normal flow that drives bulk surfactant into the core of the channel.
We find ${DR}_0 = 0.650$ and ${DR}_{NS} = 0.662$, demonstrating agreement between the predictions of the 2D long-wave model and COMSOL simulations, validating the model's accuracy under a wide range of conditions.

\subsection{Comparison with experiments} 

To facilitate comparison with experiments, in table \ref{tab:fernando_comp}, we convert ${DR}_0$ to slip lengths ($\lambda_e$) using the non-dimensionalisation in \citet{temprano2023single}. 
\citet{temprano2023single} proposed that for microchannel applications, $\lambda_e \sim (\phi/\epsilon)^2 / L_m^2$ (using our notation), where $\phi/\epsilon = \phi \hat{P}/\hat{H}$ is half the length of the liquid--gas interface divided by the channel height, $L_m^2 = (\hat{n}\hat{R}\hat{T} \hat{K}_a^2 \hat{c}_0)/(\hat{D}\hat{\mu}\hat{K}_d^2)$ is the squared mobilisation length and $\lambda_e \sim (\phi / \epsilon) \lambda_0$ in \eqref{eq:slip}. 
The slip-length formulae in \citet{landel2020theory} and \citet{sundin2022slip} have the same dependencies in this limit. 
We find that, in all regimes, the slip length shares the same dependence on $1/L_m^{2}$, with quadratic dependence on $\phi/\epsilon$ only when $1 \ll \alpha \ll 1/\epsilon^{2}$ and $\gamma \gg 1$ (region $M^\textsl{1D}$). 
However, when $\alpha \ll 1$ and $\gamma \gg \alpha^{1/6}$, $\lambda_e$ has a stronger (8/3 power) dependence on $\phi/\epsilon$ (region $M^\textsl{2D}$), and when $\alpha \gg 1/\epsilon^2$ and $\gamma \gg \epsilon^2 \alpha$, we find a weaker (linear) dependence (region $M^\textsl{1D}$). 
The remaining asymptotic solutions for ${DR}_0$ are summarised in figure \ref{fig:dr_shems}. 
When $1-{DR}_0 \ll 1$, the slip length simplifies to $\lambda_e = (\Delta p_I - \Delta p_U) / (\epsilon \Delta p_I \Delta p_U)$ ($A^\textsl{1D}$ and $A^\textsl{2D}$).
Hence, outside of the specific microchannel applications considered in \citet{temprano2023single}, whose predictions agree with those in region $M^\textsl{1D}$ (as discussed below), we have found new parameters in regions $M^\textsl{2D}$ that influence surfactant impairment of SHS drag reduction.
As diffusion weakens, the deviation between the drag-reduction predictions in $M^\textsl{1D}$ and $M^\textsl{2D}$ will increase.

\begin{table}
\centering
\resizebox{\columnwidth}{!}{%
    \centering
    \begin{tabular}{c c c c}
    & $M^\textsl{2D}$ $\left(\displaystyle \alpha \ll 1\right)$ & $M^\textsl{1D}$ $\left(\displaystyle 1\ll \alpha \ll \frac{1}{\epsilon^2}\right)$ & $M^\textsl{1D}$ and $M^\textsl{1D}_E$ $\left(\displaystyle \alpha \gg \frac{1}{\epsilon^2}\right)$ \\ [12pt]
    
    Drag reduction scaling  & $\displaystyle \frac{\phi^{5/3}\alpha^{2/3}}{\gamma} = \frac{\phi^{5/3}\hat{P}^{5/3}\hat{H}^{1/3} \hat{Q}^{1/3}}{\hat{L}_m^2 \hat{D}^{1/3}}$ & $\displaystyle\frac{\phi}{\gamma}=\frac{\phi \hat{P} \hat{H} \hat{Q}}{\hat{L}_m^2 \hat{D}} $ & $\displaystyle\frac{\epsilon^2\alpha}{\gamma}= \frac{\hat{H}^2}{\hat{L}_m^2}$ \\[12pt]
    Slip length scaling & $ \displaystyle\frac{b^{1/3} (\phi/\epsilon)^{8/3}}{L_m^2} $ & $\displaystyle \frac{b (\phi/\epsilon)^2}{L_m^2}$ & $\displaystyle\frac{\phi/\epsilon}{L_m^2}$ \\[12pt]
    \end{tabular}
    }
    \caption{
    Summary of the leading-order drag reduction $({DR}_0)$ scaling in regions $M^\textsl{2D}$, $M^\textsl{1D}$ and $M^\textsl{1D}_E$ and the corresponding leading-order slip length ($\lambda_e$) scaling using the non-dimensionalisation in \citet{temprano2023single}, in terms of the mobilisation length $L_m$.
    Hats denote dimensional quantities.
    The drag reduction is converted to the slip length using \eqref{eq:slip} for ${DR}_0 \ll 1$ and $\lambda_e \sim (\phi/\epsilon) \lambda_0$, where $\hat{L}_m^2 = (\hat{n}\hat{R}\hat{T} \hat{H}^2 \hat{K}_a^2 \hat{c}_0)/(\hat{D}\hat{\mu}\hat{K}_d^2)$, $\phi/\epsilon = \phi \hat{P}/\hat{H}$, $L_m = \hat{L}_m/\hat{H}$ and $b = \hat{Q}/\hat{D}$.}
    \label{tab:fernando_comp}
\end{table}

For laboratory experimental studies documented in the literature, we can use our 2D long-wave model to estimate surfactant effects via the drag reduction, employing parameters typical of microchannel applications in laminar flows.
Estimates for these parameters, based on the surfactant SDS, using models and experimental data from \citet{temprano2023single}, are given in table \ref{tab:exp_para}.
The remaining parameters $\hat{P}$, $\hat{Q}$, $\hat{H}$ and $\hat{K}$ vary across experiments. 
\citet{peaudecerf2017traces} use $\alpha \approx 1$ and $4.7\times 10^{1} \lessapprox \gamma \lessapprox 2.9\times10^4$ (at the $M^\textsl{1D}/M^\textsl{2D}$ boundary, figure \ref{fig:dr_shems}); and \citet{temprano2023single} use $\alpha \approx 25$ and $10^{-1} \lessapprox \gamma \lessapprox 3.7\times 10^{2}$ (at the $A^\textsl{1D}/M^\textsl{1D}$ boundary, figure \ref{fig:dr_shems}). 
Qualitatively, all of these studies report surfactant effects, which is consistent with their location in the non-dimensional parameter space.
All of these studies operate within the strong-exchange regime for the values of $\hat{K}_a$ and $\hat{K}_d$ given in table \ref{tab:exp_para}.
However, as $\hat{K}_a$ and $\hat{K}_d$ are merely estimates based on fitting to microchannel experiments, weak exchange could be achieved, for instance, when $\hat{K}_a = 3.4\times10^{-8}\SI{}{\metre \squared \per \second}$ and $\hat{K}_d = 7.5\times10^{-5}\SI{}{\per \second}$.
\st{Quantitatively, we compare predicted values of slip velocity using our theoretical model with experimental results in table \ref{tab:exp_comp}. 
Using the experimental parameters from \citet{temprano2023single}, where $\alpha \approx 25$ and $\epsilon \approx 0.002$, we confirm $1\ll \alpha \ll 1/\epsilon^2$ and they are within the regime $M^\textsl{1D}$.
From (\ref{eq:m}\textit{b}), we calculate ${DR}_0$, convert to $\lambda_e$ using table \ref{tab:fernando_comp}, and then to $u_{Ic}/u_{Ic}^\textmd{clean}$ following \citet{temprano2023single}.
Agreement is observed across $\phi/\epsilon = 253$, 418 and 587, which is remarkable as our theory does not include any fitting parameters. 
The deviation from \citet{temprano2023single} at  $\phi/\epsilon = 762$ could be attributed to experimental error, which is consistent with the expected behaviour of increasing slip with grating length.}

\begin{table}
    \centering
    \begin{tabular}{c c c c}
    Quantity & Symbol & Units & Value \\ [6pt]
    Adsorption rate & $\hat{K}_a$ & \SI{}{\metre \squared \per \second} & $3.4\times 10^{-4}$ \\ [6pt]
    Desorption rate & $\hat{K}_d$ & \SI{}{\per \second} & 0.75 \\ [6pt]
    Surface activity & $\hat{A}$ & \SI{}{\joule \per \mole} & 4913.6 \\ [6pt]
    Dynamic viscosity & $\hat{\mu}$ & \SI{}{\kilogram \per \metre \per \second} & $8.9\times10^{-4}$ \\ [6pt]
    Surface diffusivity & $\hat{D}$ & \SI{}{\metre \squared \per \second} & $7\times10^{-10}$ \\ [6pt]
    Bulk diffusivity & $\hat{D}_I$ & \SI{}{\metre \squared \per \second} & $7\times10^{-10}$ \\ [6pt]
    \end{tabular}
    \caption{A summary of the parameters in the dimensional problem, \eqref{eq:dimensional_domain}--\eqref{eq:dimensional_drag}, with their values based on the surfactant SDS, models and experimental data from \citet{temprano2023single}.}
    \label{tab:exp_para}
\end{table}

\begin{table}
\centering
    \centering
    \begin{tabular}{c c c c c}
     & $\phi/\epsilon = 253$ & $\phi/\epsilon = 418$ & $\phi/\epsilon = 587$ & $\phi/\epsilon = 762$ \\ [6pt]
     Peaudecerf  \textit{et al.} (2017) & - & - & $0.1410$ & -  \\ [6pt]
     Temprano-Coleto \textit{et al.} (2023) & $0.0661$ & $0.1187$ & $0.1903$ & $0.1450$ \\ [6pt]
     Current & 0.0610 & 0.1164 & 0.1859 & 0.2716 \\ [6pt]
    \end{tabular}
    \caption{\st{A comparison between the slip velocity normalised by the clean (surfactant-free) value, $u_{Ic}/u_{Ic}^\textmd{clean}$, predicted using the current model \eqref{eq:c_bvp_1_wd}--\eqref{eq:c_bvp_6_wda}, and experimental data from \citet{peaudecerf2017traces} and \citet{temprano2023single}. 
    The parameter $\phi/\epsilon$ is half the length of the liquid--gas interface divided by the channel height.}}
    \label{tab:exp_comp}
\end{table}

\section{Discussion} \label{discussion}

In this study, we develop a long-wave theory to analyse the behaviour of a 2D laminar pressure-driven channel flow contaminated with soluble surfactant. 
The channel is bounded by a SHS with streamwise-periodic grooves and a solid wall. 
We linearise the equation of state and adsorption--desorption kinetics, solving Stokes and advection--diffusion equations in the long-wave limit.
Our investigation focuses primarily on regimes where concentration gradients in the streamwise and cross-channel directions are comparable.
By numerically solving the 2D long-wave model, we delineate asymptotic regions within the parameter space. 
Under conditions of strong cross-channel diffusion, our findings align with the 1D results in \citet{tomlinson2023laminar}.
Conversely, when cross-channel diffusion is weak, we unveil new regions of the parameter space, where the drag reduction exhibits a non-trivial dependence on the thickness of the bulk-concentration boundary layer and surfactant strength (discussed below).
We derive asymptotic solutions for the boundary-layer problem, validating them against both the numerical solution of the 2D long-wave model \eqref{eq:c_bvp_1_wd}--\eqref{eq:c_bvp_6_wda} and the full Stokes and advection--diffusion equations \eqref{eq:nondimensional_equations}--\eqref{eq:nd_velocity_flux} in COMSOL.
These complementary methodologies, ranging from asymptotic solutions to long-wave numerical simulations, give physical insight for applications where numerical simulations can be computationally prohibitive.

We have explored how the interfacial concentration ($\Gamma_0$), bulk concentration ($c_0$) and leading-order drag reduction (${DR}_0$) are influenced by bulk diffusion ($\alpha$), surface advection ($\beta$), surfactant strength ($\gamma$), interfacial diffusion ($\delta$) and bulk--surface exchange ($\nu$) (figures \ref{fig:12} and \ref{fig:11}); the dimensionless parameters $\alpha$, $\beta$, $\gamma$, $\delta$ and $\nu$ are given in table \ref{tab:1}.
In cases where cross-channel diffusion dominates, we recover the Marangoni-dominated ($M^\textsl{1D}$ and $M^\textsl{1D}_E$), advection-dominated ($A^\textsl{1D}$ and $A^\textsl{1D}_E$) and diffusion-dominated ($D^\textsl{1D}$ and $D^\textsl{1D}_E$) regions identified in \citet{tomlinson2023laminar} (see figure~\ref{fig:dr_shems}).  
However, under weak cross-channel diffusion, we unveil new subregions dominated by Marangoni effects ($M^\textsl{2D}$ and $M^\textsl{2D}_E$) and advection ($A^\textsl{2D}$ and $A^\textsl{2D}_E$).
In these regions, a concentration boundary layer forms in the bulk, influencing the drag reduction when bulk--surface exchange is strong.
When Marangoni effects dominate, the interface is immobilised and the surfactant distribution along it is approximately linear.
The asymptotic solutions for ${DR}_0$ are summarised in figure \ref{fig:dr_shems}. 

When both bulk diffusion and Marangoni effects are weak, part of the interface is shear-free and the surfactant distribution forms either what we call a quasi-stagnant cap (when bulk--interface exchange is strong) or a classical stagnant cap (when exchange is weak).  
\st{The classical stagnant cap for an insoluble surfactant-contaminated SHS has been described in \citet{baier2021influence} and \citet{mayer2022superhydrophobic}.
The quasi-stagnant cap has received much less attention and has a particularly intricate structure, illustrated in figure \ref{fig:asym_shems}(\textit{b}) above.  
Surface stretching and compression accommodate weak adsorption from, and desorption to, the bulk across thin concentration boundary layers, `remobilising' the interface (analogous to a mechanism described by \cite{crowdy2023fast} at low $\Pen$ --- and so without the involvement of bulk boundary layers --- for a linear extensional flow).} 
The upstream `slip' region has a bulk-concentration boundary layer of thickness $O(\alpha^{1/2})$, which is to be expected above an almost fully mobile interface.  
Weak stretching draws surfactant from the bulk to the interface via diffusive adsorption.   
The slip region transitions abruptly to a quasi-stagnant region, which grows to a thickness $O(\gamma^{1/5}\alpha^{3/10})$. 
This differs from a classical stagnant cap (of a fully insoluble surfactant) in having weak surface compression at a rate that balances diffusive desorption.  
The transition from the mobile to the quasi-immobile interface takes place across two nested regions at the tip of the quasi-stagnant cap: a short deceleration region, where the surface velocity falls abruptly, displacing the bulk boundary layer upwards towards the core; and a slightly longer transition region, across which shear in the boundary layer balances weakening surface advection (figure~\ref{fig:asym_shems}$b$). 
Shear then dominates in the bulk-concentration boundary layer along the quasi-stagnant region.

It is likely that the physical balances arising in these regions may emerge in other flow configurations involving soluble surfactant transport near confined interfaces at high P\'{e}clet numbers, such as the cap forming at the rear of a rising drop or bubble.  
For example, computations by \citet{oguz1988effects} and \citet{tasoglu2008effect} revealed quasi-stagnant caps at large but finite $\Pen$.  
Interfacial flux balances between diffusion-limited adsorption (slip) and desorption (stagnant region) were given by \cite{harper2004} and \cite{palaparthi2006theory}, treating the size of the cap as a parameter. 
Here, we determine the size of the cap by solving the bulk-concentration boundary layer, which leads (for example) to the interfacial surfactant concentration being of size $\beta/(\gamma^{2/5}\alpha^{1/10})$ in the $A^\textsl{2D}$ regime.

In summary, our study compares asymptotic and numerical solutions for a 2D laminar pressure-driven channel flow, confined by a streamwise-periodic SHS and a solid wall, and contaminated with soluble surfactant. 
While numerical solutions demand significant computational resources, our asymptotic solutions provide a cost-effective alternative, particularly in regimes at high P\'eclet numbers where accurate numerical solutions of very thin boundary layers are computationally very demanding. 
These asymptotic solutions offer accurate predictions devoid of empirical fitting coefficients and provide physical insights into the mechanisms governing drag reduction across a large part of the parameter space, including in particular high-P\'eclet-number flow regimes.

\section*{Acknowledgements}

We acknowledge support from CBET--EPSRC (EPSRC Ref. EP/T030739/1, NSF \#2054894), as well as partial support from ARO MURI W911NF-17-1-0306.
For the purpose of open access, the authors have applied a Creative Commons Attribution (CCBY) licence to any Author Accepted Manuscript version arising.

\section*{Declaration of interests}

The authors report no conflict of interest.

\appendix

\section{Numerical solutions for weak cross-channel diffusion} \label{sec:numerical}

The numerical solution to \eqref{eq:c_bvp_1_wd}--\eqref{eq:c_bvp_6_wda} can be divided into two subroutines, employing Chebyshev collocation techniques as outlined by \citet{trefethen2000spectral}. 
First, given $\Gamma_0^{\text{old}}$, we evaluate $c_0^{\text{new}}$ in $\mathcal{D}_1$ and $\mathcal{D}_2$. 
The subdomains of the periodic cell are mapped to the 2D canonical Chebyshev collocation domain, $\mathcal{D}_n = \{\xi\in[-1, \, 1]\} \times \{\eta\in[-1, \, 1]\}$, using the transformations
\begin{equation}  
\label{eq:transformations}
    (\xi_1, \, \xi_2, \, \eta) = \left( (x + \phi)/\phi - 1, \, (x-\phi)/(1-\phi) - 1, \, y - 1\right).
\end{equation}
We then discretise using $N = (N_\xi + 1)\times(N_\eta + 1)$ nodes in each subdomain, at points
\begin{equation} 
\label{eq:points}
    (\xi_{1,\,i}, \, \xi_{2,\,i}, \, \eta_j) = (\cos(i \pi / N_\xi), \, \cos(i \pi / N_\xi), \, \cos(j \pi / N_\eta)),
\end{equation}
where $i = 0, \, 1, \, ..., \, N_\xi$ and $j = 0, \, 1, \, ..., \, N_\eta$.
Continuity boundary conditions \eqref{eq:nondimensional_periodicity_0_wd} are enforced on $c_0^{\mathrm{new}}$ at $i=0$ and $N_\xi$, supplemented with continuity of bulk diffusive surfactant flux between subdomains
\refstepcounter{equation}
\begin{equation} 
    c_{0, \, 1}^{\text{new}} = c_{0, \, 2}^{\text{new}}, \quad \frac{\partial c_{0, \, 1}^{\text{new}}}{\partial \xi_1} = \frac{\partial c_{0, \, 2}^{\text{new}}}{\partial \xi_2}. \tag{\theequation\textit{a,\,b}}
\end{equation}
Boundary conditions \eqref{eq:c_bvp_2_wd}--\eqref{eq:c_bvp_4_wd} are enforced at $j = 0$ and $N_\eta$, ensuring continuity of surfactant flux at the liquid-gas interface, no-flux of surfactant at the solid wall in $\mathcal{D}_1$ and no-flux of surfactant at the solid walls in $\mathcal{D}_2$.
Discretising the bulk surfactant equation \eqref{eq:c_bvp_1_wd} in conservative form in domains $\mathcal{D}_1$ and $\mathcal{D}_2$ and incorporating the discretised matching conditions (represented by $\mathsfbi{M}_{12}$ and $\mathsfbi{M}_{21}$) gives 
\begin{equation} 
\begin{pmatrix}
    \mathsfbi{A}_1 & \mathsfbi{M}_{12} \\
    \mathsfbi{M}_{21} & \mathsfbi{A}_{2}
\end{pmatrix} \begin{pmatrix}
    \boldsymbol{c}_{0, \, 1}^{\text{new}} \\
    \boldsymbol{c}_{0, \, 2}^{\text{new}}
\end{pmatrix} = \begin{pmatrix}
    \boldsymbol{f}_{1} \\
    \boldsymbol{0}
\end{pmatrix},
\end{equation}
where $\boldsymbol{f}_{1}$ is the forcing due to the interfacial surfactant concentration $\Gamma_0^{\text{old}}$ in (\ref{eq:c_bvp_2_wd}\textit{a}).

Second, given a $c_0^{\text{new}}$, we evaluate $\Gamma_0^{\text{new}}$ for $x\in[-\phi, \, \phi]$ and $y=0$. 
The liquid--gas interface is mapped to the 1D canonical Chebyshev collocation domain, $\mathcal{D}_i = \{\xi\in[-1, \, 1]\}$, using $\xi_1$ in \eqref{eq:transformations}, discretised using $N_\xi + 1$ nodes at points $\xi_{1,\,i}$ in \eqref{eq:points}.  
Discretising the interfacial surfactant equation (\ref{eq:c_bvp_2_wd}\textit{b}) for $x\in[-\phi, \, \phi]$ and $y=0$ and incorporating the discretised no-flux conditions \eqref{eq:c_bvp_3_wd}, results in
\begin{equation} 
    \mathsfbi{B}(\boldsymbol{\Gamma}_{0}^{\text{old}}) \boldsymbol{\Gamma}_{0}^{\text{new}} = \boldsymbol{g},
\end{equation}
where $\boldsymbol{g}$ is the forcing due to the bulk surfactant concentration $c_0^{\text{new}}$ in (\ref{eq:c_bvp_2_wd}\textit{b}).

Therefore, to evaluate ${DR}_0$ for a given $\alpha$, $\beta$, $\gamma$, $\delta$, $\epsilon$, $\nu$ and $\phi$, we choose an initial guess based on the 1D long-wave model discussed in \S\ref{subsec:Strong cross-channel diffusion and moderate exchange}.
The initial guess can be substituted into the forcing $\boldsymbol{f}_1$ in the bulk concentration subroutine and the linear system inverted to calculate $\boldsymbol{c}_0^{\text{new}}$.
The bulk concentration can then be substituted into the forcing $\boldsymbol{g}$ in the interfacial concentration subroutine and the linear system inverted to calculate $\boldsymbol{\Gamma}_{0}^{\text{new}}$.
We then set $\boldsymbol{\Gamma}_{0}^{\text{old}}=\boldsymbol{\Gamma}_{0}^{\text{new}}$ and repeat the above steps.
This procedure of updating $\boldsymbol{c}_0$ and $\boldsymbol{\Gamma}_{0}$ is then continued until convergence is achieved for both the bulk and interfacial concentration fields. 

\section{Key results of the 1D long-wave model} \label{sec:1d_asym}

Here we outline key limits of the 1D long-wave model (\ref{eq:c_bvp_6_sd}, \ref{eq:model3}), exploiting analogous findings as in \cite{tomlinson2023laminar}.
For strong exchange, with $\alpha \gg 1$ and $\gamma \gg \max(1, \, \epsilon^2 \alpha)$, a region dominated by Marangoni effects ($M^{\textsl{1D}}$, figure \ref{fig:dr_shems}\textit{a}) is characterised by the immobilisation of the liquid--gas interface.
The surfactant distribution and drag reduction are given by
\refstepcounter{equation}  \label{eq:m}
\begin{equation} 
    \bar{c}_0 \approx \bar{\Gamma}_0 \approx 1 + \frac{3\beta}{2\gamma}\left(x - \frac{\phi(E+1)}{(E-1)}\right), \quad {DR}_0 \approx \frac{2}{\gamma}\left(\epsilon^2(2\alpha+\delta) + \frac{\phi(E+1)}{(E-1)}\right), \tag{\theequation\textit{a,\,b}}
\end{equation}
where $E\equiv\exp((1-\phi)/(\epsilon^2\alpha))$.
Thus, ${DR}_0 \approx 2\phi/\gamma$ for $1\ll \alpha \ll 1/\epsilon^2$ and ${DR}_0\approx 2\epsilon^2(2\alpha + \delta)/\gamma$ for $\alpha \gg 1/\epsilon^2$.
Region $M^{\textsl{1D}}$ transitions into a diffusion-dominated region ($D^{\textsl{1D}}$, figure \ref{fig:dr_shems}\textit{a}) when $\alpha \gg 1/\epsilon^2$ and $\gamma \sim\epsilon^2\alpha$, and into an advection-dominated region ($A^{\textsl{1D}}$, figure \ref{fig:dr_shems}\textit{a}) when $1 \ll \alpha \ll 1/\epsilon^2$ and $\gamma \sim 1$. 
For $\gamma \ll \epsilon^2 \alpha$ and $\alpha \gg 1/\epsilon^2$ (region $D^{\textsl{1D}}$),
\refstepcounter{equation} \label{eq:d}
\begin{equation} 
    \bar{c}_0 \approx \bar{\Gamma}_0 \approx \frac{4\alpha +2\delta (1 -\phi)}{\alpha(4+3\beta\phi) +2\delta(1-\phi)}, \ \ \ {DR}_0 \approx 1 - \frac{\gamma(1-\phi)}{\epsilon^2(\alpha(4+3\beta\phi) +2\delta(1-\phi))}, \tag{\theequation\textit{a,\,b}}
\end{equation}
and for $\gamma \ll 1$ and $1 \ll \alpha \ll 1/\epsilon^2$ (region $A^{\textsl{1D}}$),
\refstepcounter{equation} \label{eq:a}
\begin{equation} 
    \bar{c}_0 \approx \bar{\Gamma}_0 \approx \frac{4}{3\beta + 4} + \frac{3\beta}{3\beta+4}\exp\left(\frac{(4+3\beta)(x-\phi)}{4\epsilon^2(2\alpha + \delta)}\right), \quad {DR}_0 \approx 1 - \frac{\gamma}{\phi (3\beta+4)}. \tag{\theequation\textit{a,\,b}}
\end{equation}
The onset of 2D effects in the 1D long-wave model was estimated in \citet{tomlinson2023laminar} to arise via the emergence of shear dispersion, where $\alpha \sim 1$ and $\gamma \ll 1$ or $\alpha \sim 1/\gamma$ and $\gamma \gg 1$.  
We refine such estimates by solving the 2D long-wave model, which reveals regions $M^\textsl{2D}$ and $A^\textsl{2D}$ shown in figure \ref{fig:dr_shems}(\textit{a}).

A similar picture emerges when bulk--surface exchange is weak. 
A region dominated by Marangoni effects ($M^\textsl{1D}_{E}$, figure \ref{fig:dr_shems}\textit{b}) arises for $\delta \gg 1$ and $\gamma \gg \max(1,\,\epsilon^2 \alpha)$.  
Here, $\bar{c}_0$ decouples from $\bar{\Gamma}_0$ and is approximately equal to its background value (i.e. $\bar{c}_0 \approx 1$). 
$\bar{\Gamma}_0$ varies around this value to fulfil the net adsorption--desorption condition, \eqref{eq:nd_fluxes_d1_0_wd}, decreasing the drag reduction compared to the strong-exchange problem 
\refstepcounter{equation}  \label{eq:m_we}
\begin{equation} 
    \bar{c}_0 \approx 1, \quad \bar{\Gamma}_0 \approx 1 + \frac{3\beta x}{2\gamma}+O(\epsilon^2), \quad {DR}_0 \approx \frac{2\epsilon^2 \delta}{\gamma}. \tag{\theequation\textit{a--c}}
\end{equation}
The region $M^\textsl{1D}_{E}$ transitions to a diffusion-dominated region ($D^\textsl{1D}_{E}$, figure \ref{fig:dr_shems}\textit{b}) when $\gamma \sim \epsilon^2 \delta$ and $\delta \gg 1/\epsilon^2$ and an advection-dominated region ($A^\textsl{1D}_{E}$, figure \ref{fig:dr_shems}\textit{b}) when $\gamma \sim 1$ and $1\ll \delta \ll 1/\epsilon^2$.
For $\gamma \ll  \epsilon^2 \delta$ and $\delta \gg 1/\epsilon^2$ (region $D_E^{\textsl{1D}}$),
\refstepcounter{equation} \label{eq:d_we}
\begin{equation} 
    \bar{c}_0 \approx 1, \quad \bar{\Gamma}_0 \approx 1 + \frac{3\beta x}{4\epsilon^2 \delta}, \quad {DR}_0 \approx 1 - \frac{\gamma}{2\epsilon^2\delta}, \tag{\theequation\textit{a--c}}
\end{equation}
and for $\gamma \ll 1$ and $1\ll \delta \ll 1/\epsilon^2$ (region $A_E^{\textsl{1D}}$),
\refstepcounter{equation} \label{eq:a_we_1D}
\begin{equation} 
    \bar{c}_0 \approx 1, \quad \bar{\Gamma}_0 \approx \begin{cases} \ 0 \hspace{1.92cm} \text{for} \quad x\in[-\phi,\, x_0], \\ \displaystyle \ \frac{3\beta}{2\gamma}(x-x_0) \quad \text{for} \quad x\in[x_0,\,\phi], \end{cases} {DR}_0 \approx 1 - \left(\frac{2\gamma}{3\phi \beta}\right)^{1/2}, \tag{\theequation\textit{a--c}}
\end{equation}
where $x_0 = \phi - (8 \phi\gamma /(3\beta))^{1/2}$ gives the length of the stagnant cap.

\section{Asymptotic solutions for weak cross-channel diffusion} \label{sec:asymptotic}

We derive asymptotic solutions to the 2D long-wave model, \eqref{eq:c_bvp_1_wd}--\eqref{eq:c_bvp_6_wda}, when bulk diffusive effects are weak and confined to a boundary layer along $y=0$.
In this limit, streamwise diffusion terms do not appear at leading order, which means that we do not resolve inner regions around $x = -\phi$, $\phi$ and $2-\phi$.
First, we consider the strong-exchange regime, where Marangoni effects (\st{Appendix \ref{sec:asymptotic_m}}, figure~\ref{fig:asym_shems}$a$) and advection (\st{Appendix \ref{sec:asymptotic_a}}, figure~\ref{fig:asym_shems}$b$) dominate.  
Second, we consider the weak-exchange regime (Appendix~\ref{subsec:MA_WE}).
Throughout, we assume $\beta=O(1)$ and $\phi = O(1)$.

Assuming that $c_0 \approx 1$ in the core of the channel, we anticipate that the transport equations \eqref{eq:c_bvp_1_wd} for the bulk surfactant involve a diffusive boundary layer near $y=0$.
The diffusive boundary layer is made up of components over the interface and ridge, which are governed by
\begin{subequations} 
\label{eq:bl_1}
\begin{align}
\alpha c_{0yy} - \left(\frac{3}{4}+\frac{\gamma}{\beta}\left(-\frac{1}{2} + y\right)\Gamma_{0x}\right)c_{0x} - \frac{\gamma}{\beta}\left(\frac{y}{2} - \frac{y^2}{2}\right)\Gamma_{0xx} c_{0y} &=0 \quad \text{in} \quad \mathcal{D}_1, \\
\alpha c_{0yy} - {3y} c_{0x}/2 &=0  \quad \text{in} \quad \mathcal{D}_2,
\end{align}
\end{subequations}
where we have assumed $y\ll 1$ and have neglected second (third) order terms in $O(y^2)$ ($O(y^3)$) in the streamwise (wall-normal) velocity. 
For $x\in[-\phi, \, \phi]$ and $y=0$, the bulk--interface flux condition and the interfacial transport equation for the surfactant in \eqref{eq:c_bvp_2_wd} give
\begin{equation} 
\label{eq:bl_2}
    \alpha c_{0y} = \nu(c_0 - \Gamma_0) = \beta\bigg[\left(\frac{3}{4}-\frac{\gamma}{2\beta}\Gamma_{0x}\right) \Gamma_0 \bigg]_x. 
\end{equation}
At $x=\pm\phi$ and $y=0$, the no-surfactant-flux conditions at the stagnation points \eqref{eq:c_bvp_3_wd} become
\begin{equation} 
\label{eq:bl_3}
    \left(\frac{3}{4}-\frac{\gamma}{2\beta}\Gamma_{0x}\right) \Gamma_0 =0.   
\end{equation}
For $x\in[\phi, \, 2 - \phi]$ and $y=0$, the no-flux condition at the solid wall \eqref{eq:c_bvp_4_wd} gives
\begin{equation} 
\label{eq:bl_4}
	c_{0y} =0.
\end{equation}
For $x\in[-\phi, \, 2 - \phi]$ and $y\rightarrow1$ (shorthand for $y$ values in the core of the channel, far outside the boundary layer along $y=0$), the core condition is
\begin{equation} 
\label{eq:bl_5}
	c_0 \rightarrow 1.
\end{equation}
The boundary-layer equations \eqref{eq:bl_1}--\eqref{eq:bl_5} can be integrated across the boundary layer to derive excess-surfactant-flux conditions. 
The total flux constraint (\ref{eq:c_bvp_6_wda}\textit{b,\,c}) is satisfied at leading order in the core of the channel, implying that 
\begin{subequations} 
\label{eq:bl_6}
\begin{align}
\beta\left(\frac{3}{4}-\frac{\gamma}{2\beta}\Gamma_{0x}\right) \Gamma_0 + \int_{0}^1 \left(\frac{3}{4}+\frac{\gamma}{\beta}\left(-\frac{1}{2} + y\right)\Gamma_{0x}\right)\left(c_0-1\right) \, \text{d}y &= 0 \quad \text{in} \quad \mathcal{D}_1, \\
\int_{-\phi}^\phi c_{0y} \, \text{d} x = \int_{-\phi}^\phi (c_0-\Gamma_0) \, \text{d} x &= 0 \quad \text{in} \quad \mathcal{D}_1, \\
\int_{0}^1 \frac{3}{2}y\left(c_0-1\right) \, \text{d}y &= 0  \quad \text{in} \quad \mathcal{D}_2,
\end{align}
\end{subequations}
provided $c_0\rightarrow 1$ sufficiently quickly where the boundary layer meets the core flow. 
Equation (\ref{eq:bl_6}\textit{b}) is derived by integrating \eqref{eq:bl_2} and using \eqref{eq:bl_3}. 
The condition in (\ref{eq:bl_6}\textit{a}) indicates that the extra surfactant flux carried by the flow at the interface balances the flux deficit in the boundary layer (where $c_0\leq 1$), which happens over the upstream part of the interface. 
Conversely, over the downstream part of the interface, the reduced interfacial surfactant flux is compensated by a surplus of bulk surfactant in the boundary layer (where $c_0\geq 1$). 
This can be seen in the boundary-layer profiles at small $\alpha$ values, as depicted in figures \ref{fig:12}(\textit{a,\,f}) and \ref{fig:11}(\textit{a,\,f}). 
The condition in (\ref{eq:bl_6}\textit{b}) specifies that the total amounts of surfactant adsorbed and desorbed along the interface must match, with continuity between the bulk diffusive flux (first integral) and the adsorption--desorption flux (second integral).  
Motivated by the simulations and scaling arguments provided in \S\ref{sec:strong exchange}, we propose the asymptotic structures for regions $M^\textsl{2D}$ and $A^\textsl{2D}$ outlined in figure \ref{fig:asym_shems} and derive asymptotic solutions of the surfactant distributions in the following subsections for the different regions in the parameter space where boundary layers exist.

\subsection{Strong Marangoni effect and strong bulk--surface exchange: region $M^\textsl{2D}$} \label{sec:asymptotic_m}

We first consider the boundary-layer equations, \eqref{eq:bl_1}--\eqref{eq:bl_6}, in the limit where bulk diffusion is weak and bulk--surface exchange is strong, i.e. $\alpha \ll 1$ and $\nu$ large. 
Suppose the interfacial surfactant gradient leads to a significant adverse Marangoni effect, immobilising the interface.  
Surfactant adsorbs onto the interface along its upstream part and desorbs along the downstream part. 
Motivated by the scaling arguments given in \S\ref{sec:strong exchange}, we set
\refstepcounter{equation} \label{eq:bl_m_exp}
\begin{equation} 
    \Gamma_0 = M_1 + \frac{3 \beta x}{2 \gamma} - \alpha^{2/3} G(x) + ..., \quad c_0 = C(x, \, Y), \tag{\theequation\textit{a,\,b}}
\end{equation}
where $y = \alpha^{1/3} Y$ balances advection and diffusion in the bulk-surfactant equation \eqref{eq:bl_1} and $M_1$ is a constant to be determined. 
Substituting \eqref{eq:bl_m_exp} into \eqref{eq:bl_1}--\eqref{eq:bl_6}, \eqref{eq:bl_1} becomes, at leading order,
\begin{equation} 
\label{eq:bl_1_m}
C_{YY} - 3Y C_x/2  =0.
\end{equation}
For $x\in[-\phi, \, \phi]$ and $Y=0$, we enforce $\Gamma_0 = c_0$ as bulk--surface exchange is strong, so that (\ref{eq:bl_2}) becomes
\begin{equation} 
\label{eq:bl_2_m}
    C_{Y} - \frac{\gamma}{2}\left[G_x\left(\frac{3\beta x}{2\gamma} + M_1 \right)\right]_x= 0.   
\end{equation}
At $x=\pm\phi$ and $Y=0$, the no-flux conditions \eqref{eq:bl_3} at the stagnation points are
\begin{equation} 
\label{eq:bl_3_m}
    G_x =0,   
\end{equation}
and for $x\in[\phi, \, 2 - \phi]$ and $Y=0$, the no-flux condition \eqref{eq:bl_4} at the solid wall is
\begin{equation} 
\label{eq:bl_4_m}
	C_{Y} =0.
\end{equation}
In the bulk of the periodic domain, for $x\in[-\phi, \, 2 - \phi]$ and $Y\rightarrow\infty$, the core condition \eqref{eq:bl_5} becomes 
\begin{equation} 
\label{eq:bl_5_m}
	C \rightarrow 1.
\end{equation}
The excess-surfactant-flux conditions \eqref{eq:bl_6} reduce to 
\begin{subequations} 
\label{eq:bl_6_m}
\begin{align}
\int_{0}^\infty \frac{3 Y}{2 }\left(C-1\right) \, \text{d}Y + \frac{G_x}{2}\left(\frac{3\beta x}{2} + \gamma M_1 \right) &= 0 \quad \text{in} \quad \mathcal{D}_1, \\
\int_{0}^\infty \frac{3Y}{2}\left(C-1\right) \, \text{d}Y &= 0  \quad \text{in} \quad \mathcal{D}_2.
\end{align}
\end{subequations}

We construct a similarity solution to \eqref{eq:bl_1_m}--\eqref{eq:bl_6_m}, assuming that the length of the interface is long compared to the boundary-layer thickness, and that there is no interaction between adjacent plastrons.
The bulk concentration is split into two contributions that grow from the leading edge of the plastron,
\begin{equation}  
\label{eq:ss_m}
    C = 1 - \left(1 - M_1 + \frac{3\beta \phi}{2\gamma} \right)F(\eta) + \frac{3\beta}{2\gamma}(x+\phi)H(\eta)
\end{equation}
for some functions $F$ and $H$ depending on the similarity variable $\eta = Y/(x+\phi)^{1/3}$.  
The function $F$ captures the response of the bulk field to the sudden drop in concentration along $Y=0$ at $x=-\phi$; $H$ captures the response to the linear profile of $\Gamma_0$ in $x>-\phi$.
Substituting \eqref{eq:ss_m} into \eqref{eq:bl_1_m}--\eqref{eq:bl_5_m} leads to the boundary-value problems
\refstepcounter{equation} \label{eq:f_m}
\begin{equation} 
\vspace{-.6cm}
    F_{\eta\eta} + \eta^2 F_\eta / 2 = 0, \quad F(0)=1, \quad F(\infty) =0, \tag{\theequation\textit{a--c}}
\end{equation}
\refstepcounter{equation} \label{eq:g_m}
\begin{equation} 
    H_{\eta\eta} + \eta^2 H_\eta / 2 - 3 \eta H/2 = 0, \quad H(0)=1, \quad H(\infty) =0, \tag{\theequation\textit{a--c}} 
\end{equation}
which have solutions
\refstepcounter{equation} \label{eq:FandG}
\begin{equation}  
    F = \frac{\Gammaup(\tfrac{1}{3}, \, \eta^3/6)}{\Gammaup(\tfrac{1}{3})}, \quad H = \frac{ (4 + \eta^3) \Gammaup(\tfrac{1}{3}, \, \eta^3/6) -6^{2/3} \eta \exp(-\eta^3/6) }{4 \Gammaup(\tfrac{1}{3})}. \tag{\theequation\textit{a,\,b}}
\end{equation}
Here $\Gammaup(z)$ is the gamma function and $\Gammaup(s,\,z)$ is the upper incomplete gamma function. 
We then use the surfactant-flux conditions \eqref{eq:bl_6_m} to calculate the response of the interfacial concentration field to the bulk field \eqref{eq:ss_m},
\begin{multline} 
\label{eq:bl_f_x}
    \frac{\gamma G_x}{2}\left(\frac{3\beta x}{2 \gamma} + M_1\right) = \frac{3}{2} (x+\phi)^{2/3}\left(1 - M_1 + \frac{3\beta\phi}{2\gamma}\right)\int_{0}^\infty \eta F \, \text{d}\eta \\ - \frac{9\beta}{4\gamma}(x+\phi)^{5/3} \int_{0}^\infty \eta H \, \text{d}\eta.
\end{multline}
The total flux of surfactant must be continuous between domains. 
The no-flux condition \eqref{eq:bl_3_m} at $x=\phi$ requires that
\begin{equation} 
\label{eq:bl_m_def}
    M_1 = 1 + \frac{3\beta\phi}{2\gamma}\left(\int_{0}^\infty \eta F \, \text{d}\eta- 2 \int_{0}^\infty \eta H \, \text{d}\eta\right)\bigg/\int_{0}^\infty \eta F \, \text{d}\eta = 1 - \frac{3\beta \phi}{10\gamma}.
\end{equation}
Integrating the total-flux condition \eqref{eq:bl_f_x} across the plastron reveals that
\begin{multline} 
    \Delta G = \int_{-\phi}^{\phi}\bigg\{\frac{27\beta\phi}{10\gamma} (x+\phi)^{2/3}\int_{0}^\infty \eta F \, \text{d}\eta \\ - \frac{9\beta}{4\gamma}(x+\phi)^{5/3} \int_{0}^\infty \eta H \, \text{d}\eta\bigg\}\bigg/\left[\frac{\gamma}{2}\left(1 + \frac{3\beta}{2 \gamma}\left(x - \frac{\phi}{5}\right)\right)\right]\,\text{d}x.
\end{multline}
Hence, in region $M^\textsl{2D}$, substituting \eqref{eq:bl_m_exp} into \eqref{eq:dr_def}, the leading-order interfacial concentration and drag reduction are given by
\refstepcounter{equation}
\label{eq:bl_m_gamma}
\begin{equation} 
    \Gamma_0 = 1 + \frac{3\beta}{2\gamma}\left(x - \frac{\phi}{5}\right) + ..., \quad 
    {DR}_0 = \frac{\gamma\alpha^{2/3}\Delta G}{3\phi\beta}. \tag{\theequation\textit{a,\,b}}
\end{equation}
The leading-order interfacial surfactant distribution $\Gamma_0$ in region $M^\textsl{2D}$ is a linear profile with mean value close to 1, as found numerically (see figure~\ref{fig:12_b}\textit{a}). This approximation requires $\Gamma_0>0$ at $x=-\phi$, so that $\gamma>9\beta\phi/5$.
For $\gamma \gg 1$, (\ref{eq:bl_m_gamma}) reduces to 
\begin{equation} 
    \Delta G= \frac{81\beta (2\phi)^{8/3}}{50\gamma^2}\int_{0}^\infty \eta F \, \text{d}\eta - \frac{27\beta (2\phi)^{8/3}}{16\gamma^2} \int_{0}^\infty \eta H \, \text{d}\eta \equiv \frac{\Delta J}{\gamma^2},
\end{equation}
\refstepcounter{equation}
\label{eq:bl_m_dr_gg}
\begin{equation} 
    \Gamma_0 = 1 + \frac{3\beta}{2\gamma}\left(x - \frac{\phi}{5}\right) + ..., \quad
    {DR}_0 = \frac{\alpha^{2/3}\Delta J(\beta, \, \phi)}{3\phi\beta\gamma} \equiv \frac{m_1 \alpha^{2/3}\phi^{5/3}}{\gamma}, \tag{\theequation\textit{a,\,b}}
\end{equation}
where $m_1 = 81 (3/2)^{2/3}/(50 \mathrm{\Gamma}(\tfrac{1}{3})) \approx 0.79$, as in \eqref{eq:m_e}. 
This expression for the leading-order drag reduction (\ref{eq:bl_m_dr_gg}\textit{b}) assumes strong Marangoni effects ($\gamma\gg 1$), weak diffusion ($\alpha\ll 1$) and strong bulk--surface exchange, and shows how drag reduction in this limit arises from differences between the two self-similar components $F$ and $H$ of the bulk concentration field. As mentioned previously, this asymptotic prediction agrees with the numerical simulations in the $M^\textsl{2D}$ region shown in figure~\ref{fig:12}(\textit{a,\,c,\,d,\,e}).

\subsection{Weak Marangoni effect and strong bulk--surface exchange: region $A^\textsl{2D}$} \label{sec:asymptotic_a}

We now consider the boundary-layer equations \eqref{eq:bl_1}--\eqref{eq:bl_6} when $\gamma \ll 1$ and $\alpha \ll 1$.
Weak Marangoni effects imply that the interface cannot support any stress over most of its length.  
Consequently, there exists a slip flow along most of the plastron, characterised by a slip velocity in (\ref{eq:bl_1}\textit{a}) close to $3/4$. A short transition region, mediated by surface diffusion, exists between a no-slip and a slip flow at the upstream contact line, but this is disregarded here. 
Surfactant adsorbs onto the interface in the slip region (in $-\phi<x<x_0$, say) and accumulates (and desorbs) in a region at the downstream end of the plastron (in $x_0<x<\phi$). 
Across this ``quasi-stagnant" region, the surface velocity smoothly decreases to zero at the downstream contact line.   
Motivated by simulations and the scaling arguments given in \S\ref{sec:strong exchange}, we propose the asymptotic structure illustrated in figure \ref{fig:asym_shems}(\textit{b}), in which ``deceleration" and ``transition" regions are nested at the tip of the quasi-stagnant region over length scales indicated.   
The formal requirement that the deceleration, transition and quasi-stagnant layers are nested ($\gamma \alpha^{1/2}\ll \gamma^{3/4}\alpha^{1/8}\ll \gamma^{3/5}/\alpha^{1/10} \lesssim 1$)
is $\gamma \ll \alpha^{-3/2}$. 
We now discuss these regions in turn.

\subsubsection{The slip region $(-\phi<x<x_0)$} \label{subsubsec:slip}

In the slip region, surfactant adsorbs onto the interface, causing the interface to carry a spatially growing component of the overall flux.
Consequently, the bulk advective flux falls with $x$. 
Here, we set 
\refstepcounter{equation} \label{eq:scaling_slip}
\begin{equation} 
    \Gamma_0=\alpha^{1/2}G(x), \quad c_0=C(x,Y), \tag{\theequation\textit{a,\,b}}
\end{equation} 
where $y=\alpha^{1/2}Y$ balances advection and diffusion in the bulk-transport equation \eqref{eq:bl_1}. 
When we substitute \eqref{eq:scaling_slip} into \eqref{eq:bl_1}--\eqref{eq:bl_6}, the bulk-transport equation \eqref{eq:bl_1} in $\mathcal{D}_1$ becomes
\begin{equation} 
\label{eq:bl_1_a}
C_{YY} - 3 C_x/4 =0.
\end{equation}
At the interface $\alpha^{1/2} G(x) = C(x,0)$ in the strong-exchange limit and the flux condition \eqref{eq:bl_2} reduces to
\begin{equation} 
\label{eq:bl_2_a} 
    C_{Y} (x,0)= 3\beta G_x/4.  
\end{equation}
The no-flux condition at the upstream stagnation point \eqref{eq:bl_3} is
\begin{equation} 
\label{eq:bl_3_a} 
    G(-\phi) =0.   
\end{equation}
For $x\in[-\phi, \, x_0]$ and $Y\rightarrow\infty$, the core condition \eqref{eq:bl_5} is
\begin{equation} 
\label{eq:bl_5_a}
	C \rightarrow 1.
\end{equation}

We construct a similarity solution to \eqref{eq:bl_1_a}--\eqref{eq:bl_5_a} by expanding the surfactant fields for $\alpha \ll 1$ as
\refstepcounter{equation} \label{eq:slip_surf}
\begin{equation} 
    C(x,Y)=C_0(\eta)+(\alpha(x+\phi))^{1/2}C_1(\eta)+..., \quad G(x)=(x+\phi)^{1/2}C_1(0), \tag{\theequation\textit{a,\,b}}
\end{equation}
with the similarity variable given by $\eta=Y/(\phi+x)^{1/2}$.
Substituting \eqref{eq:slip_surf} into \eqref{eq:bl_1_a}--\eqref{eq:bl_5_a} yields the boundary-value problems
\refstepcounter{equation} \label{eq:ss_a_1}
\begin{equation} 
\vspace{-.6cm}
C_0''+3\eta C_0'/8=0, \quad C_0(0)=0, \quad C_0(\infty) = 1, \tag{\theequation\textit{a--c}} 
\end{equation}
\refstepcounter{equation} \label{eq:ss_a_2}
\begin{equation} 
C_1''+3\eta C_1'/8-3C_1/8=0, \quad 3\beta C_1(0) = 8C_0'(0), \quad C_1(\infty)=0. \tag{\theequation\textit{a--c}}
\end{equation}
These are solved by
\refstepcounter{equation} \label{eq:erf}
\begin{equation} 
C_0=\text{erf}(\sqrt{3}\eta/4), \quad C_1=4\text{ierfc}(\sqrt{3}\eta/4)/(\sqrt{3}\beta), \tag{\theequation\textit{a,\,b}}
\end{equation}
where $\text{erf}(z)$ is the error function, $\text{erfc}(z)$ is the complementary error function and $\mathrm{ierfc}(z) \equiv e^{-z^2}/\sqrt{\pi}-z\,\text{erfc}(z)$ is the integrated complementary error function. 
We find that the interfacial concentration field in the slip region is given by
\begin{equation} \label{eq:gamma_slip}
\Gamma_0=\frac{4}{\beta} \left(\frac{\alpha(x+\phi)}{3 \pi}\right)^{1/2}+...,
\end{equation}
contributing to (\ref{eq:a_e}$a$).  
The interfacial surfactant distribution $\Gamma_0$ in region $A^\textsl{2D}$ is a non-linear boundary layer profile scaling as $\alpha^{1/2}x^{1/2}$ in the slip region of the interface (up to $x_0$), as found numerically (see figure~\ref{fig:12_b}\textit{b}). 
There is a large vertical concentration gradient across the boundary layer.  
Adsorption of surfactant onto the interface is accommodated by the horizontal gradient of the advective flux $\beta u_0 \Gamma_0 \sim 3\beta\Gamma_0/4$.
Marangoni effects remain subdominant, so the slip flow is uniform to this order, but the concentration rises with $x$ to accommodate the adsorbed material.  
The total flux adsorbed across the slip region, up to the location $x=x_0$ (to be determined), is given by
\begin{equation} 
\label{eq:surf_ads}
\int_{-\phi}^{x_0} c_{0y}(x,0)\,\text{d}x= \left(\frac{9\alpha(x_0+\phi)}{3 \pi}\right)^{1/2}+...,
\end{equation}
which is subdominant to the flux in the core.
The slip region terminates abruptly, meeting a very short deceleration region near $x=x_0$ nested at the start of the quasi-stagnant region (figure~\ref{fig:asym_shems}\textit{b}).  

\subsubsection{Deceleration region $( x-x_0 =O(\gamma\alpha^{1/2}))$}\label{sec:decelerationregion}

In the deceleration region  (figure~\ref{fig:asym_shems}\textit{b}), $u_0(x,\,0)$ varies by $O(1)$ over a short horizontal scale $\Delta x$ and $\Gamma_0\sim\alpha^{1/2}$ (where $\sim$ denotes `scales like'), based on continuity with the slip region.
Thus, $\Delta x\sim \gamma\alpha^{1/2}$ for $\gamma \Gamma_{0x}$ to vary by $O(1)$ in \eqref{eq:bl_2}.  
Bulk diffusion balances advection by $u_0$ over a vertical lengthscale $y\sim (\alpha \Delta x)^{1/2}\sim \gamma^{1/2}\alpha^{3/4}$ using \eqref{eq:bl_1}.  
Diffusion therefore influences the concentration field 
near the interface, while advection dominates transport in the upper part of the boundary layer. 
At the downstream limit of this region, $u_0$ falls towards zero, with $\Gamma_{0x}\approx 3\beta/(2\gamma)$.
The deceleration region is too short for there to be appreciable desorption so flux conservation requires that $u_0\sim \alpha^{1/2}/\Gamma_0\sim\alpha^{1/2} \gamma/(x-x_0)$ at the outlet of the region.
We write the interfacial flux as
\refstepcounter{equation} \label{eq:lambda}
\begin{equation} 
u_0(x,\,0) \Gamma_0 =\frac{3}{\beta}\left(\frac{\alpha(x_0+
\phi)}{3\pi}\right)^{1/2} \equiv \frac{3 \alpha^{1/2}}{4} \lambda,\quad \text{where} \quad \lambda = \frac{4}{\beta}\left(\frac{x_0+\phi}{3\pi}\right)^{1/2},\tag{\theequation\textit{a,\,b}}
\end{equation} 
as $u_0 = 3/4 + ...$ and $\Gamma_0$ is \eqref{eq:gamma_slip} in the slip region.
We then rescale, using 
\refstepcounter{equation} \label{eq:scalings_decel}
\begin{equation}  
    x=x_0+\gamma\alpha^{1/2} X, \ \ \  y=\gamma^{1/2}\alpha^{3/4} Y, \ \ \ \Gamma_0=\alpha^{1/2}\lambda G(X), \ \ \ c_0=\alpha^{1/2}\lambda C(X,Y). \tag{\theequation\textit{a--d}} 
\end{equation}
Substituting \eqref{eq:scalings_decel} into  \eqref{eq:bl_1}--\eqref{eq:bl_6}, the bulk-surfactant equation \eqref{eq:bl_1} in $\mathcal{D}_1$ is given by
\begin{equation} 
\label{eq:bl_1_a_dec}
C_{YY} - \left(\frac{3}{4}-\frac{\lambda}{2\beta} G_X\right) C_X  - \frac{\lambda Y}{2 \beta}G_{XX} C_Y =0,
\end{equation} 
neglecting terms of $O(\alpha^{3/4})$.
We have $G(X) = C(X,0)$ in the strong-exchange limit and the interfacial-surfactant equation \eqref{eq:bl_2} simplifies to
\begin{equation} 
\label{eq:bl_2_a_dec} 
    \left[G\left(\frac{3}{4} - \frac{\lambda}{2\beta} G_X \right)\right]_X  = 0.
\end{equation}
Integrating \eqref{eq:bl_2_a_dec} and matching it to the slip region, where $G \rightarrow 1$ for $X\rightarrow -\infty$, gives
\begin{equation} 
\label{eq:bl_3_a_dec}
     \frac{3}{4}G - \frac{\lambda}{2\beta} G G_X   = \frac{3}{4},
\end{equation}
which can be solved directly to give 
\begin{equation} \label{eq:GGG}
    G+\log(G-1)=(3\beta/(2\lambda))(X-X_0)
\end{equation}
for some $X_0$ (as in \cite{jensen1998stress1}, for example).
Substituting $G$ into (\ref{eq:lambda}\textit{a}), we compare $u_0(x,\,0)$ with the COMSOL simulations in figure \ref{fig:12_b}(\textit{d}), showing how $u_0$ falls steeply from $3/4$. 
The core condition \eqref{eq:bl_5} becomes
\begin{equation} 
\label{eq:bl_4_a_dec}
	C \rightarrow \infty\quad \mathrm{as}\quad Y\rightarrow \infty.
\end{equation}

Equations \eqref{eq:bl_1_a_dec}--\eqref{eq:bl_4_a_dec} constitute a nonlinear boundary-layer problem that describes the abrupt ejection of bulk concentration towards the core of the channel in figure \ref{fig:12}(\textit{f}).  
We only consider the downstream limit of \eqref{eq:bl_1_a_dec}--\eqref{eq:bl_4_a_dec} here.  
The upstream condition on $C$ is provided from the near-wall component of the concentration in the slip region, which grows in amplitude with $Y$ in \eqref{eq:erf}.  
The downstream limit has self-similar form
\refstepcounter{equation} \label{eq:dowsntream_ss}
\begin{equation} 
G\approx \frac{3\beta}{2\lambda}(X-X_0)+..., \quad C\approx \frac{3\beta}{2\lambda}(X-X_0)\hat{C}(\eta)+.... \tag{\theequation\textit{a,\,b}}
\end{equation}
where the similarity variable is given by $\eta={Y}/(X-X_0)$ for some $X_0$ and $\hat{C}$ satisfies $\hat{C}_{\eta\eta}=(\lambda/2\beta)\hat{C}$, which has solutions $\exp[\pm \sqrt{\lambda/2\beta}\eta]$.  Concentration contours are constant along lines $Y=\sqrt{2\beta/\lambda}(X-X_0)$, representing ejection of bulk surfactant away from the interface.  
Expressed in terms of the original variables using \eqref{eq:scalings_decel}, the downstream limit \eqref{eq:dowsntream_ss} of this region yields 
\refstepcounter{equation} \label{eq:decout2}
\begin{equation} 
\Gamma_0 \approx \frac{3\beta}{2\gamma}(x-x_0)\quad u_0(x,\,0) \approx \frac{\lambda \alpha^{1/2} \gamma}{2\beta(x-x_0)} \quad \mathrm{for}\quad \gamma\alpha^{1/2}\ll x-x_0\ll \gamma^{3/4}\alpha^{1/8}, \tag{\theequation\textit{a,\,b}} 
\end{equation}
which delivers an interfacial advective flux $3\lambda\alpha^{1/2}/4$ into a very short transition region near $x=x_0$.  
Surface diffusion is a subdominant effect in this region provided $\epsilon^2 \ll \gamma/\alpha^{1/2}$.  
The leading-order interfacial concentration profile in the deceleration region, (\ref{eq:decout2}\textit{a}), contributes to the second part of (\ref{eq:a_e}\textit{a}) (for $x_0\leq x$) and (\ref{eq:decout2}\textit{b}) explains the steep fall in surface velocity in figure~\ref{fig:12_b}(\textit{d}) at the end of the slip region; the incident boundary layer of thickness $\alpha^{1/2}$ is thickened to $\alpha^{1/2}/u_0$ as $u_0$ diminishes (see the arrow in figure~\ref{fig:12}\textit{f}).
The surface velocity in this regime, computed using \eqref{eq:lambda} with $\Gamma_0$ evaluated using \eqref{eq:bl_3_a_dec}, is shown in figure \ref{fig:12_b}(\textit{d}).

\subsubsection{Transition region $(x- x_0=O(\gamma^{3/4}\alpha^{1/8}))$}\label{sec:transitionregion}

In the transition region, there is a balance in the bulk between vertical diffusion, advection by $u_0$ (which is $o(1)$) and advection by shear (which is affected by viscous and Marangoni effects). 
Hence, over a horizontal lengthscale $\Delta x$, $\alpha/y^2\sim u_0/\Delta x\sim y/\Delta x$; thus $y\sim u_0\sim (\alpha \Delta x)^{1/3}$.  
Matching $u_0$ to the deceleration region requires $u_0\sim \alpha^{1/2}\gamma/\Delta x$, implying $\Delta x\sim \gamma^{3/4}\alpha^{1/8}$ and $y\sim u_0\sim\gamma^{1/4}\alpha^{3/8}$.  
The interfacial concentration is linear at leading order ($\Gamma_0\approx 3\beta(x-x_0)/(2\gamma))$ with a correction of size $u_0 \Delta x/\gamma\sim \alpha^{1/2}$.
The bulk concentration is set by the dominant interfacial concentration and is $O(\Delta x/\gamma)\sim \alpha^{1/8}/\gamma^{1/4}$.  
The region is again too short for there to be appreciable desorption, so the interfacial flux is still $O(\alpha^{1/2})$. 
To provide a bridge between the deceleration region and the quasi-stagnant region further downstream, we reintroduce shear $3Y/2$ to the bulk transport equation, using the scaling
\refstepcounter{equation} \label{eq:rescaling_trans}
\begin{multline} 
x=x_0+\alpha^{1/8}\gamma^{3/4}X, \quad y=\alpha^{3/8}\gamma^{1/4}Y, \\ \Gamma_0=\frac{3\beta}{2\gamma}(x-x_0)+\alpha^{1/2}G(X), \quad c_0=\frac{\alpha^{1/8}}{\gamma^{1/4}}C(X,Y). \tag{\theequation\textit{a--d}}
\end{multline}
Substituting \eqref{eq:rescaling_trans} into  \eqref{eq:bl_1}--\eqref{eq:bl_6}, the bulk transport equation \eqref{eq:bl_1} in $\mathcal{D}_1$ simplifies to
\begin{equation} 
\label{eq:transbulk}
C_{YY} + \left(\frac{G_X}{2\beta} -\frac{3Y}{2} \right)C_X- \frac{G_{XX}}{2\beta} Y C_Y = 0.
\end{equation}
Again $C(X,0)=G(X)$ and the interfacial transport equation \eqref{eq:bl_2} becomes
\begin{equation} 
\label{eq:decel}
- \left[\frac{3\beta X}{4}G_X\right]_X = 0.
\end{equation}
Using \eqref{eq:decel}, there is no exchange of surfactant with the bulk to leading order. 
Integrating \eqref{eq:decel} in $X$ and matching to the deceleration region, $XG_X \rightarrow \lambda$ as $X\rightarrow-\infty$. This can be integrated in $X$ to give $G=-\lambda \log X+A_1$ for some constant $A_1$, which we substitute into \eqref{eq:transbulk}, to obtain
\begin{equation} 
C_{YY}-\frac{\lambda}{2\beta} \left(\frac{1}{X}C_X -\frac{YC_Y}{X^2}\right) -\frac{3}{2}YC_X = 0.
\end{equation}
This linear boundary-layer problem bridges the scalings $Y\sim X$ for $X \ll 1$ and $Y\sim X^{1/3}$ for $X\gg 1$, but we do not solve the full problem here.
Expressed in terms of original variables using \eqref{eq:rescaling_trans}, the downstream limit of the transition region demonstrates that
\refstepcounter{equation} \label{eq:transout}
\begin{equation} 
\Gamma_0\approx \frac{3\beta(x-x_0)}{2\gamma}, \quad u_0(x,\,0)\approx \frac{\lambda \alpha^{1/2}\gamma}{2\beta(x-x_0)}\quad\mathrm{for} \quad \gamma^{3/4}\alpha^{1/8}\ll x-x_0\ll 1.\tag{\theequation\textit{a,\,b}}
\end{equation}
Note that the interfacial concentration profile in this sub-region, (\ref{eq:transout}\textit{a}), remains linear and exactly the same as in the deceleration region, see (\ref{eq:transout}\textit{a}), by continuity. 
Over this part of the interface, the interfacial flux is dominated by advection: $u_0(x,0)\Gamma_0\approx 3\lambda\alpha^{1/2}/4$.
This flux is delivered to the rest of the interface, in the last region that we call the quasi-stagnant region for $x_0 < x < \phi$. 
The drag reduction can only be determined once the interfacial concentration profile across the whole interface is determined.

\subsubsection{The quasi-stagnant region ($x_0<x<\phi$)}

In the quasi-stagnant region, compression of the almost-immobile interface promotes desorption, returning surfactant to the bulk. 
Therefore, the interfacial flux of surfactant decreases and the bulk flux increases with $x$.  Following the scaling arguments and numerical simulations outlined in \S\ref{sec:strong exchange}, we write 
\refstepcounter{equation} \label{eq:qs_rescaling}
\begin{equation} 
    \Gamma_0=\frac{3\beta}{2\gamma}(x-x_0)+..., \quad c_0=\frac{3\beta}{2\gamma}(x-x_0)C(\eta) + ..., \tag{\theequation\textit{a,\,b}}
\end{equation}
where the similarity variable $\eta=y/(\alpha(x-x_0))^{1/3}$ is obtained by balancing advection and diffusion in the bulk-transport equation \eqref{eq:bl_1}.
Substituting \eqref{eq:qs_rescaling} into \eqref{eq:bl_1}--\eqref{eq:bl_6}, the bulk-surfactant equation \eqref{eq:bl_1} in $\mathcal{D}_1$ simplifies  to
\refstepcounter{equation} \label{eq:c_for_main}
\begin{equation} 
C_{\eta\eta} + \eta^2 C_{\eta}/2 - 3\eta C/2  = 0, \quad C(0)=1, \quad C(\infty) = 0. \tag{\theequation\textit{a--c}} 
\end{equation}
Equation \eqref{eq:c_for_main} is the same as \eqref{eq:g_m}, whose solution is given in (\ref{eq:FandG}\textit{b}).
At the interface $x\in[x_0, \, \phi]$ and $Y=0$, the interfacial-surfactant equation \eqref{eq:bl_2} becomes
\begin{equation} 
\label{eq:bl_2_qs}
\beta \left[ u_0(x,\,0)(x-x_0) \right]_x = (\alpha(x-x_0))^{2/3}C'(0).
\end{equation}
Integrating \eqref{eq:bl_2_qs} and applying the no-flux condition $u_0(\phi,\,0)=0$, gives
\begin{equation} 
\label{eq:bl_3_qs}
\beta  u_0(x,\,0) = - \frac{3\alpha^{2/3} C'(0)}{5} \left(\frac{(\phi-x_0)^{5/3}}{x-x_0}-(x-x_0)^{2/3}\right),
\end{equation} 
where $C'(0) = - 3^{5/3}/(2^{4/3}\mathrm{\Gamma}(\tfrac{1}{3}))\approx - 0.92$.  
This profile successfully captures the COMSOL prediction in figure \ref{fig:12_b}(\textit{d}).  
In figure \ref{fig:12_b}(\textit{d}), the small discrepancy between the asymptotic results (\ref{eq:lambda}, \ref{eq:bl_3_qs}) and the 2D long-wave model arises from the retention of surface diffusion in its numerical approximation.
Matching \eqref{eq:bl_3_qs} to the surface velocity in the transition region \eqref{eq:transout} yields
\begin{equation} 
5\lambda \gamma= -6 \alpha^{1/6} C'(0) (\phi-x_0)^{5/3}.
\label{eq:findx0}
\end{equation}
The constants $\lambda$ and $x_0$ have to be evaluated numerically for a given $\alpha$, $\beta$, $\gamma$ and $\phi$ using \eqref{eq:lambda} and \eqref{eq:findx0}.
Thus, in the quasi-stagnant region of $A^\textsl{2D}$, the interfacial concentration is given by
\begin{equation}  
\label{eq:gamma_qs_cap}
    \Gamma_0 \approx \frac{3\beta}{2\gamma}\left(x - x_0\right) = \frac{3\beta}{2\gamma}\left(x - \phi + \left(\frac{- 5\lambda \gamma}{6\alpha^{1/6}C'(0)}\right)^{3/5} \right),
\end{equation}
using \eqref{eq:findx0}.
As anticipated, the leading-order interfacial surfactant concentration \eqref{eq:gamma_qs_cap} obtained in the quasi-stagnant cap region is linear and equal to the profiles obtained in the two previous short sub-regions: the deceleration  (\S\ref{sec:decelerationregion}) and the transition regions (\S\ref{sec:transitionregion}). 
This linear profile agrees with our numerical results in region $A^\textsl{2D}$ for the quasi-stagnant part of the interface, i.e. for $x_0< x<\phi$, as depicted in figure~\ref{fig:12_b}(\textit{b}).
Only now can we substitute \eqref{eq:gamma_qs_cap} into \eqref{eq:dr_def}, to obtain the leading-order drag reduction in region $A^\textsl{2D}$, which is given by
\begin{equation} 
\label{eq:a1_def}
    {DR}_0 = 1 - \frac{1}{2\phi}\left(\frac{-5\lambda}{6 C'(0)} \right)^{3/5}\frac{\gamma^{6/10}}{\alpha^{1/10}} \equiv 1 - \frac{a_1 \gamma^{6/10}}{\alpha^{1/10}},
\end{equation}
as in (\ref{eq:a_e}\textit{b}).  
Equations \eqref{eq:gamma_qs_cap} and \eqref{eq:a1_def} show how bulk diffusion, captured in the non-dimensional coefficient $\alpha^{1/10}$, has a weak but non-zero influence on the interfacial surfactant profile that determines $DR_0$.
The solution is valid as long as the layers are nested ($\gamma \lesssim \alpha^{-3/2}$) and $x_0 > -\phi$, such that $1-{DR}_0 < 1$.
In order to determine $x_0$, $\lambda$ and $a_1$, we solve \eqref{eq:lambda} and \eqref{eq:findx0} numerically to get $\lambda$ and $x_0$ for a given $\alpha$, $\beta$, $\gamma$ and $\phi$, from which, we can evaluate $a_1 = (-5\lambda/(6C'(0)))^{3/5}/(2\phi)$ and therefore ${DR}_0$.

In summary, to acquire the solution in the quasi-stagnant region, where the surfactant gradient decelerates the flow towards the downstream stagnation point, \eqref{eq:gamma_qs_cap}--\eqref{eq:a1_def}, we have resolved a nested asymptotic structure which includes deceleration (\S\ref{sec:decelerationregion}) and transition (\S\ref{sec:transitionregion}) regions.
The nested asymptotic structure has been matched to a slip region at the upstream end of the plastron (\S\ref{subsubsec:slip}), where the surfactant gradient has a weak effect on the flow and slips from the upstream stagnation point.
These four asymptotic regions are detailed in figure \ref{fig:asym_shems}(\text{b}), which we used to describe and explain the numerical solution that we see in figure \ref{fig:12}(\textit{f}).
Using the expressions for $u_0(x,\,0)$ and $\lambda$ in the quasi-stagnant and slip regions, \eqref{eq:bl_3_qs} and (\ref{eq:lambda}\textit{b}),  respectively, the surfactant desorbed across the quasi-stagnant region is given by 
\begin{equation} 
    \int_{x_0}^\phi c_{0y}(x,\,0)\,\text{d}x = - \left(\frac{9 \alpha (x_0+\phi)}{3\pi}\right)^{1/2}+...,
\end{equation}
which matches the total amount of surfactant adsorbed in the slip region in \eqref{eq:surf_ads}, as required by the conservation of surfactant.

\subsection{Weak Marangoni effect and weak bulk--surface exchange: region $A^\textsl{2D}_E$}
\label{subsec:MA_WE}

In region $A^\textsl{2D}_E$, we simplify the boundary-layer equations \eqref{eq:bl_1}--\eqref{eq:bl_6}, where bulk diffusion and bulk--surface exchange are weak compared to Marangoni effects and advection, i.e. $\alpha \ll 1$ and small $\nu$.
In this regime, there is a one-way coupling of the bulk concentration onto the interfacial concentration, with $c_0 \approx 1$ everywhere, and the interfacial surfactant distribution is in the classical stagnant-cap regime \citep{he1991size}, with a streamwise-averaged concentration imposed by the bulk concentration that is approximately equal to 1.  
Here, the surfactant gradient generated by the flow renders the downstream end of the interface no slip and the upstream end of the interface shear-free. 
We expand the surfactant field as follows
\refstepcounter{equation} \label{eq:am_expansion}
\begin{equation} 
    c_0 = 1 + \alpha C_1(x,\,y) + ..., \quad \Gamma_0 = G_0(x) + \alpha G_{1}(x) + ..., \tag{\theequation\textit{a,\,b}}
\end{equation}
and substitute \eqref{eq:am_expansion} into  \eqref{eq:bl_1}--\eqref{eq:bl_6}.
In the interfacial surfactant equation \eqref{eq:bl_2}, there is no flux of bulk surfactant onto the interface, such that $c_{0y}=0$. 
For $x\in[-\phi, \, \phi]$ and $y=0$, we have
\begin{equation} 
\label{eq:ame_2}
    \frac{3\beta}{4} G_{0} - \frac{\gamma}{2} G_{0x}G_{0} = 0, 
\end{equation} 
at leading order, and where we used the no-flux of interfacial surfactant condition \eqref{eq:bl_3}.
The surfactant flux condition (\ref{eq:bl_6}\textit{b}) in $\mathcal{D}_1$ with $c_0 \sim 1$ reduces to 
\begin{equation} 
\label{eq:ame_4} 
\int_{-\phi}^\phi G_{0} \,\text{d}x  = 2\phi.
\end{equation}
A piecewise linear solution to the nonlinear ordinary differential equation \eqref{eq:ame_2} exists for $2 \phi > 4\gamma / 3\beta$, such that in $A^\textsl{2D}_E$ the interfacial-surfactant distribution is given by 
\begin{equation} 
\label{eq:we_surf}
G_{0} =
\begin{cases}
    \ 0 \hspace{1.92cm} \text{for} \quad -\phi < x < x_0, \\ \displaystyle \ \frac{3\beta}{2\gamma} (x - x_0) \quad \text{for} \quad x_0 < x < \phi.
\end{cases}
\end{equation}
This piecewise linear stagnant-cap profile for the interfacial surfactant concentration agrees with our numerical results shown in figure~\ref{fig:11_b}(\textit{b}). 
We evaluate $x_0$ using the no-net-flux condition \eqref{eq:ame_4}, such that 
\begin{equation} \label{eq:we_x_0}
    x_0 = \phi - \left(\frac{8 \phi \gamma }{3\beta}\right)^{1/2}, 
\end{equation}
where $x_0>-\phi$ to ensure existence of the no-surfactant part of the profile in \eqref{eq:we_surf}.
Hence, for $2\phi > 4\gamma / 3\beta$, the leading-order drag reduction calculated by substituting \eqref{eq:we_surf} into \eqref{eq:dr_def} is given by
\begin{equation} 
\label{eq:dr_am}
    {DR}_0 = 1 - \left(\frac{2 \gamma }{3 \phi \beta}\right)^{1/2}. 
\end{equation}  
This solution for the $A^\textsl{2D}_E$ region, already given in (\ref{eq:a_we}\textit{c}), is equivalent to that in region $A^\textsl{1D}_E$ in the strong cross-channel diffusion problem \citep{tomlinson2023laminar}. It agrees with numerical results showed in figure~\ref{fig:11}.

\bibliographystyle{jfm}
\bibliography{jfm-instructions}

\end{document}